\documentclass[a4paper,fleqn,usenatbib]{mnras}

\usepackage{newtxtext,newtxmath}

\usepackage[T1]{fontenc}
\usepackage{ae,aecompl}

\usepackage{graphicx}	
\usepackage{amsmath}	
\usepackage{amssymb}	
\usepackage{bm}		
\usepackage{hyperref}
\usepackage{subfigure}
\usepackage{romannum}

\newcommand{\diff}{{\rm d}}

\title[Relativistic TDE disc]{Relativistic accretion disc in tidal disruption events}

\author[Mageshwaran \& Bhattacharyya]{
T. Mageshwaran$^{1}$\thanks{E-mail: t.mageshwaran@tifr.res.in}
and Sudip Bhattacharyya$^{1}$
\\
$^{1}$Department of Astronomy and Astrophysics, Tata Institute of Fundamental Research, Mumbai 400005, India
}

\date{Accepted 2020 June 03. Received 2020 May 29; in original form 2020 March 24}

\pubyear{2020}

\begin{document}
\label{firstpage}
\pagerange{\pageref{firstpage}--\pageref{lastpage}}
\maketitle


\begin{abstract}

We construct a time-dependent relativistic accretion model for tidal disruption events (TDEs) with an $\alpha-$viscosity and the pressure dominated by gas pressure. We also include the mass fallback rate $\dot{M}_f$ for both full and partial disruption TDEs, and assume that the infalling debris forms a seed disc in time $t_c$, which evolves due to the mass addition from the infalling debris and the mass loss via accretion onto the black hole. Besides, we derive an explicit form for the disc height that depends on the angular momentum parameter in the disc. We show that the surface density of the disc increases at an initial time due to mass addition, and then decreases as the mass fallback rate decreases, which results in a decrease in the disc mass $M_{\rm d}$ with a late-time evolution of $M_{\rm d} \propto t^{-1.05}$ and $M_{\rm d} \propto t^{-1.38}$ for full and partial disruption TDEs respectively, where $t$ is the time parameter. The bolometric luminosity $L$ shows a rise and decline that follows a power-law at late times given by $L \propto t^{-1.8}$ and $L \propto t^{-2.3}$ for full and partial disruption TDEs respectively. Our obtained luminosity declines faster than the luminosity inferred using $L \propto \dot{M}_f$. We also compute the light curves in various spectral bands.

\end{abstract}




\section{Introduction}

A tidal disruption event is a phenomenon where a star is shredded into debris when the black hole's tidal gravity exceeds the star's self-gravity \citep{1982ApJ...262..120L,1988Natur.333..523R}. The critical radius below which TDE occurs is the tidal radius $r_t= (M_{\bullet}/M_{\star})^{1/3} R_{\star}$, where $M_{\bullet}$ is the black hole mass, and $M_{\star}$ and $R_{\star}$ are the stellar mass and radius respectively \citep{1975Natur.254..295H,1976MNRAS.176..633F}. The fraction of debris bound to the black hole depends on the star's orbital energy and angular momentum before the disruption \citep{2015ApJ...814..141M}. The bound debris is assumed to follow a Keplerian orbit and returns to the pericenter with a mass fallback rate of $\dot{M}_f \propto t^{-5/3}$, where $t$ is the orbital period of the debris \citep{1989IAUS..136..543P,2009MNRAS.392..332L}. The infalling debris interacts with the outflowing debris resulting in the exchange of the angular momentum leading to the circularization and formation of an accretion disc \citep{2015ApJ...804...85S,2016MNRAS.455.2253B}. If the infalling debris loses its energy and angular momentum on a timescale shorter than the orbital period of debris, they are accreted on to the black hole with the mass accretion rate equal to the mass fallback rate, and thus the luminosity is then given by $L \propto \dot{M}_f \propto t^{-5/3}$ \citep{2002ApJ...576..753L,2009MNRAS.392..332L}. 

\citet{2009MNRAS.392..332L} developed the mass fallback model for the tidal disruption of stars with zero stellar rotation. \citet{2019ApJ...872..163G} included the stellar rotation and found that the prograde stellar spin increases the peak of mass fallback rate whereas the retrograde stellar spin decreases the peak of mass fallback rate. The late time evolution of mass fallback rate is close to $t^{-5/3}$ evolution. \citet{2013ApJ...767...25G} studied through numerical simulation the variation in mass fallback rate with penetration factor $\beta = r_t /r_p$, where $r_p$ is the pericenter of the stellar orbit. They found that there is a critical $\beta_d$ such that the stars are fully disrupted if $\beta > \beta_d$ and for a polytropic star, $\beta_d \sim 0.9~(\Gamma = 5/3)~{\rm and}~2.0~(\Gamma = 4/3)$, where $\Gamma$ is the polytrope index. For partial disrupted TDEs ($\beta < \beta_d$), the mass fallback rate declines faster than $t^{-5/3}$ and the declining rate varies with $\beta$. In partial TDEs, the fraction of star mass that forms the surviving core has an impact on the mass fallback rate. \citet{2019ApJ...883L..17C} constructed the mass fallback model from the Lagrangian dynamics and showed that the late time decline of mass fallback rate increases with an increase in the core mass. We use this simple model of partial TDEs in constructing our accretion dynamics with fallback.

A TDE provides an excellent opportunity to study the accretion phenomenon. A non-relativistic steady accretion model for TDE was constructed by \citet{2009MNRAS.400.2070S} with a mass accretion rate that follows the mass fallback rate. They also constructed a model for spherical outflows for a super-Eddington disc where the ratio of mass outflowing rate to mass accretion rate is constant. However, \citet{2015ApJ...814..141M} included the time-varying ratio of mass outflowing rate to mass accretion rate obtained using the simulation results of \citet{2011MNRAS.413.1623D}. \citet{2014ApJ...784...87S} constructed a model of a TDE disc with and without fallback from the disrupted debris. They developed a self-similar structure of a non-radiative, advective disc with an outflowing wind and used the self-similar solution of \citet{1990ApJ...351...38C} for a radiative thin disc with the constant total angular momentum. The self-similar solution of \citet{1990ApJ...351...38C} results in bolometric luminosity $L \propto t^{-1.2}$ and hence show a slower decline than $t^{-5/3}$ evolution. A time-dependent accretion model with a mass fallback and $\beta-$viscosity was constructed by \citet{2011ApJ...736..126M}, and they showed that the luminosity follows $t^{-5/3}$ dependence at late times. 

The ratio of tidal radius to black hole horizon ($\displaystyle{r_{\rm H}\propto G M_{\bullet}/c^2}$) is given by $\displaystyle{r_t/r_{\rm H} \propto M_{\bullet}^{-2/3}}$, and thus the relativistic effects can be significant for the higher mass black holes. The relativistic accretion model is also important in understanding the effect of black hole spin on TDE light curves. \citet{2018MNRAS.481.3348B} constructed a relativistic thin disc model for TDEs with a viscous stress function of radius and showed that the increase in black hole spin increases the luminosity declining rate. The obtained late time luminosity declining rate is given by $L \propto t^{n}$ with $n \in \{-1,~0\}$. Using an $\alpha-$viscosity to derive the viscous stress that depends on both surface density and radius, \citet{2019MNRAS.489..132M} obtained $n=-1.14$ for a Thompson opacity and $n=-1.25$ for a Kramers opacity, which are close to the solution of \citet{1990ApJ...351...38C}. These models have assumed that all the bound debris has formed an initial Gaussian disc which evolves due to the viscosity. \citet{2019MNRAS.483..565C} performed a GRMHD simulation to study the post-fallback super-Eddington accretion disc and showed that the non-jetted TDEs have optical/UV emissions from the torus, while the emission from heated wind peaks in UV/Soft X-rays. These accretion models have not included the effect of infalling debris on the accretion disc.

We construct a time-dependent relativistic accretion model with a source term due to the mass fallback at a constant outer radius, which provides a continuous mass supply (a similar assumption was considered for non-relativistic disc by \citet{2011ApJ...736..126M}). We consider a thin disc in the equatorial plane with an angular momentum of a circular orbit. The viscosity is assumed to be an $\alpha-$viscosity given by $\nu=\alpha c_s H$, where $\alpha$ is a constant, $c_s$ is the sound speed and $H$ is the disc height. We consider the pressure in the disc to be dominated by the gas pressure and using energy conservation, we derive the viscous stress. We assume that the infalling debris forms a seed accretion disc that evolves due to both mass fallback rate and viscous accretion onto the black hole. The time for the formation of an initial disc is taken to be a free parameter and is constrained by $\alpha$. We derive an explicit form for the vertical height which depends on the angular momentum and the radial velocity in the disc, and reduces to the result of \citet{1973A&A....24..337S} in the non-relativistic limit. We self-consistently obtain the initial and boundary conditions for the disc. We construct the accretion model for both full and partial disruption TDEs. Our solution satisfies the thin structure assumption and the luminosity shows a rise and decline. The late-time evolution of luminosity is roughly $L \propto t^{-1.8}$ for full disruption TDEs and $L \propto t^{-2.3}$ for partial disruption TDEs. The luminosity decline in our model is sharper than that obtained from an accretion model without fallback by \citet{2019MNRAS.489..132M}. 

In section \ref{tdd}, we review the TDE disruption dynamics and discuss the mass fallback rate for both full and partial disruption TDEs. We also discuss the circularization timescale and radius. Section \ref{add} presents our time-dependent accretion model where the basic assumptions and conditions are discussed. The detailed derivation of mass and momentum conservation equations are given in appendix \ref{dbe}. The evolution of the surface density obtained by solving the accretion dynamics and the corresponding evolution in disc mass and luminosity is shown in section \ref{result}. We discuss our results and their implications in section \ref{discus}. A summary of the paper and the conclusions are presented in section \ref{sac}.

\section{TDE disruption dynamics}
\label{tdd}

As the star enters in the tidal sphere, the black hole's tidal force exceeding the star's binding energy break the star and the star's fluid elements move on geodesic trajectories where the bound debris returns to the tidal radius following an orbit. \citet{2013MNRAS.435.1809S} showed that the debris energy freezes in at the tidal radius rather than the pericenter which is also confirmed through numerical simulation by \citet{2013ApJ...767...25G}. We consider the stellar orbital pericenter $r_p= r_t= (M_{\bullet}/M_{\star})^{1/3}R_{\star}$ in our calculations, and the debris returns to the pericenter following an orbit. We consider an impulse approximation where the star is assumed to be frozen until it reaches the pericenter where a short duration impulse of tidal potential disrupts the star \citep{2009MNRAS.392..332L}. The mass fallback rate of the debris depends on the black hole mass, star mass, the fraction of star mass bound to the black hole and the stellar rotation. For a star on a parabolic orbit, half of the star mass is bound to the black hole if the star is completely disrupted \citep{1988Natur.333..523R}. For a partial TDE, the amount of bound debris depends on the surviving core mass which also affects the late time power-law decline of mass fallback rate.

We consider three mass fallback models given by MFR1, MFR2 and MFR3.  The stars experience a full disruption in models MFR1 and MFR2. In MFR1, the stars have zero stellar spins at an initial time and the tidal interaction spin-up the stars, whereas in MFR2, the stars have non-zero stellar spin which may be prograde or retrograde and the tidal spin-up is neglected. The stars experience a partial disruption in MFR3 where the mass of surviving core affects the late time decline in the mass fallback rate. Next, we present the mass fallback rate for various TDE scenarios. 

\subsection{MFR1: full disruption of a star}
\label{fds}

We assume that the centre of star follows a nearly parabolic orbit and is disrupted at the pericenter $r_p$ taken to be $r_t$. The specific energy of the disrupted debris is governed by the variation of the black hole potential across the star and the tidal spin-up of the star as a result of the tidal interaction. The spin-up of a star due to the tidal interaction is a complex process \citep{1992ApJ...385..604K} and depends on the stellar structure. \citet{2001ApJ...549..948A} through a linear perturbation theory showed that the spin-up angular velocity ($\omega_s$) via tidal interaction for a solar mass main-sequence star with a polytrope of $1.5$ is  $\omega_s \approx 0.86 \omega_p$, where $\omega_p$ is the angular velocity of the star at the pericenter. They also showed via numerical simulations that the nonlinear effects lead to a larger energy transfer from the orbit to the star and a larger spin-up ($\omega_s \approx \omega_p$) than predicted by linear theory. By including the tidal interaction, \citet{2002ApJ...576..753L} formulated the energy of the disrupted debris given by $E_{\rm d}= -k G M_{\bullet} \Delta R/ r_t^2$, where the tidal spin up factor is $k=1$ for no tidal spin up and $k=3$ for tidal spin up. In the non-relativistic limit, the debris follows a Keplerian orbit and the time period of innermost debris is given by \citep{1988Natur.333..523R}

\begin{equation}
t_m= 40.8~{\rm days}~ M_6^{1/2} m^{1/5} k^{-3/2},
\label{tmt}
\end{equation}

\noindent where $M_6= M_{\bullet}/[10^6 M_{\odot}]$, $m=M_{\star}/M_{\odot}$ and the radius of star $R_{\star}=R_{\odot} m^{0.8}$ \citep{1994sse..book.....K}. The mass fallback rate is given by \citep{1989IAUS..136..543P}

\begin{equation}
\dot{M}_f = \frac{1}{3} (2 \pi G M_{\bullet})^\frac{2}{3}\frac{\diff M}{\diff E_{\rm d}} t^{-\frac{5}{3}},
\label{mfb}
\end{equation}  

\noindent where $M$ is the debris mass with energy $E_{\rm d}$. The $\diff M/\diff E_{\rm d}$ depends on the density distribution within the star and following an impulse approximation \citep{2009MNRAS.392..332L}, it is given by

\begin{equation}
\frac{\diff M}{\diff E_{\rm d}}= \frac{M_{\star} r_t^2}{k G M_{\bullet} R_{\star}}2 \pi b \int_{x}^{1} \theta^{u}(x') x' \, \diff x',
\label{med}
\end{equation}

\noindent where $\theta(x)$ is the solution of Lane-Emden equation for a polytropic index $\Gamma= 1+ 1/u$, $b$ is the ratio of central to mean density of the star \citep{1943ApJ....97..255C} and $x=\Delta R / R_{\star}= \tau^{-2/3}$ with $\tau = t/t_m$. Thus, the mass fallback rate is given by

\begin{equation}
\dot{M}_f= \frac{4 \pi  b}{3} \frac{M_{\star}}{t_m} \tau^{-5/3} \int_{x}^{1} \theta^{u}(x') x' \, \diff x'.
\label{mfbn}
\end{equation}

\noindent The integral is nearly constant at late times and thus the mass fallback rate follows $t^{-5/3}$ decline. We include the tidal spin up contribution by taking $k=3$ and the polytropic index $\Gamma= 5/3$ that results in $u=3/2$. The relativistic effect of black holes may have an impact on the mass fallback rate.

\citet{2012PhRvD..86f4026K} studied the relativistic effect on the mass fallback rate  and found that the peak mass fallback rate increases by a factor of two with an increase in the black hole spin. The late-time evolution of the mass fallback rate is similar to the equation (\ref{mfbn}). A similar study for TDE by a Schwarzschild black hole is done by \citet{2014PhRvD..90f4020C} and showed that the late time profile for the main sequence star is close to the Newtonian $t^{-5/3}$ law. A relativistic smoothed particle hydrodynamic simulation by \citet{2019MNRAS.487.4790G} showed that the disrupted debris in deep encounters (pericenter is at a few event horizon radii) experiences strong precessions and the fluid can be launched into a plunging orbit that results in a decrease in the fallback rate. Thus, the relativistic effects have an impact on the peak mass fallback rate but the late time profile is similar to a Keplerian rate. Here, we do not consider the TDEs with deep encounters, and assume the mass fallback rate given by equation (\ref{mfbn}) in our calculations, as the relativistic effects have a small contribution to the mass fallback rate. 

\subsection{MFR2: full disruption of a star with stellar rotation}
\label{fdsr}

In the previous section \ref{fds}, we have not included the effect of initial stellar rotation on the mass fallback. \citet{2019ApJ...872..163G} have shown that the peak of mass fallback rate increases for a prograde stellar spin and decreases for a retrograde stellar spin. We use their impulse approximation model to obtain the mass fallback rate. If the stellar spin is along the $z$-direction, the energy of the disrupted debris is given by \citet{2019ApJ...872..163G}

\begin{equation}
E_{\rm d} = - (1+ \sqrt{2} \lambda) \frac{G M_{\bullet} \Delta R}{r_t^2},
\label{edmfr2}
\end{equation}

\noindent where the stellar rotation frequency $\Omega = \lambda \sqrt{G M_{\star}/R_{\star}^3}$ and $\lambda$ is taken to be a constant. The positive $\Omega$ corresponds to an alignment between the orbital angular momentum vector of the stellar centre of mass and the angular velocity of the star. Following a Keplerian orbit, the time period of innermost debris is given by  

\begin{equation}
t_m= 40.8~{\rm days}~ (1+ \sqrt{2} \lambda)^{-3/2} M_6^{1/2} m^{1/5}.
\label{tmtr}
\end{equation}

\noindent Thus, for a prograde stellar spin, the energy of the debris increases which increases the peak of mass fallback rate. Following the procedure similar to that in section \ref{fds}, the mass fallback rate is given by equation (\ref{mfbn}), where $t_m$ is given by equation (\ref{tmtr}).

\subsection{MFR3: partial disruption of a star}
\label{pds}

\begin{figure}
\begin{center}
\subfigure[]{\includegraphics[scale=0.25]{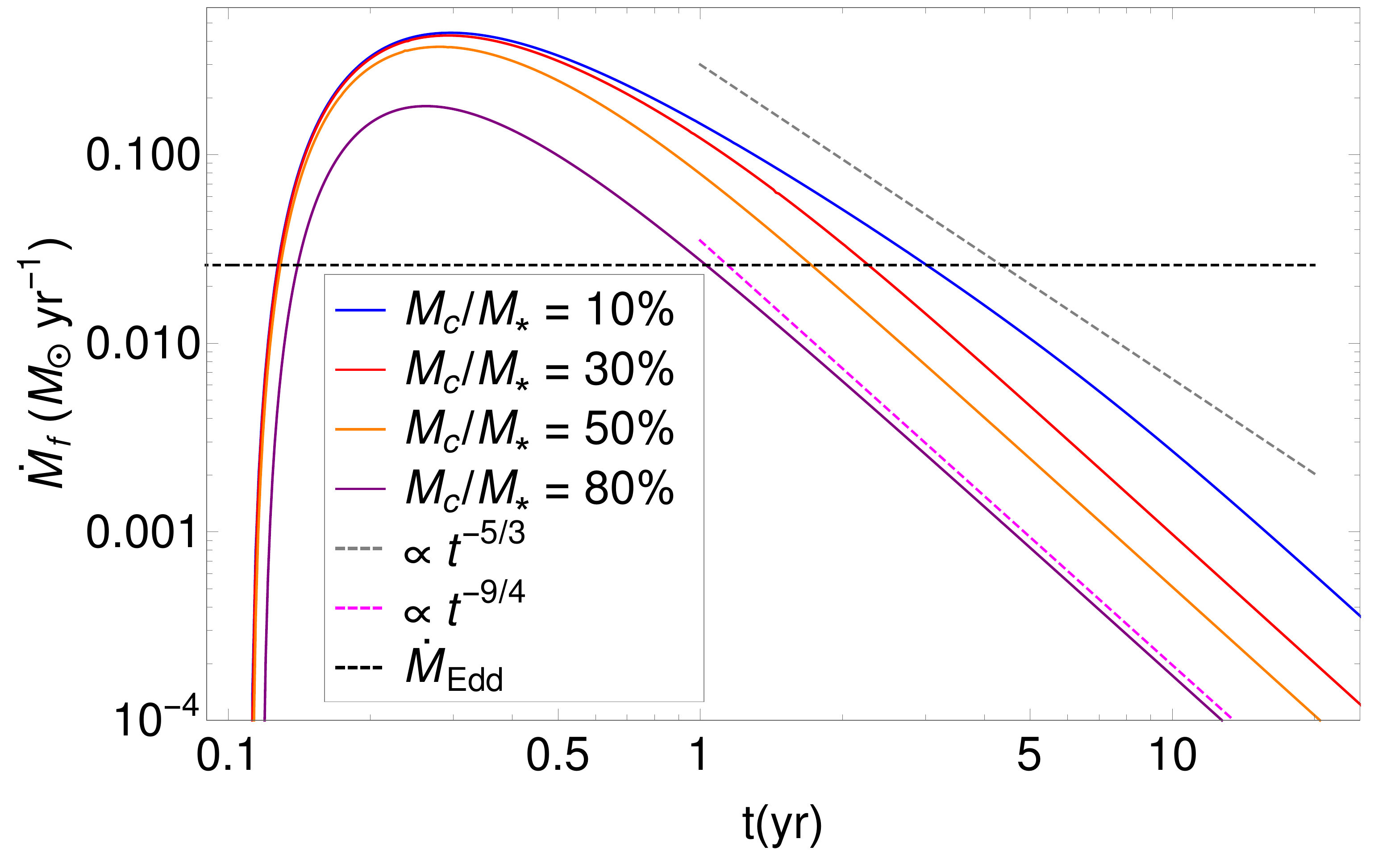}}
\subfigure[]{\includegraphics[scale=0.27]{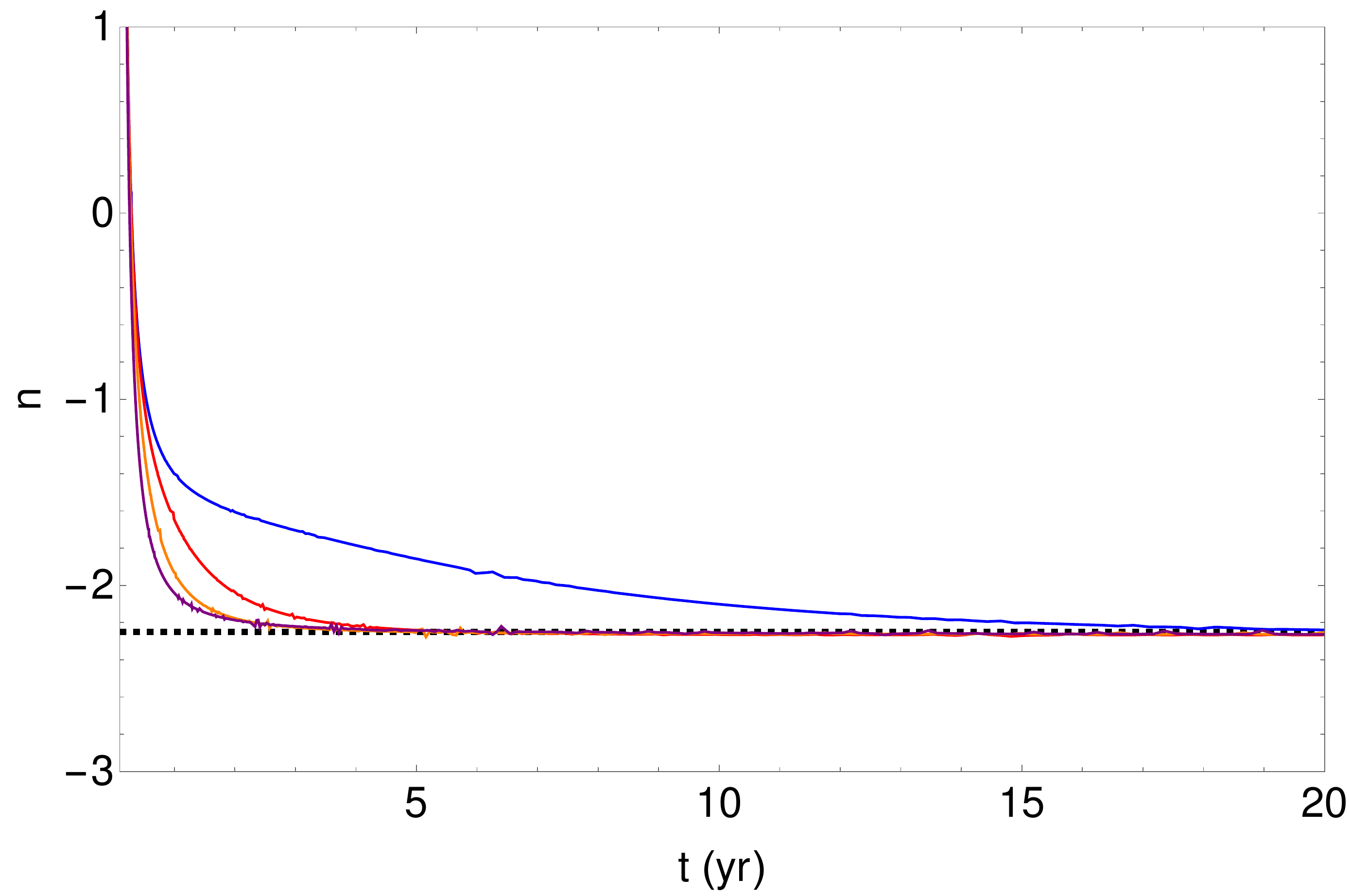}}
\end{center}
\caption{(a) The mass fallback rate obtained for a partial TDE with black hole mass given by $M_{\bullet}=10^6 M_{\odot}$, star mass given by $M_{\star}=M_{\odot}$ and for various values of the core mass $M_c$ is shown. The gray and magenta dashed lines correspond to the late time evolution of $t^{-5/3}$ and $t^{-9/4}$ respectively \citep{2019ApJ...883L..17C}. The Eddington mass accretion rate (black dashed line) is calculated for a radiative efficiency of 0.1. (b) The time evolution of power-law index of the mass fallback rate given by $n = \diff \ln(\dot{M}_f)/\diff \ln(t)$ is shown for the mass fallback rate given in (a). The black dashed line corresponds to $n_{\infty} \sim -9/4$. See section \ref{pds} for more details. }
\label{mfpTDE}
\end{figure}

In the previous sections (\ref{fds}, \ref{fdsr}), we assume that the star is completely disrupted. However, in certain cases, a star can be partially disrupted where the surviving core induces an additional potential to the debris orbit and results in a power-law decline steeper than that for full disruption. \citet{2013ApJ...767...25G} through numerical simulations showed that the late time evolution of mass fallback rate for a partial TDE is $\sim t^{-2.2}$. \citet{2019ApJ...883L..17C} developed a model using Lagrangian dynamics for the fallback of tidally disrupted debris from a partial TDE and they obtained the late time mass rate to be $\dot{M}_f \propto t^{n_{\infty}}$, where $n_{\infty}$ in case of $\mu_c \leq 10^{-5}$ is given by

\begin{equation}
n_{\infty}= -2.257 - 0.690 \mu_c^{1/3} + \mathcal{O}(\mu_c^{2/3}),
\label{ninf}
\end{equation}

\noindent where $\mu_c = M_c/M_{\bullet}= 10^{-6} (M_c/M_{\star})~ m~ M_{6}^{-1}$ and $M_c$ is the mass of the surviving core. The dependence of $n_{\infty}$ on the core mass is weak as $\mu_c \sim 10^{-6}$ and decreases with an increase in the black hole mass. The $\mu_c$ increases with the stellar mass but is still small. Hence, the first term in equation (\ref{ninf}) dominates and the $n_{\infty} \sim -2.26 \sim -9/4$. The approximated late time evolution of the mass fallback rate is $\dot{M}_f \propto t^{-9/4}$. We follow the method outlined in \citet{2019ApJ...883L..17C} to calculate the mass fallback rate. The mass fallback rate obtained is shown in Fig. \ref{mfpTDE}a and the time evolution of power-law index of the mass fallback rate given by $n = \diff \ln(\dot{M}_f)/\diff \ln(t)$ is shown in Fig. \ref{mfpTDE}b. The $n \approx n_{\infty} \sim -9/4$ at late times implies the fallback rate is $\dot{M}_f \propto t^{-9/4}$.

The disrupted debris circularizes to form an accretion disc and the outer radius of the formed disc is taken to be the circularization radius. For a star on a parabolic orbit, the specific angular momentum at the pericenter is $J= \sqrt{2 G M_{\bullet} r_t}$, and the specific angular momentum for a circular orbit is $J_c= r^2 \omega= \sqrt{G M_{\bullet} r_c}$, where $\omega=\sqrt{G M_{\bullet}/r_c^3}$ is the Keplerian angular frequency and $r_c$ is the circularization radius. By conserving the specific angular momentum of the debris, the circularization radius is $r_c= 2 r_t$ \citep{1999ApJ...514..180U,2009MNRAS.400.2070S}, and is used by \citet{2011ApJ...736..126M} as the outer radius to construct their time-dependent accretion model. \citet{2016MNRAS.455.2253B} showed through numerical simulation around a non-rotating black hole for a full disruption TDE that the debris stream circularizes to form a narrow ring at $r_c$ which may be greater than $2 r_t$ if there is an exchange in angular momentum during circularization and the time for circularization is $1-10$ times the orbital period of the star. A numerical simulation around rotating black holes by \citet{2016MNRAS.461.3760H} showed that the circularization time increases with an increase in the black hole prograde spin and the circularization radius is close to the Newtonian radius $r_c$. An increase in the angular momentum exchange rate due to stream interactions could lead to a higher circularization radius. Since the outer radius of the formed disc is close to $r_c= 2 r_t$, and to avoid ambiguity in calculating $r_c$, we consider $r_c= 2 r_t$ as the constant outer radius in our calculation where the infalling mass is added to the disc. The effect of a surviving core in a partial TDE on the circularization of the debris is uncertain. So, for the simplicity, we assume that the debris circularizes similar to that for full disruption TDEs with the circularization radius $r_c= 2 r_t$.
In the next section, we discuss our relativistic accretion model.

\section{Accretion disc dynamics}
\label{add}

We transform the metric in Boyer-Lindquist coordinate $\{t,~r,~\theta,~\phi\}$ to a cylindrical coordinate $\{t,~R,~\phi,~z\}$, using $z=r \cos\theta$ and $R=r \sin \theta$, and solve the mass and momentum conservation equations. \citet{2018MNRAS.481.3348B} have also considered the cylindrical coordinate to construct their accretion model without a fallback. In our accretion model, we are taking the vertical flow to be zero and this assumption is simple to apply for equations in the cylindrical coordinate. The mathematical derivation of the equations is given in appendix \ref{dbe}. We employ their reduced form near the equatorial plane, as the disc equations involve vertically averaged quantities since the disc we are considering is not thick. It is also simple to derive an explicit form for the disc height using the vertical momentum equation in cylindrical coordinate. The time evolution of surface density $\Sigma$ for a thin disc ($H/R \ll 1$, where $H$ is the disc height) with circular angular momentum is given by  

\begin{equation}
\frac{\partial \Sigma}{\partial t} = \frac{c}{G M_{\bullet} R_g} \frac{\Delta_k^{1/2}}{A_k^{1/2}} \frac{1}{\gamma_L} \frac{\partial}{\partial x} \left[\left(\frac{\partial \ell_{\rm K}}{\partial x}\right)^{-1} \frac{\partial}{\partial x} \left(x \bar{S}_{\phi}^R\right)\right],
\label{beq}
\end{equation}

\noindent where $x=R/R_g$, $R_g=G M_{\bullet}/c^2$, $\Delta_k= x^2-2 x+j^2$, $A_k= x^4 + x^2 j^2 +2 x j^2$, angular momentum $\mathcal{L} = \ell (G M_{\bullet}/c)$, $j$ is the black hole spin, $\ell_{\rm K}$ is dimensionless angular momentum of a circular orbit, and the vertically averaged viscous stress tensor is

\begin{equation}
\bar{S}_{\phi}^R= -R_g \left(\frac{G M_{\bullet}}{c R_g^2}\right) \nu \Sigma \frac{\Delta_k^{1/2} A_k^{3/2} \gamma_L^3}{x^5} \frac{\partial \omega_1}{\partial x},
\label{sphir}
\end{equation}

\noindent where $\omega_1$ is given by equation (\ref{omega1}). In a thin disc geometry, we assume that the velocity is dominated by azimuthal flow which is subsonic, and the radial velocity defined in a co-rotating frame \citep{1996ApJ...471..762A}, $V \ll 1$, such that the equation (\ref{gammald}) reduces to $\gamma_L^2 = 1 + x^2 \ell^2 / A_k$. The angular momentum of a circular orbit is given by \citep{1972ApJ...178..347B}

\begin{equation}
\ell_{\rm K}= \pm \frac{x^2 \mp 2 j \sqrt{x} + j^2}{x^{3/4} (x^{3/2}-3 \sqrt{x} \pm 2 j)^{1/2}},
\label{llk}
\end{equation}

\noindent and its derivative is given by

\begin{equation}
\frac{\partial \ell_{\rm K}}{\partial x} = \frac{(x^2 - 6 x \pm 8 j \sqrt{x} -3 j^2)(j \pm x^{3/2})}{2 x^{7/4} (x^{3/2}-3 \sqrt{x} \pm 2 j)^{3/2}},
\label{dllk}
\end{equation}

\noindent where the upper sign is for prograde orbits and the lower sign represents retrograde orbits. We solve the vertical momentum equation neglecting any force due to radiative energy flux and after vertical integration, the height of the disc is given in appendix \ref{htd} and in terms of dimensionless variables in equation (\ref{heightd}). Using $\ell_{\rm K}$ from equation (\ref{llk}), the height of the disc is given by

\begin{equation}
H = \left(\frac{R_g}{c}\right)^2 \frac{2 P}{\Sigma} x^3 \left[1- \frac{3}{x} \pm \frac{2 j}{x^{3/2}} \right] \left[1 \mp \frac{4 j}{x^{3/2}} + \frac{3 j^2}{x^2}\right]^{-1}.
\label{heigd}
\end{equation}  

\noindent For a non-rotating black hole, spin $j=0$, which results in $H = (R_g/c)^2 (2 P/\Sigma) x^3 (1- 3/x)$ and for $x \gg 3$, it results in $H = (R_g/c)^2 (2 P/\Sigma) x^3$. The speed of sound is given by $c_s = \sqrt{P/\rho}= \sqrt{2 H P/\Sigma} $, and by incorporating $P/\Sigma$, it reduces to $c_s= H \sqrt{G M_{\bullet}/R^3}$, which is the solution for a non-relativistic thin disc \citep{1973A&A....24..337S,2002apa..book.....F}. 

The viscous stress in the accretion disc results in heating of the disc. The viscous process in the disc leads to an exchange of the angular momentum and thus an inflow of matter leading to accretion by black holes, and the viscous mechanism in the accretion disc is affected by the pressure in the disc. The dominant form of viscosity in the disc is an $\alpha-$viscosity which is due to local turbulence and for a subsonic flow, it is given by $\nu= \alpha c_s H$. Assuming a non-relativistic fluid approximation where the internal energy is neglected compared to the fluid mass energy \citep{1996ApJ...471..762A}, and the flows are subsonic, the dissipation rate $Q^{+}$ along the equatorial plane after vertical integration is given by

\begin{equation}
Q^{+}=\frac{G^2 M_{\bullet}^2}{c^2 R_g^4} \frac{\gamma_L^4 A_k^2}{x^6} \left(\frac{\partial \omega_1}{\partial x}\right)^2 \nu \Sigma.
\label{qheat}
\end{equation} 

\noindent The heat generated via viscous heating is emitted as radiation with flux $Q^{-}=Q^{+}$, where the radiative flux $Q^{-}= 8 \sigma_{\rm SB} T^4/(3 \kappa \Sigma)$ \citep{2002apa..book.....F}, where $\sigma_{\rm SB}$ is the Stefan-Boltzmann constant and $\kappa$ is the opacity in the medium which is taken to be the Thomson opacity given by $\kappa=0.34~{\rm cm^{2}~g^{-1}}$. We assume the pressure in the disc to be the gas pressure $P =  k_B \rho T/(\mu_m m_p)$, where $k_B$ is the Boltzmann constant, $\mu_m$ is the mean molecular weight taken to be ionized solar mean molecular weight of $0.65$, $m_p$ is the mass of a proton and $T$ is the temperature in the disc. Following the energy conservation and an $\alpha-$viscosity, the pressure in the disc is given by 

\begin{multline}
P = \frac{1}{2}\left(\frac{3 \kappa}{8 \sigma_{\rm SB}}\right)^{\frac{1}{6}} \left(\frac{k_B}{\mu_m m_p}\right)^{\frac{2}{3}} \left(\frac{c}{R_g}\right)^{\frac{5}{6}} \left(\frac{G M_{\bullet}}{c R_g^2}\right)^{\frac{1}{3}} \alpha^{\frac{1}{6}} \frac{\gamma_L^{2/3} A_k^{1/3}}{x} \left|\frac{\partial \omega_1}{\partial x}\right|^{\frac{1}{3}} \\ \chi_1(x,~j)^{\frac{5}{12}} \Sigma^{\frac{4}{3}},
\label{press}
\end{multline}

\noindent where $\displaystyle{\chi_1(x,~j)= x^{-3}\left[1 \mp \frac{4 j}{x^{3/2}} + \frac{3 j^2}{x^{2}}\right] \left[1- \frac{3}{x} \pm \frac{2 j}{x^{3/2}}\right]^{-1}}$. Using equation (\ref{heigd}), the disc height is given by $H \propto \Sigma^{1/3}$ and thus decreases with a decrease in the surface density.

Using the pressure given by equation (\ref{press}), the disc height given by equation (\ref{heigd}) and the $\alpha-$viscosity, the viscous stress $\bar{S}_{\phi}^R$ using equation (\ref{sphir}) is given by

\begin{equation}
\bar{S}_{\phi}^R= - S_c~ g(x)~ \Sigma^{5/3},
\label{visst}
\end{equation}

\noindent where 

\begin{eqnarray}
S_c &=& R_g \left(\frac{c}{R_g}\right)^{-\frac{4}{3}} \left(\frac{G M_{\bullet}}{c R_g^2}\right)^{\frac{5}{3}} \left(\frac{3 \kappa}{8 \sigma_{\rm SB}}\right)^{\frac{1}{3}} \left(\frac{k_B}{\mu_m m_p}\right)^{\frac{4}{3}} \alpha^{\frac{4}{3}}, \label{suv} \\
g(x) &=& \frac{\Delta_k^{1/2} A_k^{13/6} \gamma_L^{13/3} \chi_1^{-2/3}}{x^7} \left|\frac{\partial \omega_1}{\partial x}\right|^{\frac{2}{3}} \frac{\partial \omega_1}{\partial x}.
\end{eqnarray}

\noindent By substituting the viscous stress given by equation (\ref{visst}) in equation (\ref{beq}), the surface density evolution is given by

\begin{equation}
\frac{\partial \Sigma}{\partial t} = - \frac{c~ S_c}{G M_{\bullet} R_g} Y_1 \frac{\partial }{\partial x}\left[Y_2 \frac{\partial}{\partial x}\left(Y_3 \Sigma^{5/3}\right)\right],
\label{sigev}
\end{equation}

\noindent where $Y_1= \Delta_k^{1/2}/[\gamma_L A_k^{1/2}]$, $Y_2= (\partial \ell_{\rm K}/ \partial x)^{-1}$ and $Y_3 = x ~g(x)$. Our considered disc has a mass supply from the infalling debris at the outer radius, which is taken to be the circularization radius, and in the next section, we present our methodology to incorporate this mass addition to the disc. 

\subsection{Disc with mass supply}
\label{dms}

The debris of a tidally disrupted star returns to the pericenter following a Keplerian orbit with a mass fallback rate given by equation (\ref{mfbn}) for a full disruption TDE (section \ref{fds}) and the mass fallback rate for a partial TDE is discussed in section \ref{pds}. If $t=0$ is the moment of disruption, the mass fallback of innermost bound debris occurs at a time $t=t_m$, where $t_m$ in case of full disruption TDEs is given by equations (\ref{tmt}, \ref{tmtr}) for mass fallback models MFR1 and MFR2 respectively and is obtained numerically for partial disruption TDEs. The infalling debris forms an initial accretion disc in time $t_c$ with disc mass $M_{\rm d}(t_c) = \int_{t_m}^{t_c} \dot{M}_f \, \diff t$. The evolution of the formed disc is governed by both accretion onto the black hole and the mass addition at the constant outer radius. The surface density of the formed disc is assumed to be Gaussian and is given by

\begin{equation}
\Sigma_i \equiv \Sigma(t_c,~x) = \Sigma_0 \exp\left[- \frac{(x-x_{c})^2}{\varpi}\right],
\label{sigi}
\end{equation}  

\noindent where $\sqrt{\varpi}$ is the full width half maximum of the distribution, $\Sigma_0$ is the constant and $x_c = r_c/R_g$. The area of disc in $\{R,~\phi\}$ plane is given by $\displaystyle{\diff \mathcal{A} = \sqrt{g_{\rm R R} g_{\rm zz}} \diff R \diff \phi}$, and from the equations (\ref{grr}, \ref{gpp}) calculated at the equatorial plane, we have $\diff \mathcal{A}= R_g^2 \sqrt{A_k/\Delta_k} \diff x \diff \phi $. By integrating equation (\ref{sigi}) and equating with the disc initial mass $M_{\rm d}(t_c)$, we obtain

\begin{eqnarray}
\Sigma_0 &=& \frac{M_{\rm d} (t_c)}{2 \pi R_g^2 I(x_{\rm in},~x_c,~\varpi)}, \label{sig0}\\
I(x_{\rm in},~x_c,~\varpi) &=& \int_{x_{\rm in}}^{x_c} \exp\left[- \frac{(x-x_{c})^2}{\varpi} \right] \sqrt{\frac{A_k}{\Delta_k}} \, \diff x,
\end{eqnarray}

\noindent where $x_{\rm in}= R_{\rm in}/R_g$, $R_{\rm in}=R_g Z(j)$ is the innermost stable circular orbit, and $Z(j)$ is given by \citep{1972ApJ...178..347B}:

\begin{equation}
Z(j) =3+Z_2(j)-\sqrt{(3-Z_1(j)) (3+Z_1(j)+2 Z_2(j))}, \label{zjb}
\end{equation}

\noindent where

\begin{subequations}
\begin{align}
Z_1(j) &=1+(1-j^2)^{\frac{1}{3}} \left[(1+j)^{\frac{1}{3}}+(1-j)^{\frac{1}{3}}\right]\\
Z_2(j) &=\sqrt{3 j^2+Z_1(j)^{2}}.
\end{align}
\end{subequations} 

The mass accretion rate is given by $\dot{M}= - 2 \pi c R_g \Sigma V \sqrt{\Delta_k}$ \citep{2019MNRAS.483..565C}, and using equation (\ref{svdk}), we have

\begin{equation}
\dot{M}=- \frac{2 \pi c~ R_g~S_c}{G M_{\bullet}} Y_2 \frac{\partial}{\partial x}\left(Y_3 \Sigma^{5/3}\right). 
\label{mdot}
\end{equation} 

\noindent We assume the mass accretion rate at the outer radius to be equal to the mass fallback rate and to avoid any discontinuity in surface density due to mass addition, we consider the derivative of surface density at the outer radius to be $\partial \Sigma /\partial x |_{x_c} = 0$. These conditions ensure that the accretion disc at the outer radius evolves smoothly. Then, using equation (\ref{mdot}), the surface density at the outer radius at any time $t$ is given by

\begin{equation}
\Sigma_c(t)= \left[\frac{- \dot{M}_f(t)~ c}{2 \pi S_c (Y_2 \partial_x Y_3)_{x_c}}\right]^{3/5}.
\label{sxc}
\end{equation} 

\noindent The surface density of the formed disc at outer radius is $\Sigma_i(x_{c})= \Sigma_0$ and by equating it with equation (\ref{sxc}) at time $t_c$, we get $\Sigma_0=\Sigma_c(t_c)$. Using equation (\ref{mfbn}) of full disruption, the surface density of the disc at outer radius decreases as $t^{-1}$ at late times. We have $\varpi$ and $S_c$ as the two unknown parameters, and the equations (\ref{sig0}, \ref{sxc}) give us a relation between them that is given by

\begin{equation}
S_c^{3/5}= \frac{2 \pi R_g^2}{M_{\rm d}(t_c)} I(x_{\rm in},~x_c,~\varpi)  \left[\frac{- c~ \dot{M}_f(t_c)}{2 \pi (Y_2 \partial_x Y_3)_{x_c}}\right]^{3/5}.
\label{sucal}
\end{equation}

\noindent The surface density given by equation (\ref{sigi}) needs to satisfy the surface density evolution equation, and thus by substituting equation (\ref{sigi}) in equation (\ref{sigev}) and using equation (\ref{sucal}), we obtain at $x=x_c$, 

\begin{multline}
\frac{3}{5}\frac{M_d(t_c)}{\dot{M}_f^2(t_c)} \left( \frac{\partial \dot{M}_f}{\partial t}\right)_{t_c} = \frac{I(x_{\rm in},~x_c,~\varpi)}{(Y_2 \partial_x Y_3)_{x_c}} Y_1(x_c) \left[\frac{\partial}{\partial x}\left(Y_2 \frac{\partial Y_3}{\partial x}\right)_{x_c}- \right. \\  \left. \frac{10}{3 \varpi} Y_2(x_c) Y_3(x_c)\right].
\label{varcal}
\end{multline}
 
\noindent By solving equation (\ref{varcal}) for a given $t_c$, we obtain $\varpi$ which results in $S_c$ using equation (\ref{sucal}). The $S_c$ is a function of black hole mass and $\alpha$ as can be seen from equation (\ref{suv}) and using the obtained $S_c$, we get $\alpha$. Thus, we self-consistently obtain the initial width of the disc and the constant $\alpha$ which gives us the pressure and viscosity. The $\alpha-$viscosity requires $\alpha \leq 1$ and this condition gives a constrain on $t_c$. We take only those $t_c$ for which $\alpha \leq 1$. With the initial condition given by equation (\ref{sigi}) and the boundary conditions at the outer radius given by $\partial \Sigma/\partial x|_{x_c}=0$ and $\Sigma_c(t)$, we solve the evolution equation (\ref{sigev}) to obtain the surface density.  

The heat generated due to viscosity is emitted as radiation with flux $Q^{-}=Q^{+}$, where $Q^{+}$ is given by equation (\ref{qheat}) and the effective temperature of the disc is given by $T_{\rm eff}= ( Q^{-}/\sigma_{\rm SB})^{1/4}$. The mean effective temperature obtained by integrating over disc radius is given by $\left<T_{\rm eff}\right> = \int T_{\rm eff} \, \diff \mathcal{A}/ \int \, \diff \mathcal{A}$, where $\diff \mathcal{A}$ is given below equation (\ref{sigi}). We do not consider a complicated radiative transfer model and assume a simple disc blackbody approximation to calculate the luminosity. The bolometric luminosity is given by 

\begin{multline}
L(t) = 2 \pi R_g^2 \left(\frac{c}{R_g}\right)^{\frac{4}{3}} \left(\frac{3 \kappa}{8 \sigma_{\rm SB}}\right)^{\frac{1}{3}} \left(\frac{k_B}{\mu_m m_p}\right)^{\frac{4}{3}} \alpha^{\frac{4}{3}} \cdot  \\ \int_{x_{\rm in}}^{x_c}  \frac{\gamma_L^{16/3} A_k^{19/6} \chi_1^{-2/3}(x,~j)}{x^8 \Delta_k^{1/2}} \left|\frac{\partial \omega_1}{\partial x}\right|^{\frac{8}{3}}  \Sigma^{5/3}  \, \diff x.
\label{lumeq}
\end{multline}

\noindent The relativistic effects dominate close to the inner radius and for an extended disc, most of the emissions are from the Newtonian disc region. To obtain the spectral evolution in various bands, we follow the blackbody Planck function and integrate it over the disc radius and frequency bands. In the next section, we show our results for various physical parameters. 

\section{Results}
\label{result}

The accretion disc model we have discussed in the previous section has free parameters that are black hole mass $M_{\bullet}$ and spin $j$, stellar-mass $M_{\star}$ and spin fraction constant $\lambda$, and circularization time $t_c$. We now present the results of our modelling for the three mass fallback rates discussed in section \ref{tdd}. 

\subsection{MFR1: full disruption of a star}
\label{resmfr1}

\begin{figure}
\begin{center}
\includegraphics[scale=0.34]{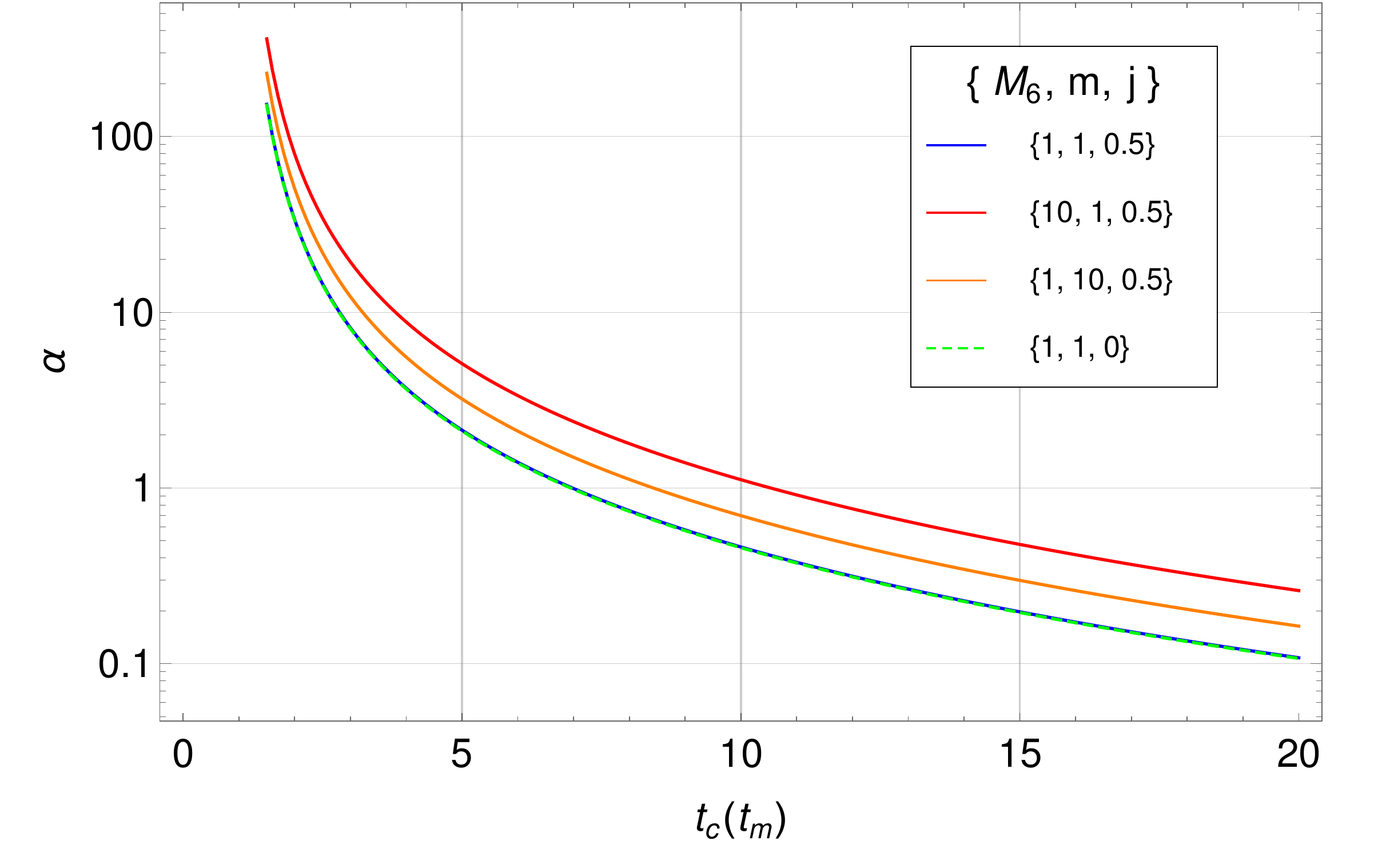}
\end{center}
\caption{The evolution of $\alpha$ as a function of $t_c$ for the mass fallback rate given in model MFR1 (section \ref{fds}). The blue and the green lines are overlapping which implies that $\alpha$ shows an insignificant variation with black hole spin $j$. The thin disc model with an $\alpha-$viscosity is valid only if $\alpha \leq 1$. See section \ref{resmfr1} for details. }
\label{alfds}
\end{figure}

\begin{figure}
\begin{center}
\subfigure[]{\includegraphics[scale=0.3]{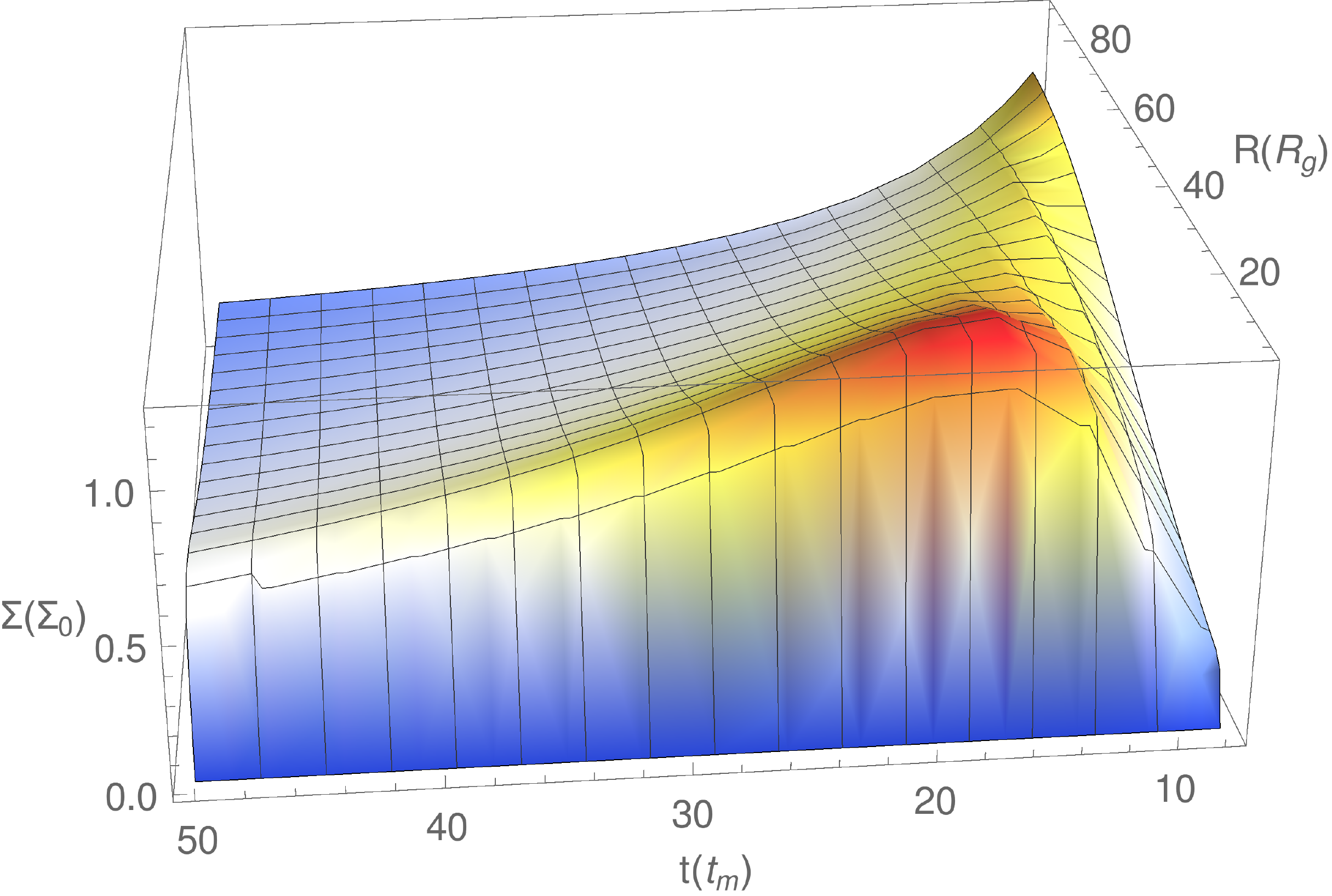}}
\subfigure[]{\includegraphics[scale=0.28]{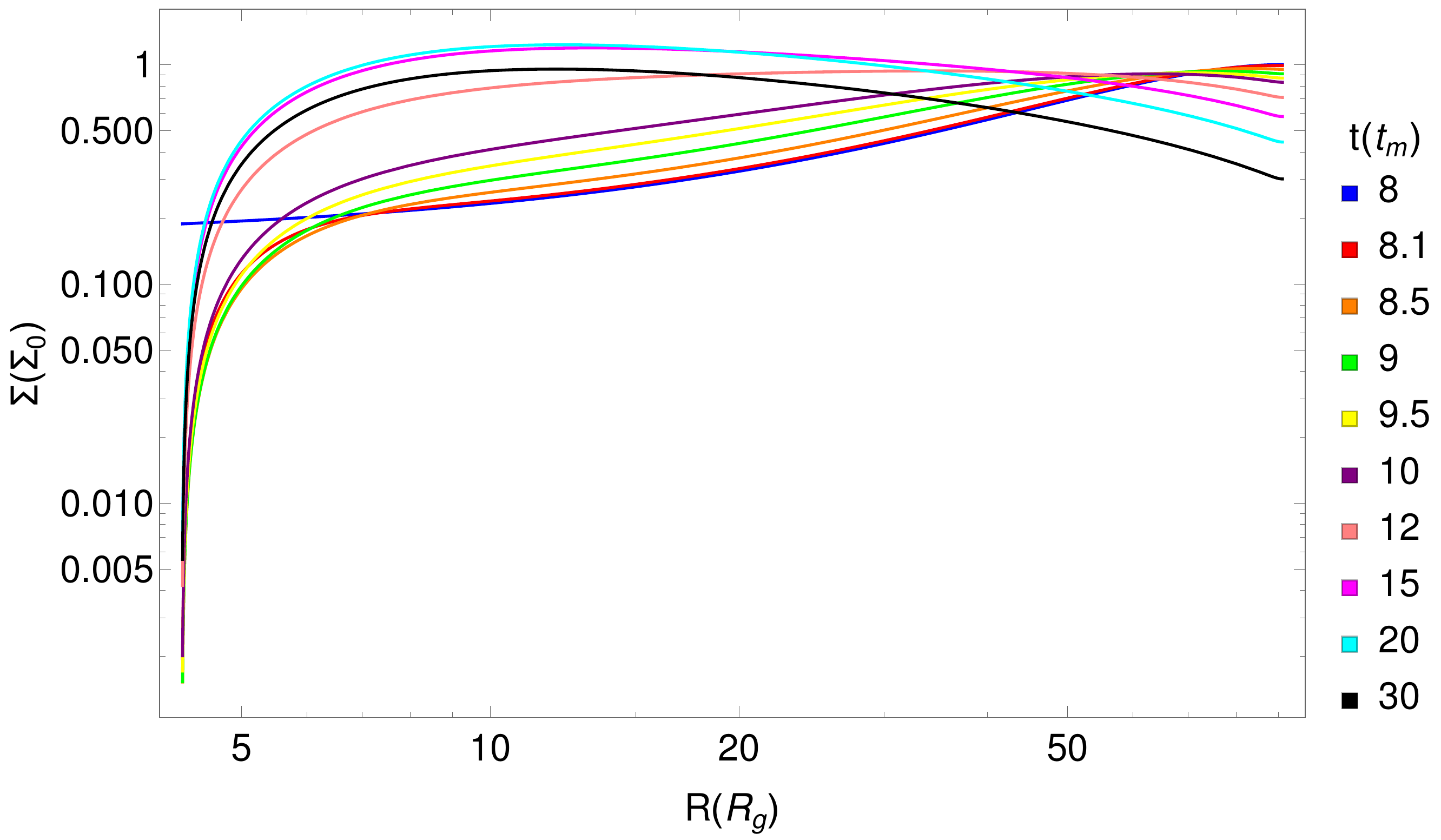}}
\end{center}
\caption{The time evolution of surface density obtained by solving equation (\ref{sigev}) for the mass fallback model MFR1 (section \ref{fds}) and the parameter set M\Romannum{1} (section \ref{resmfr1}) is shown as contour plot in (a) and for various times in (b). The normalizing constant $\Sigma_0=1.0 \times 10^{6}~{\rm g~cm^{-2}}$ and $t_m=7.82~{\rm days}$. The circularization time is taken to be $t_c = 8 t_m$.  See section \ref{resmfr1} for details. }
\label{surf}
\end{figure}

\begin{figure}
\begin{center}
\includegraphics[scale=0.3]{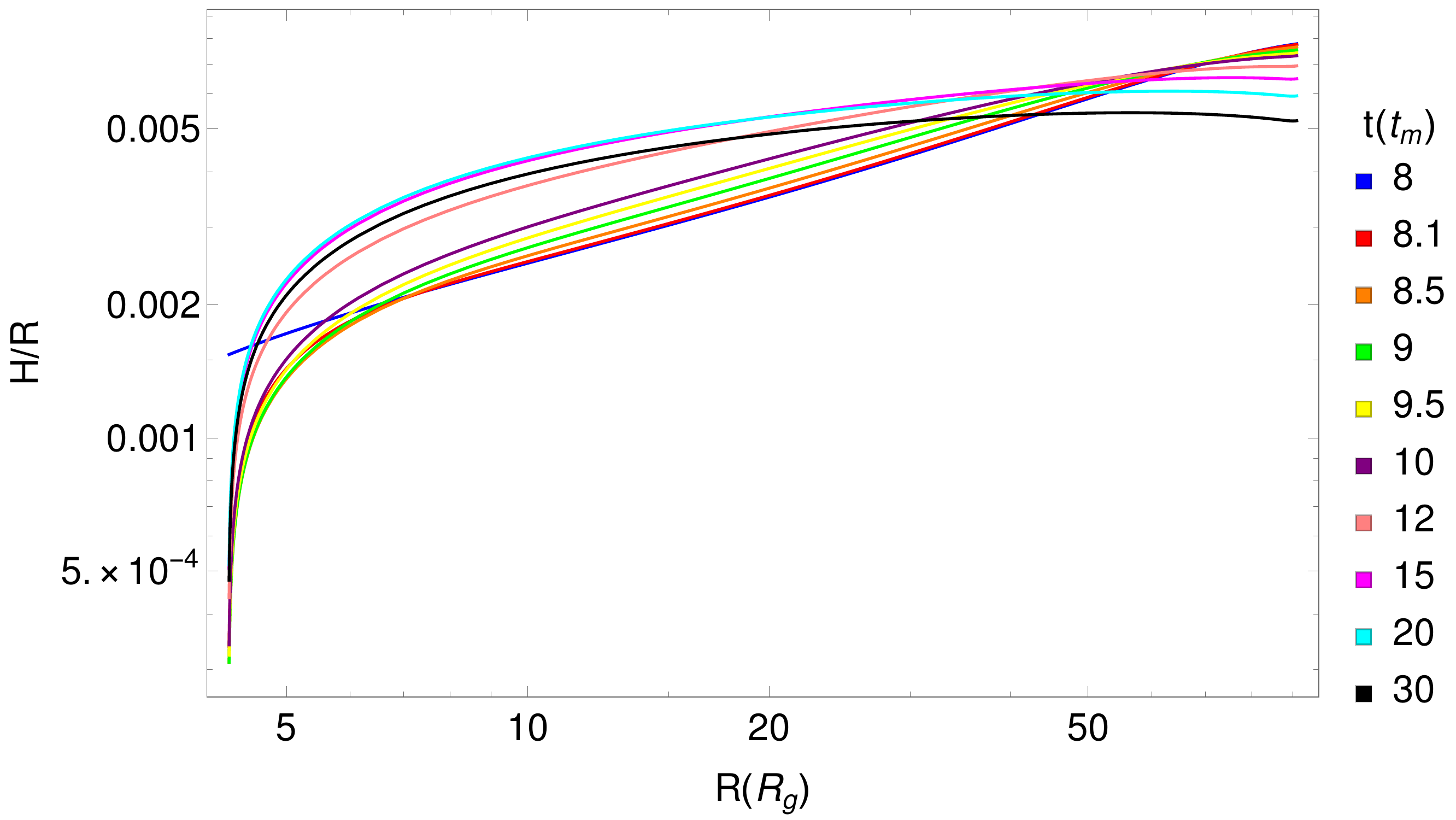}
\end{center}
\caption{The time evolution of the disc height ($H$) is shown at various times for the mass fallback model MFR1 (section \ref{fds}) and the parameter set M\Romannum{1} (section \ref{resmfr1}). The normalizing constant $t_m=7.82~{\rm days}$ and the $H/R \ll 1$ fulfills our initial assumption of thin disc (section \ref{add}). For each curve in Fig. \ref{surf}b, the radius profile of $H/R$ is given. The disc height $H \propto \Sigma^{1/3}$, and thus a decrease in the surface density with time results in a decrease in the disc height. See section \ref{resmfr1} for details. }
\label{hrp}
\end{figure}

\begin{figure}
\begin{center}
\includegraphics[scale=0.32]{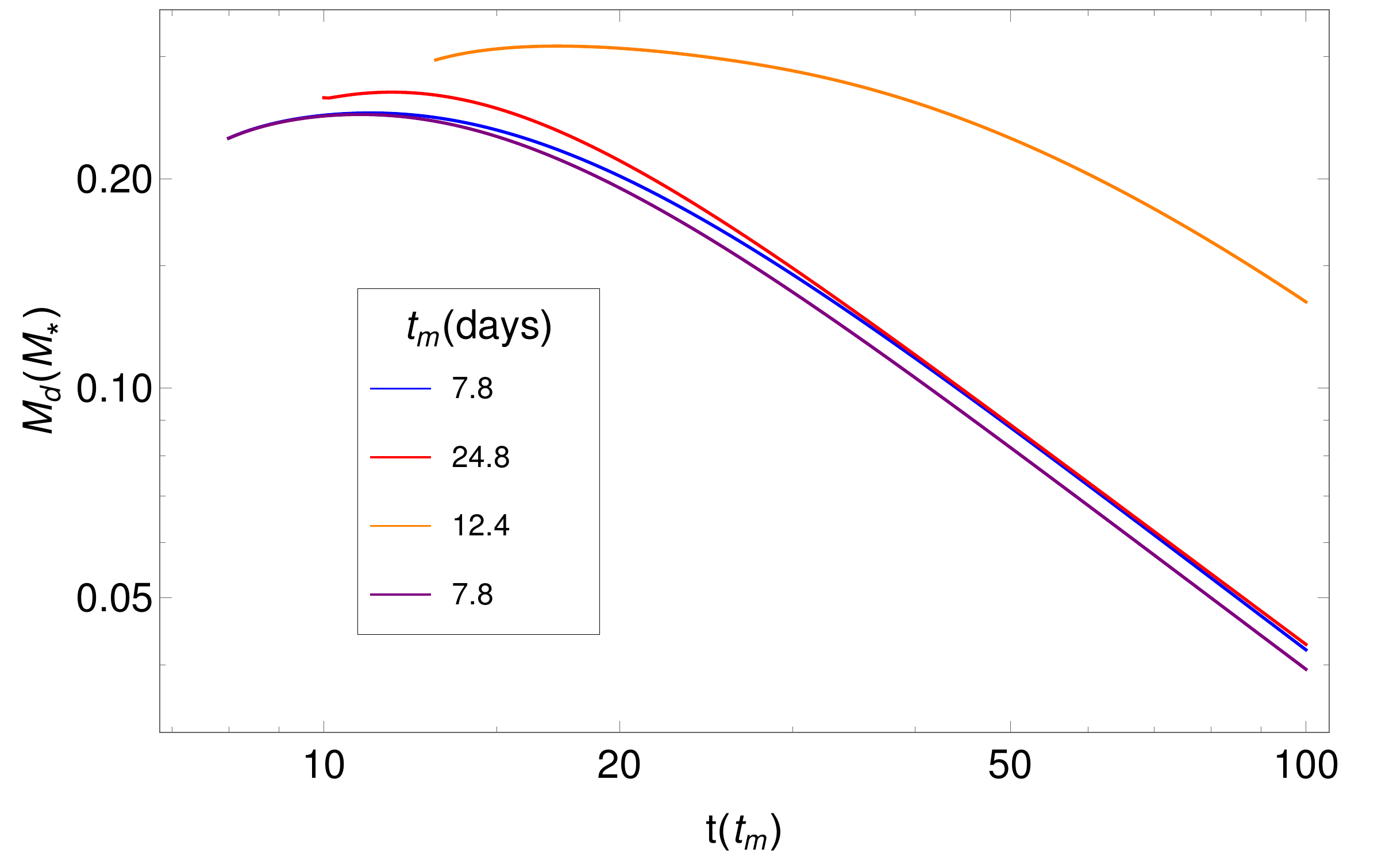}
\end{center}
\caption{The time evolution of the disc mass is shown for the mass fallback model MFR1 (section \ref{fds}). The blue line represent the parameter set M\Romannum{1} (section \ref{resmfr1}) whose surface density evolution is shown in Fig. \ref{surf}. The physical parameter that varies with red, orange and purple lines are $M_6 =10$, $m=10$ and $j =0$ respectively. While changing one parameter, we have kept all other parameters same that take the values of parameter set M\Romannum{1}. The critical time corresponding to $\alpha =1$ varies with black hole mass and star mass as can be seen from Fig \ref{alfds}. Taking in account $\alpha <1$, we take $t_c (t_m) = 8~{\rm (blue~and~purple~ curves)},~10~{\rm (red~ curve)},~13~{\rm (orange~ curve)}$. The late time evolution of disc mass is $M_{\rm d} \propto t^{p}$ where $p=-1.05~{\rm (blue)},~-1.04~{\rm (red)},~-1.1~{\rm (orange)},~{\rm and}~-1.043~{\rm (purple)}$. See section \ref{resmfr1} for details. }
\label{mdis} 
\end{figure}

\begin{figure}
\begin{center}
\subfigure[]{\includegraphics[scale=0.29]{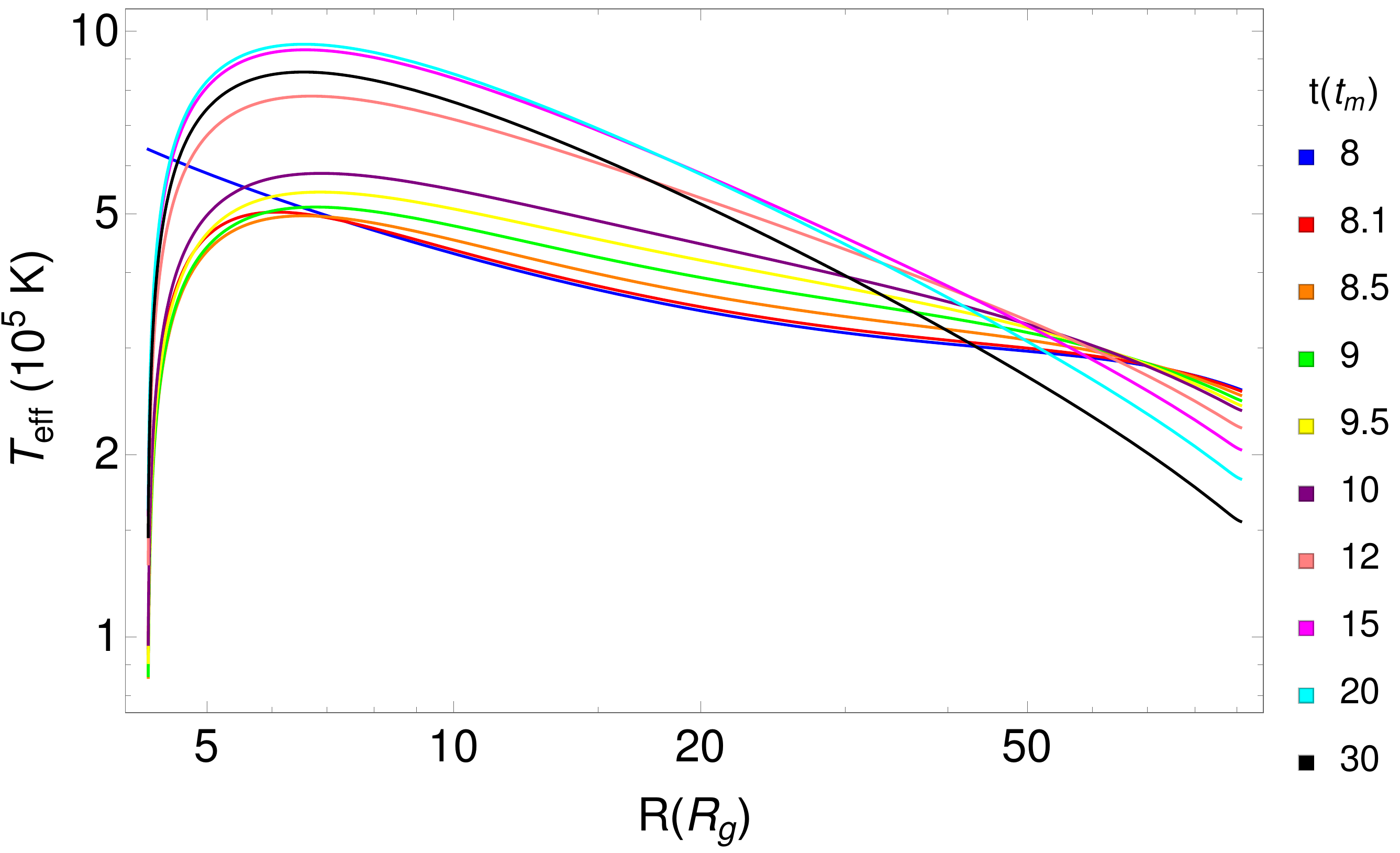}}
\subfigure[]{\includegraphics[scale=0.43]{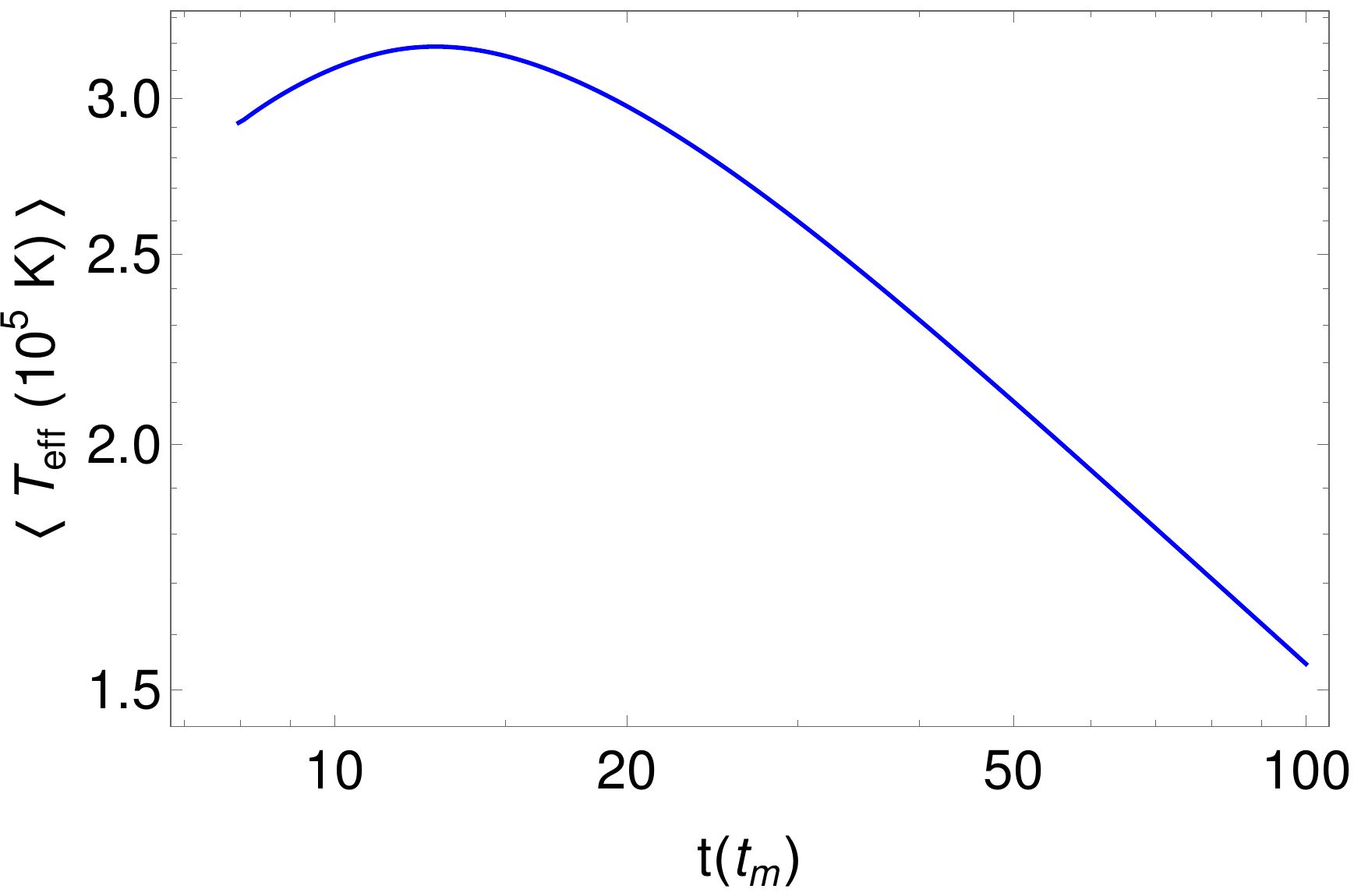}}
\end{center}
\caption{The time evolution of the effective temperature obtained for the mass fallback model MFR1 (section \ref{fds}) and the parameter set M\Romannum{1} (section \ref{resmfr1}) is shown in (a) and the mean effective temperature obtained by integrating over disc is shown in (b). The orbital time period of innermost debris is $t_m = 7.82~{\rm days}$ and the corresponding surface density is shown in Fig. \ref{surf}. The late time evolution of disc mean temperature is $\left<T_{\rm eff}\right> \propto t^{-0.43}$. See section \ref{resmfr1} for details. }
\label{temp}
\end{figure}

In case of full disruption of stars and the mass fallback given by MFR1 (section \ref{fds}), we name the prior parameter set to be M\Romannum{1} given by $\{M_6,~m,~j\} = \{1,~1,~0.5\}$, where $M_6= M_{\bullet}/[10^6 M_{\odot}]$ and $m= M_{\star}/M_{\odot}$. This results in an outer radius $x_c=90.7$ and the orbital period of innermost bound debris $t_m=7.86~{\rm days}$. We have presented in section \ref{dms} that the value of $\alpha$ depends on $t_c$ and we have shown in Fig \ref{alfds} that $\alpha$ decreases with $t_c$. Thus, there is a critical value of $t_c$ given by $t_{\rm cr}$ corresponding to $\alpha =1$ and the model is valid only for $t_c > t_{\rm cr}$. The critical time $t_{\rm cr}$ varies with black hole mass and star mass. For the parameter set M\Romannum{1}, we take $t_c/t_m = 8$ which results in $t_c = 62.8~{\rm days}$. The time evolution of the surface density in the disc is shown in Fig. \ref{surf}, where it increases with time initially and then decreases. The height to radius ratio of the disc is shown in Fig. \ref{hrp} and the $H/R \ll 1$  satisfies our initial assumption of thin disc considered to derive the equation (\ref{beq}). The time evolution of disc mass is shown in Fig. \ref{mdis} and the late time evolution is given by $M_{\rm d} \propto t^{p}$, where $p=-1.05~{\rm (blue)},~-1.04~{\rm (red)},~-1.1~{\rm (orange)},~{\rm and}~-1.043~{\rm (purple)}$. The effective temperature and the mean effective temperature of the disc are shown in Fig. \ref{temp}. The effective temperature peaks close to the inner radius and increases at an initial time due to an increase in the surface density and decreases later. The mean temperature shows a rising and a declining phase which implies that the bolometric luminosity increase and then decreases at later times.

\begin{figure}
\begin{center}
\subfigure[M\Romannum{1}a]{\includegraphics[scale=0.28]{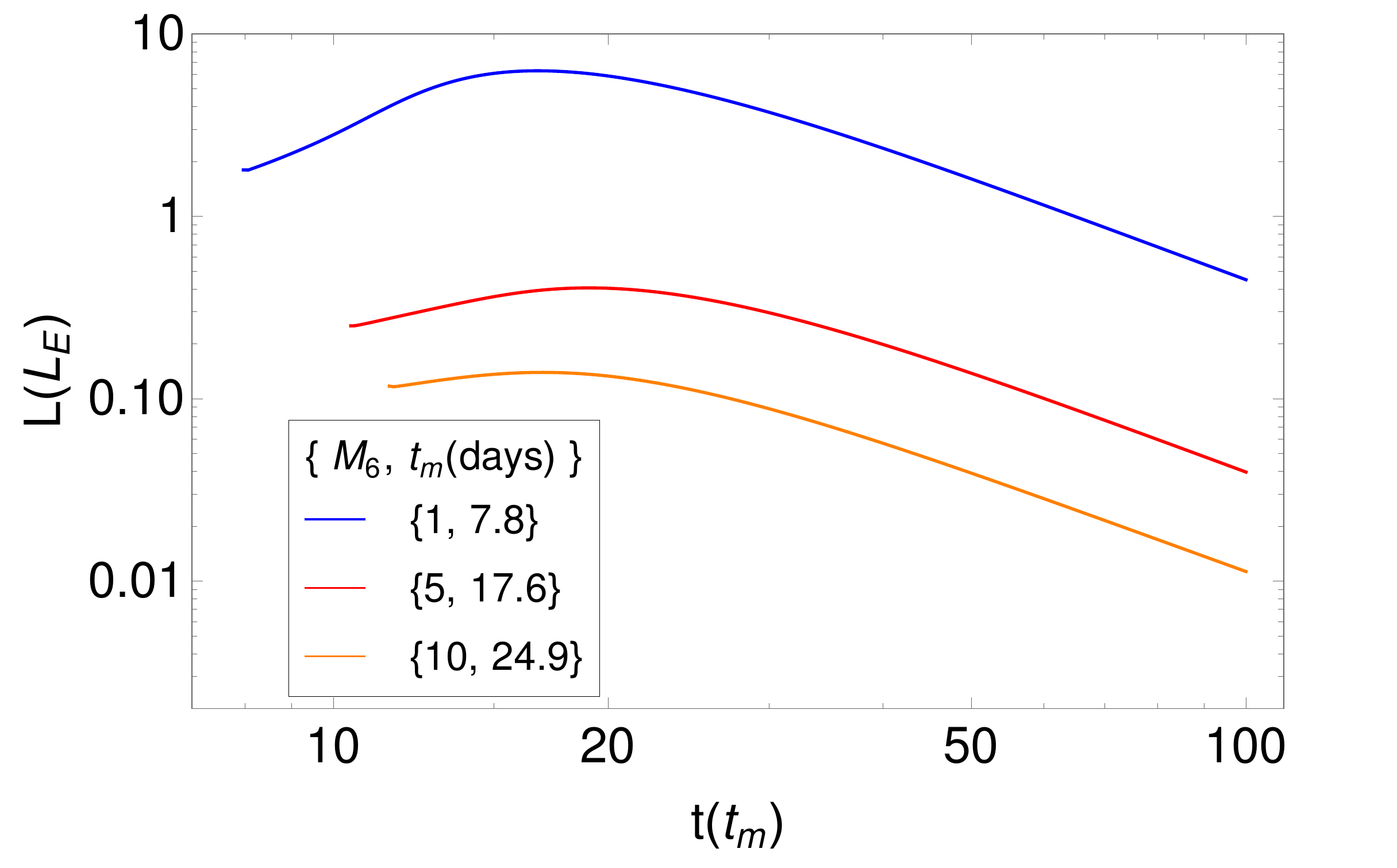}}
\subfigure[M\Romannum{1}b]{\includegraphics[scale=0.28]{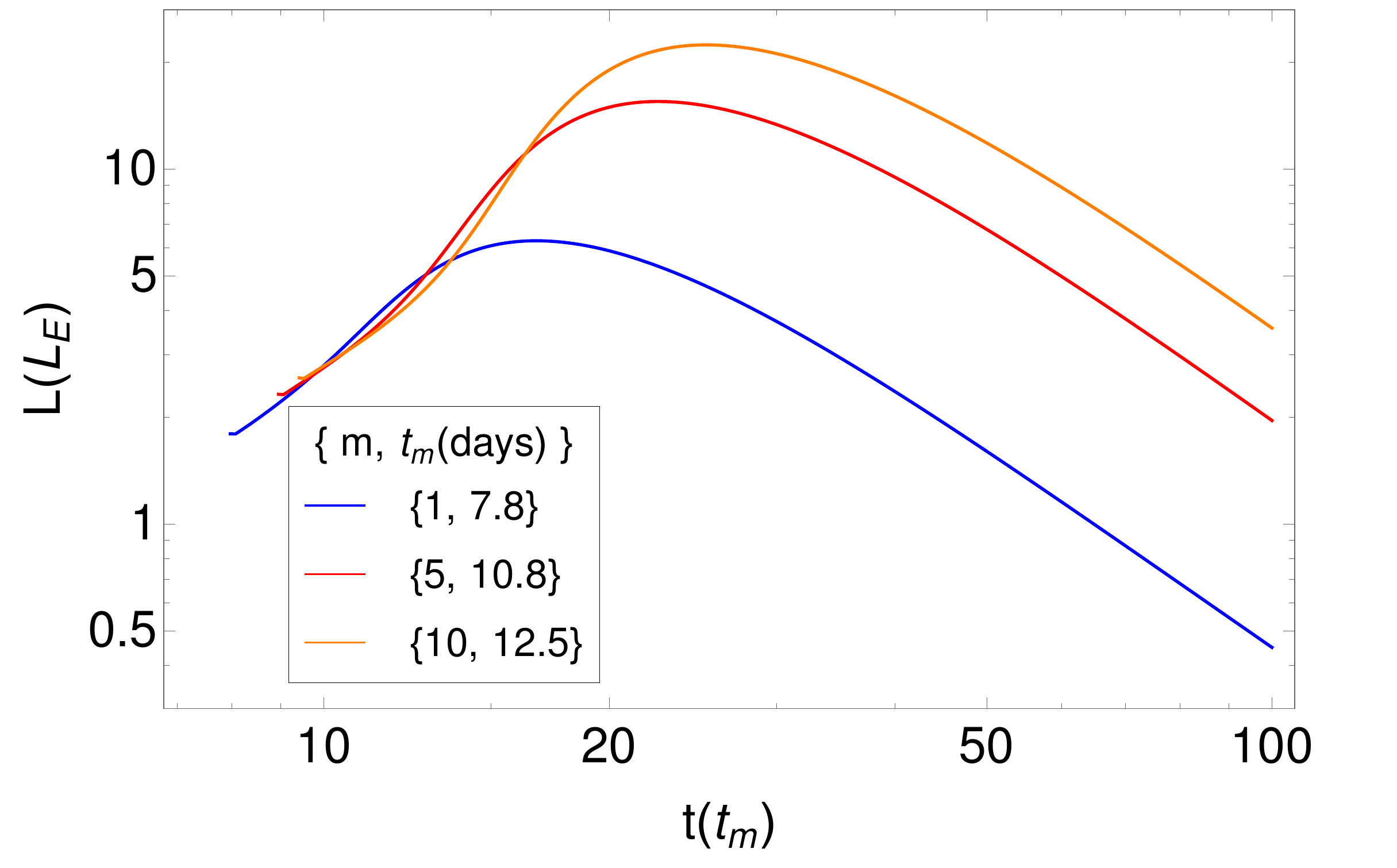}}
\subfigure[M\Romannum{1}c]{\includegraphics[scale=0.28]{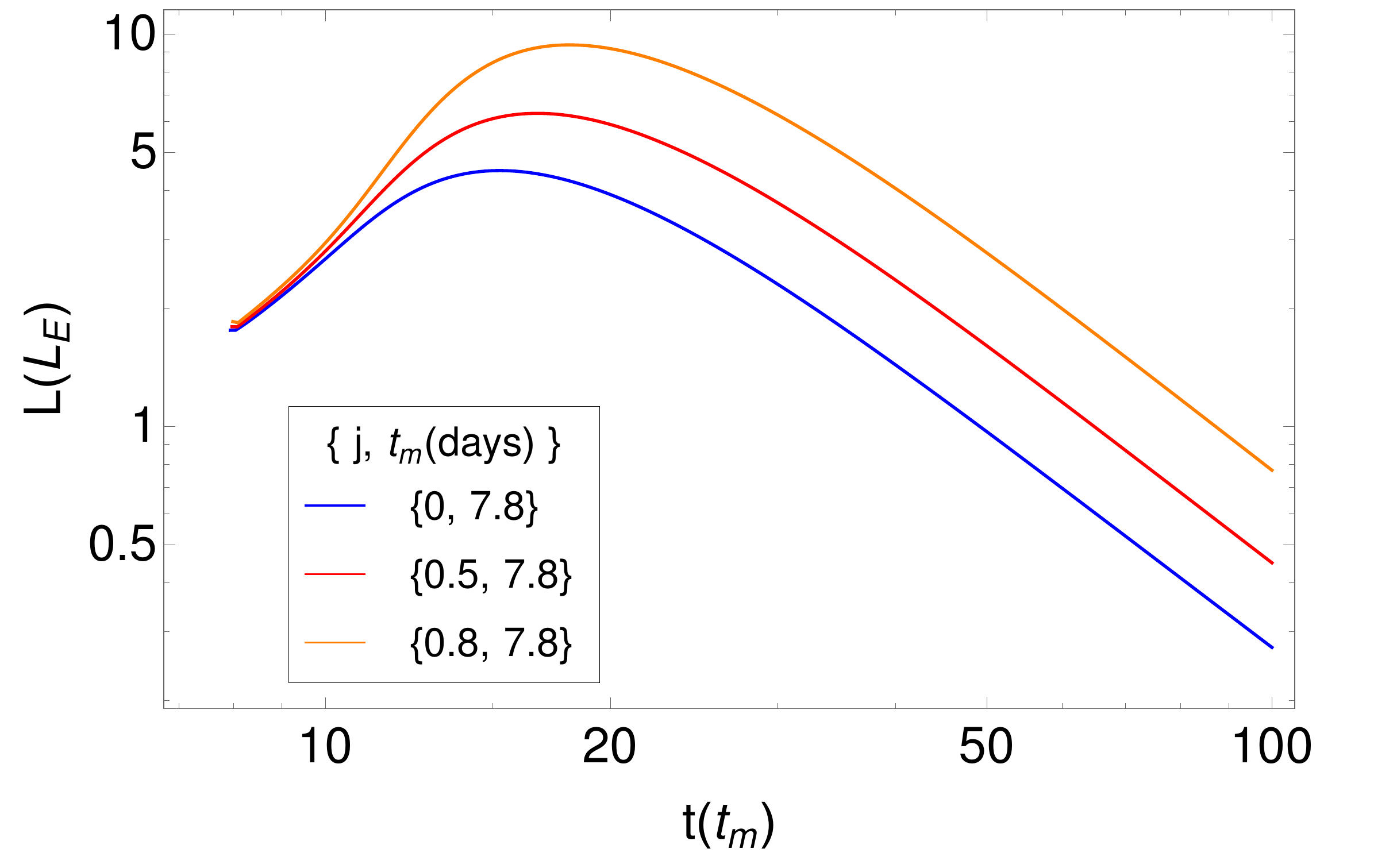}}
\subfigure[M\Romannum{1}d]{\includegraphics[scale=0.28]{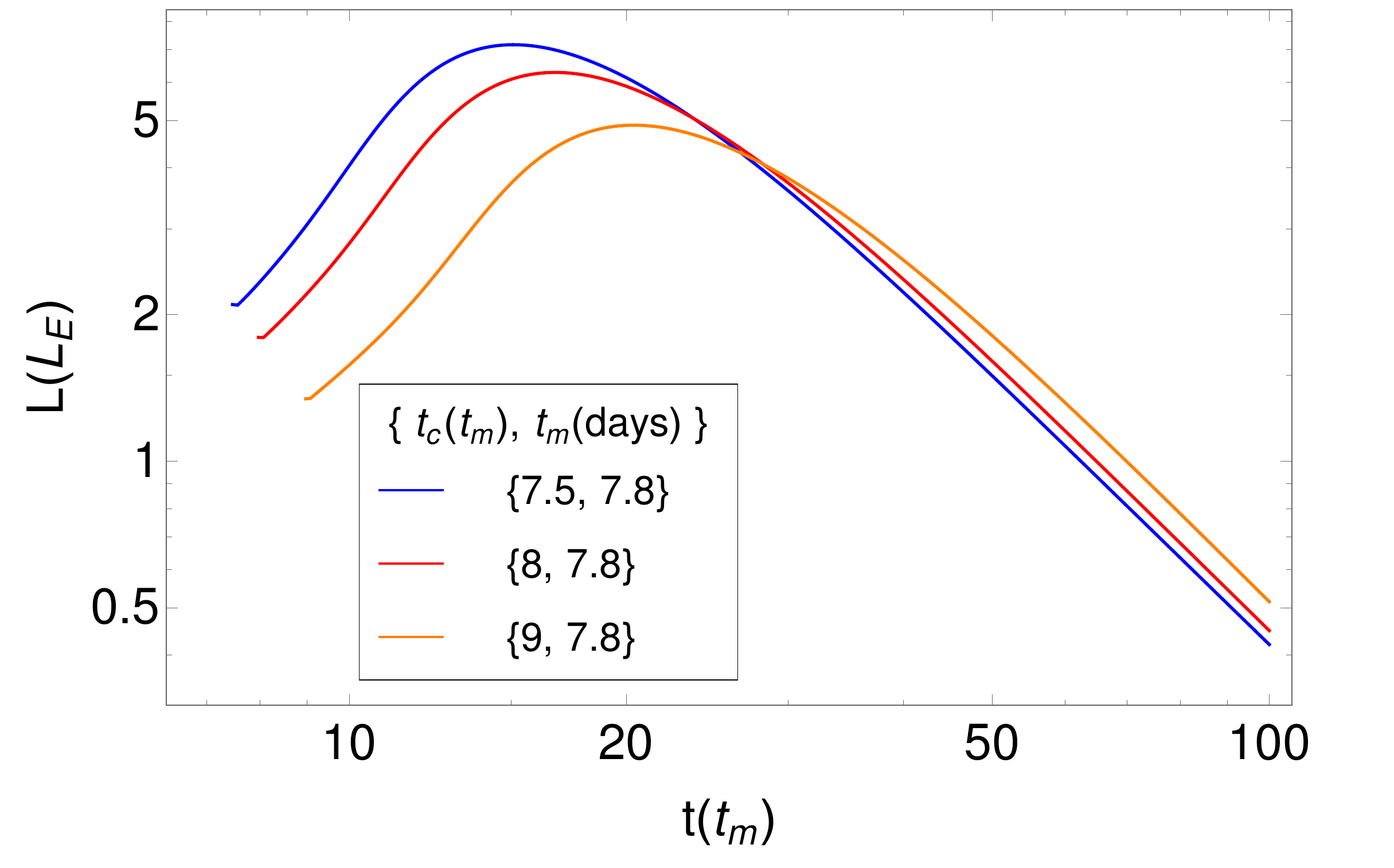}}
\end{center}
\caption{The time evolution of luminosity obtained for the mass fallback model MFR1 (section \ref{fds}) with variation in the black hole mass $M_6=M_{\bullet}/[10^6 M_{\odot}]$ in (a), stellar mass $m= M_{\star}/M_{\odot}$ in (b), black hole spin $j$ in (c) and circularization time $t_c$ in (d). While changing one parameter, we have kept all other parameters same that take the values of the parameter set M\Romannum{1} given by $\{M_6,~m,~j \} = \{1,~1,~0.5\}$. The late time profile of luminosity is given by $L \propto t^{n}$, where $n$ for various cases have been shown in Table \ref{parset}. The time evolution of $n$ is shown in Fig \ref{nev}. The Eddington luminosity is given by $L_E = 4 \pi G M_{\bullet} c/\kappa$, where $\kappa$ is the opacity due to Thomson scattering. See section \ref{resmfr1} for details. }
\label{lvary}
\end{figure}

\begin{figure}
\begin{center}
\subfigure[M\Romannum{1}a]{\includegraphics[scale=0.28]{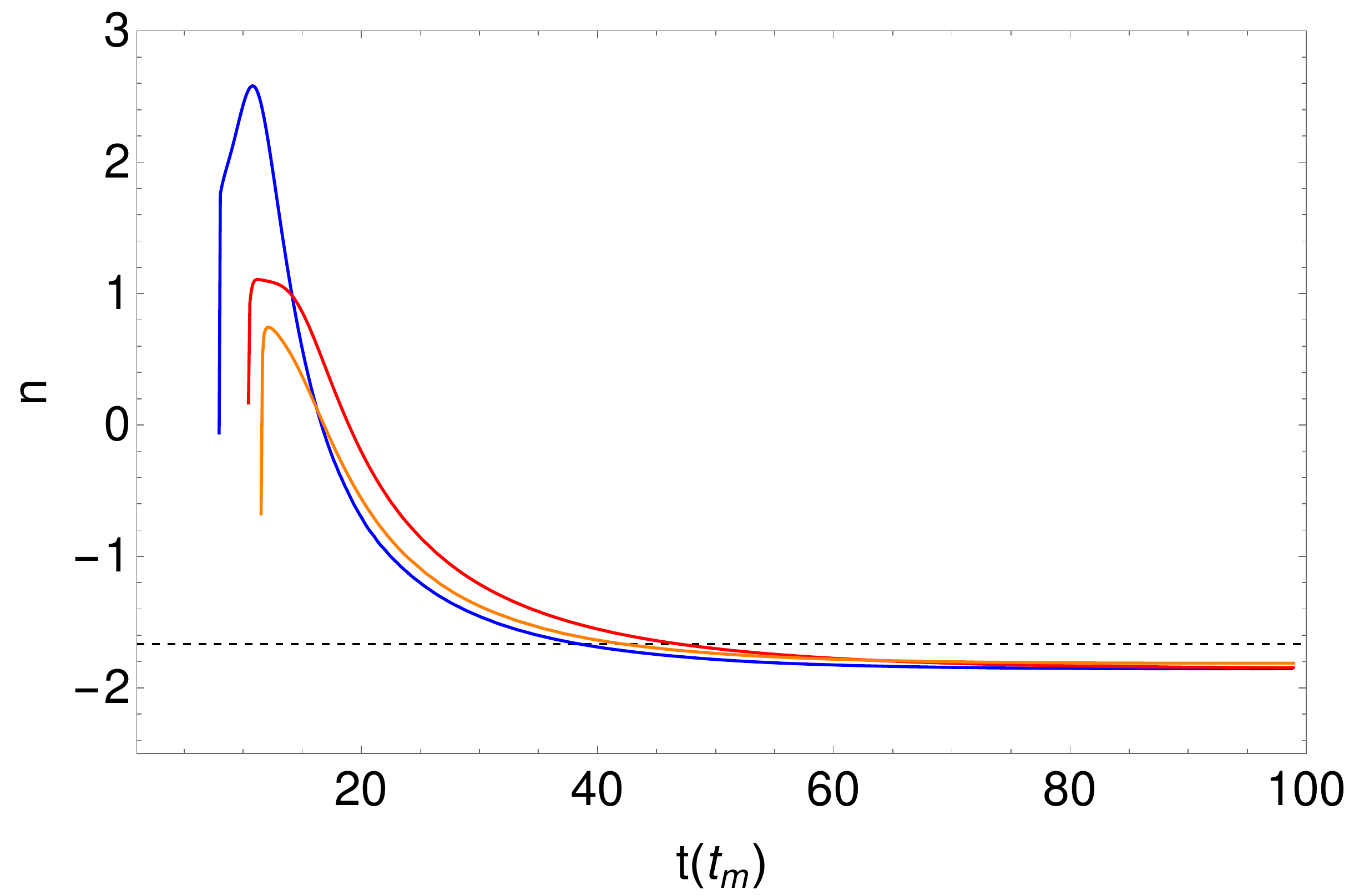}}
\subfigure[M\Romannum{1}b]{\includegraphics[scale=0.28]{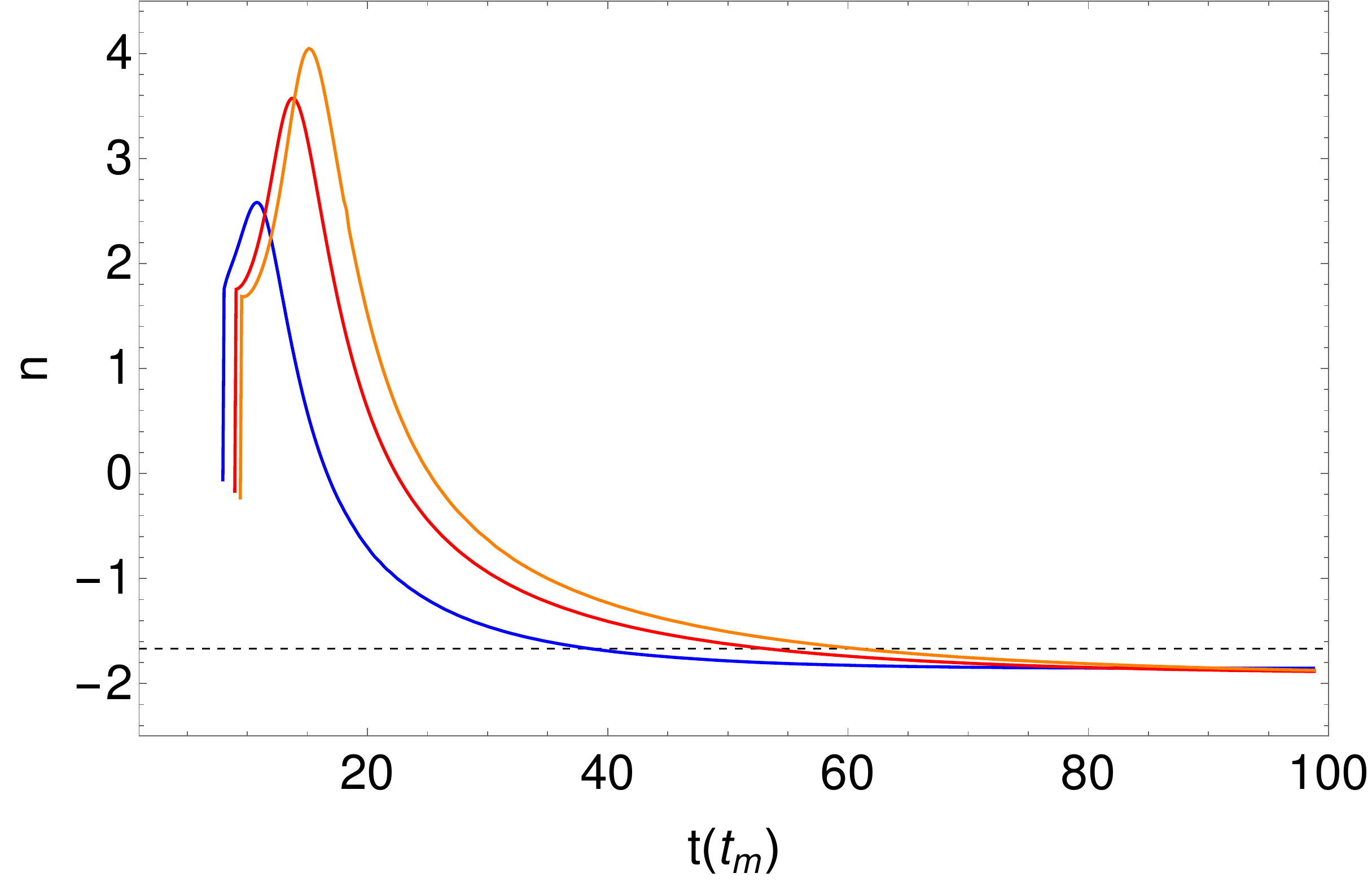}}
\subfigure[M\Romannum{1}c]{\includegraphics[scale=0.28]{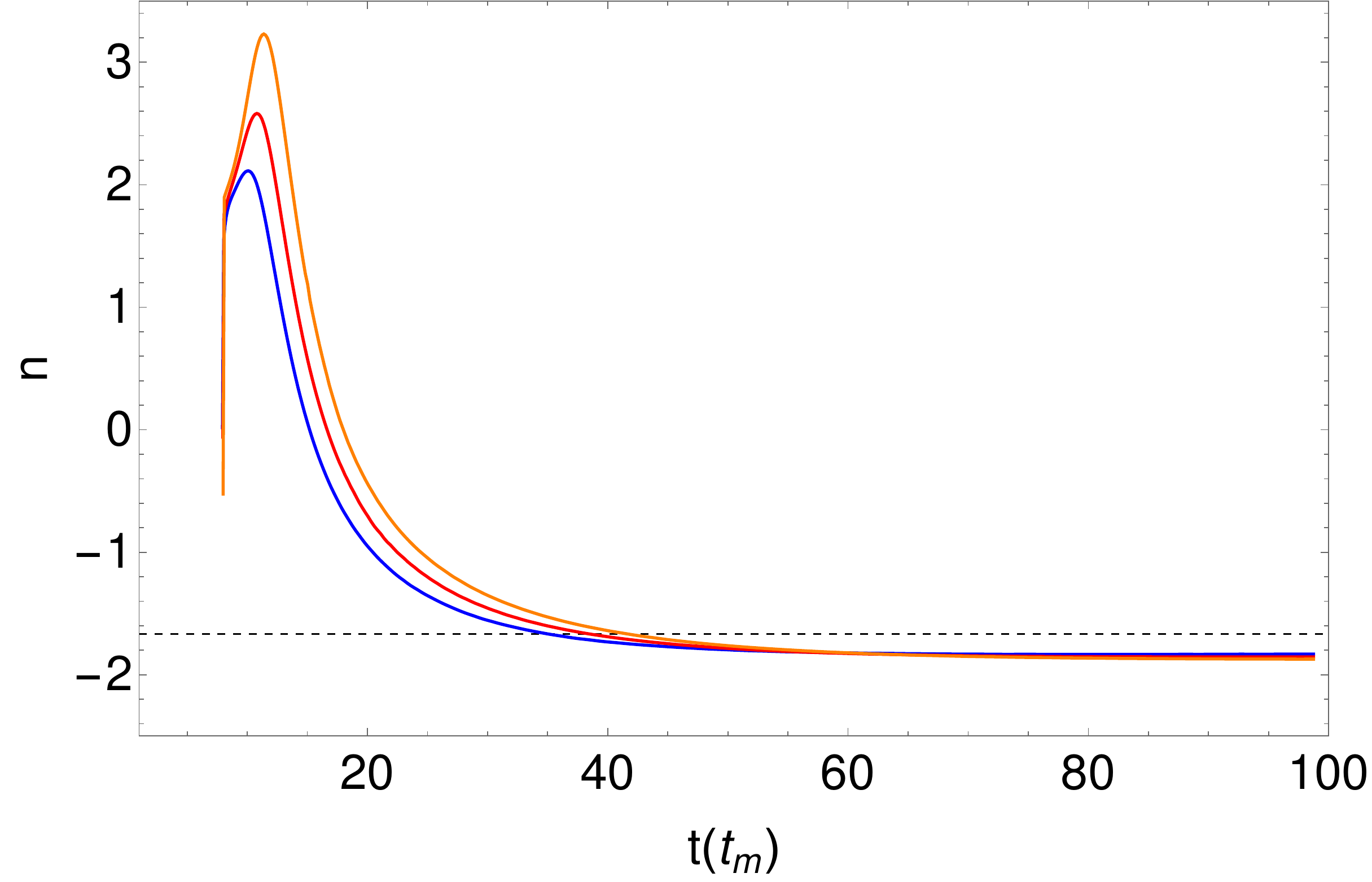}}
\subfigure[M\Romannum{1}d]{\includegraphics[scale=0.28]{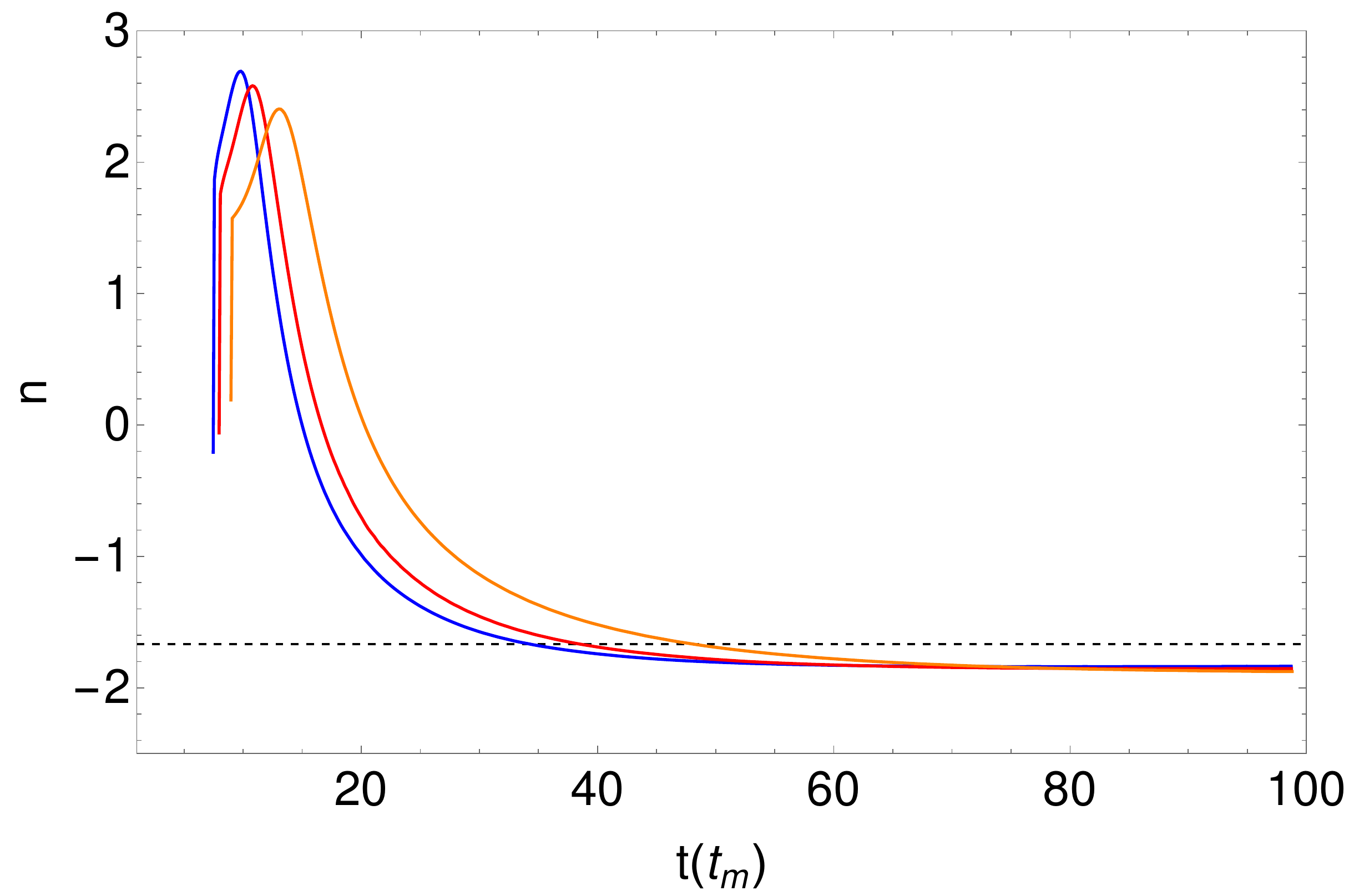}}
\end{center}
\caption{The time evolution of $n = \diff \ln L/\diff \ln t$ obtained for the mass fallback MFR1 (section \ref{fds}) with variation in the black hole mass $M_6=M_{\bullet}/[10^6 M_{\odot}]$ in (a), stellar mass $m= M_{\star}/M_{\odot}$ in (b), black hole spin $j$ in (c) and circularization time $t_c$ in (d). The luminosity and the physical parameters corresponding to various lines are given in the Fig. \ref{lvary} and Table \ref{parset}. The dashed black line corresponds to the standard declining rate of $n = -5/3$. See section \ref{resmfr1} for details. }
\label{nev}
\end{figure}

\begin{figure}
\begin{center}
\subfigure[Optical]{\includegraphics[scale=0.28]{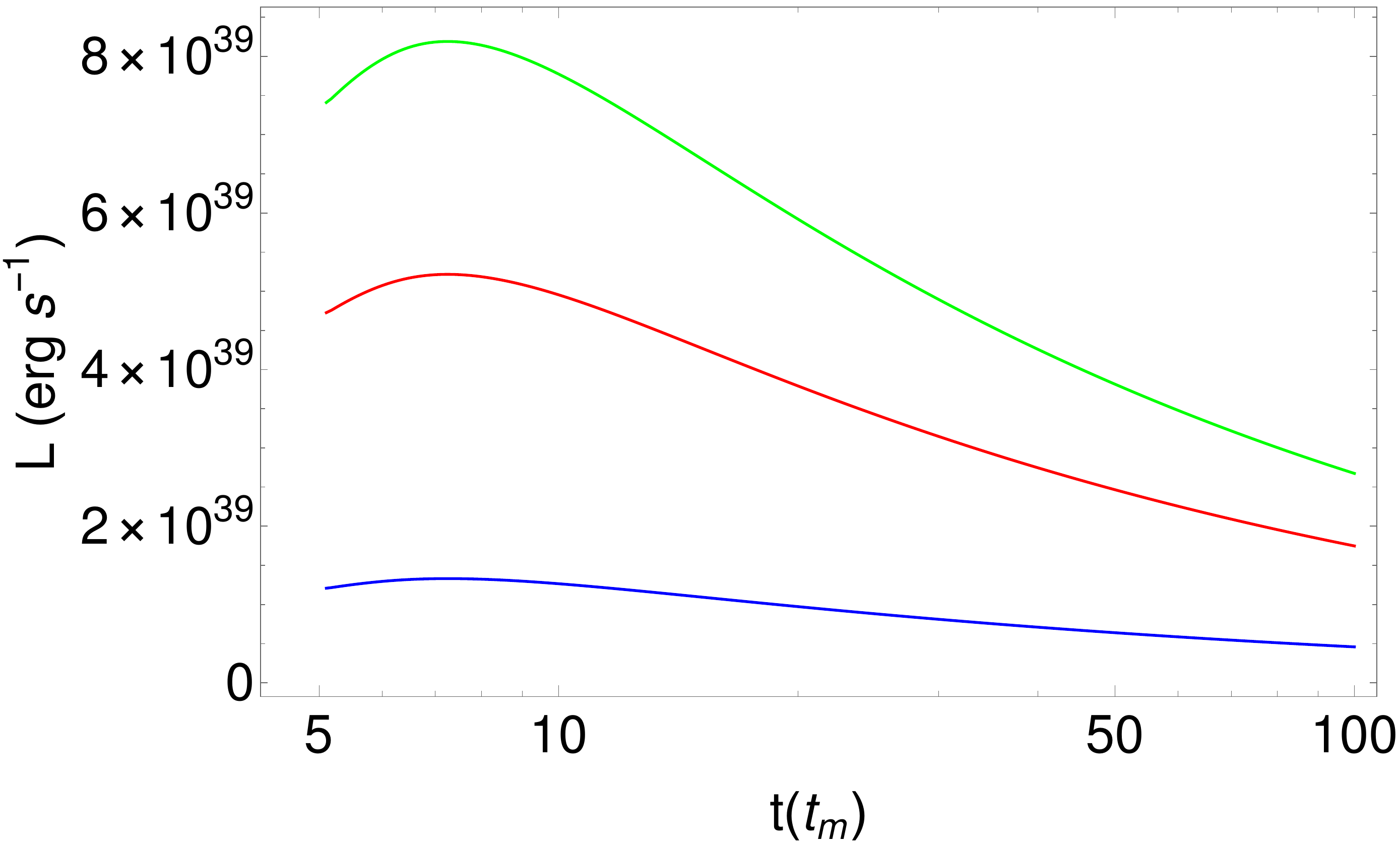}}
\subfigure[UV]{\includegraphics[scale=0.28]{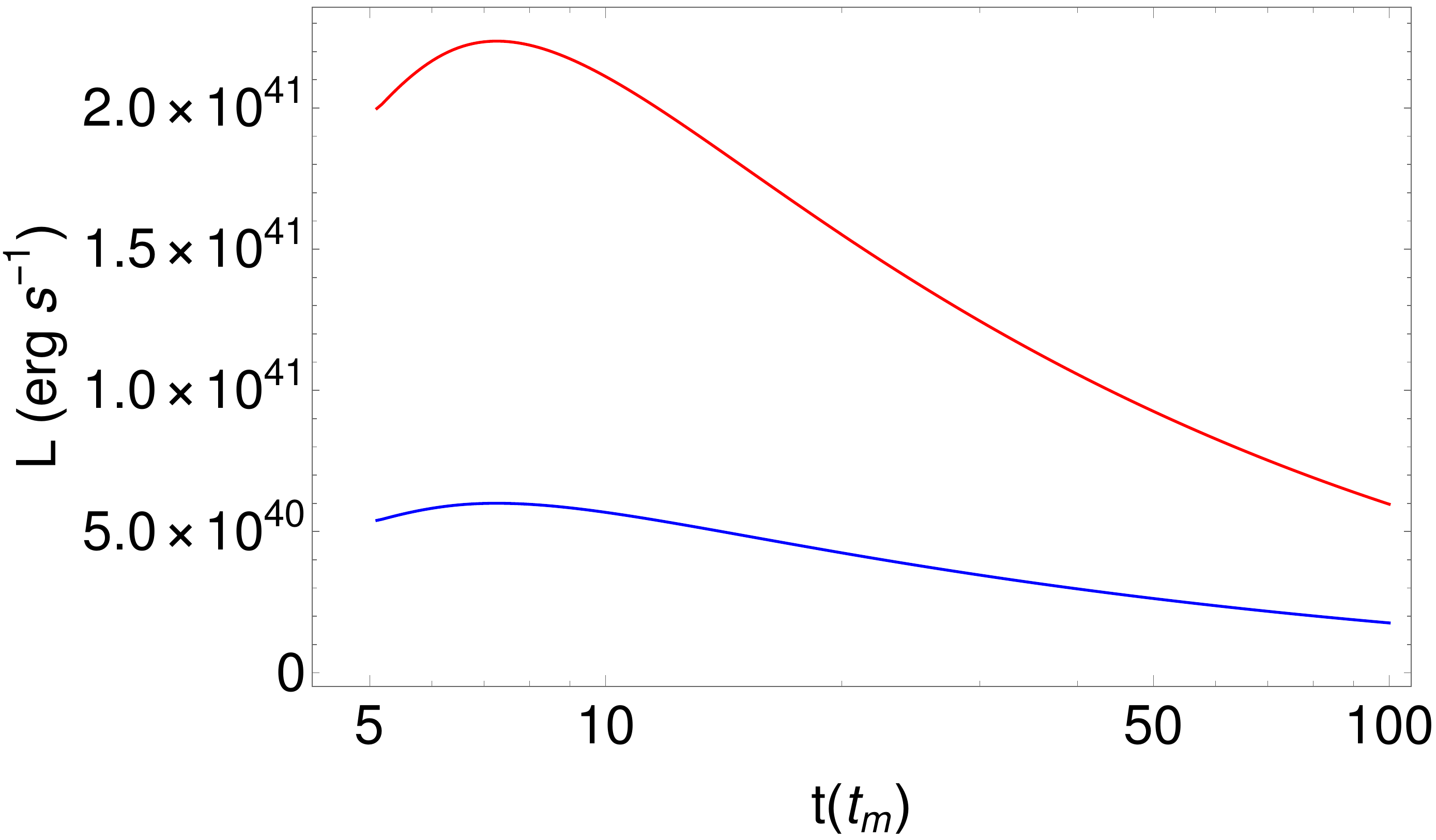}}
\subfigure[X-ray]{\includegraphics[scale=0.28]{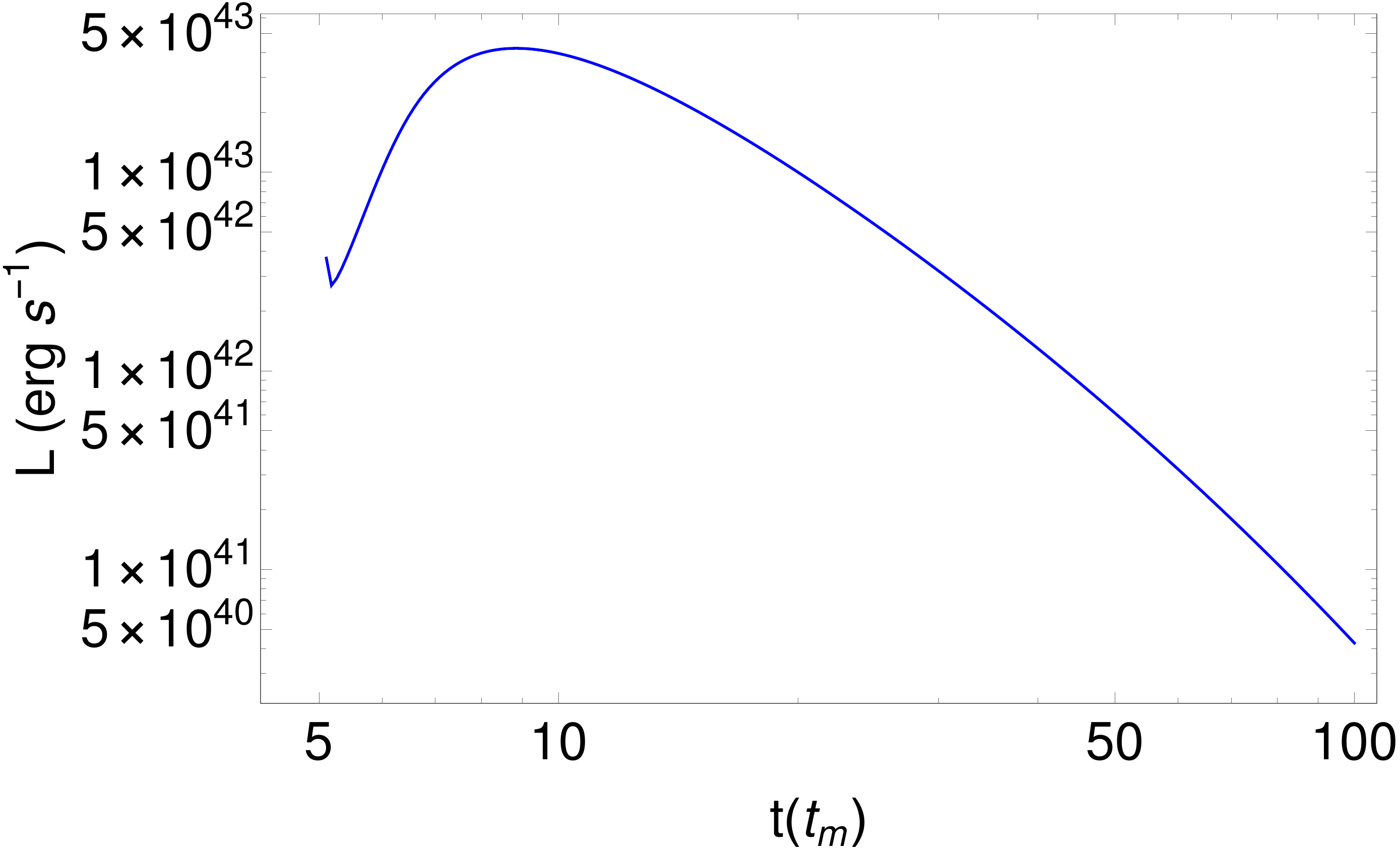}}
\end{center}
\caption{The time evolution of luminosity in various spectral bands is shown for the mass fallback MFR1 (section \ref{resmfr1}) and the parameter set M\Romannum{1} (section \ref{fds}). (a) The optical bands are V (blue), B (red) and U (green). (b) The UV bands are Swift UVM2 (blue) and UVW2 (red). (c) The X-ray band is 0.2-2 keV. See section \ref{resmfr1} for details.  }
\label{lband}
\end{figure}

The luminosity obtained using equation (\ref{lumeq}) is shown in Fig. \ref{lvary}. The luminosity increases reaching a peak value and then decreases with a late time profile given by $L \propto t^{n}$ where $n \sim -1.85$ as can be seen from Table \ref{parset}. For the luminosity shown in Fig \ref{lvary}, the time evolution of $n$ is shown in Fig \ref{nev}. The $n$ decreases to attain a steady value at late times which is lower than $n=-5/3$ implying that the luminosity declines faster than $t^{-5/3}$ at late times. The initial phase of the bolometric luminosity is affected by a change in the physical parameters, however, the late time profile shows weak changes ($L \propto t^{n}$). The outer radius of the disc increases with an increase in the black hole mass. The increase in black hole mass increases the orbital period of the debris as can be seen from equation (\ref{tmt}). The viscous heating in the disc given by $Q^{+} \propto \nu \Sigma / M_{\bullet}^2 \propto \alpha^{4/3} \Sigma^{5/3}/M_{\bullet}^{4/3}$ decreases with an increase in the black hole mass leading to a decrease in the luminosity as can be seen from Fig. \ref{lvary}a. For a star on a parabolic orbit, half of the debris is bound to the black hole after the disruption and with an increase in stellar mass $m$, the mass fallback rate increase as $\dot{M}_f \propto m^{4/5}$ (equation \ref{mfbn}). This results in an increase in the surface density and thus increases the disc luminosity (see Fig. \ref{lvary}b). The black hole spin $j$ has a dominant effect close to the inner radius of the disc. The inner radius decreases with an increase in the prograde black hole spin which makes the disc more relativistic. The peak of the bolometric luminosity increases with an increase in $j$ (see Fig. \ref{lvary}c), but $n$ at late times shows weak changes. Thus, we conclude that the peak luminosity is affected by the variations in the parameter values, but the late time profile shows minute variations and decreases faster than $L \propto \dot{M}_f \propto t^{-5/3}$. Following a blackbody disc approximation, the luminosity in optical (U, B and V bands), UV (Swift UVM2 and UVW2 bands) and X-ray bands are shown in Fig. \ref{lband} for the parameter set given by M\Romannum{1}. The declining rate of X-ray luminosity is higher than that of optical and UV. If we approximate the late time profile of V, UVW2 and X-ray bands by a power-law time evolution, we obtain the luminosity $L_{\rm V} \propto t^{-0.48}$, $L_{\rm UVW2} \propto t^{-0.6}$, and $L_{\rm X-ray} \propto t^{-3.2}$. 

\begin{table}
\scalebox{0.93}{
\begin{tabular}{|c|ccccc|}
\hline
&&&&&\\
Model & Varying parameter & Values & $t_c(t_m)$ & $n$ & Figure   \\
&&&&& \\
\hline
&&&&&\\
M\Romannum{1}a & $M_6$ & 1 & 8 & -1.85 &\\
& (Black hole mass & 5 & 10 & -1.84 & Fig. \ref{lvary}a\\
& $M_{\bullet} = M_6 10^6 M_{\odot}$ ) & 10 & 11.5 & -1.81 &\\
&&&&\\
\hline
&&&&\\
M\Romannum{1}b & $m$ & 1 & 8 & -1.85 &\\
& (Stellar mass & 5 & 9 & -1.87  & Fig. \ref{lvary}b\\
& $M_{\star}= m M_{\odot}$)& 10 & 10 & -1.87 &\\ 
&&&& \\
\hline
&&&&\\
M\Romannum{1}c & $j$ & 0 & 10 & -1.83 &\\
& (Black hole prograde & 0.5 & 8 & -1.85 & Fig. \ref{lvary}c\\
& spin)& 0.8 & 9 & -1.87 &\\
&&&& \\
\hline
&&&&\\
M\Romannum{1}d & $t_c (t_m)$ & 7.5 & & -1.84 &\\
& (Circularization time)& 8 & & -1.85 & Fig. \ref{lvary}d\\
&  & 9 & & -1.87 &\\
&&&&\\
\hline
\end{tabular}
}
\caption{The parameter set M\Romannum{1} given by $\{M_6,~m,~j\} = \{1,~1,~0.5\}$ for the mass fallback MFR1 (section \ref{fds}) is used, and one parameter is varied to observe the variation in light curve (see Fig. \ref{lvary}). The critical value of $t_c$ above which $\alpha < 1$ is a function of black hole mass and spin, and star mass (see Fig. \ref{alfds}). Thus, the value of $t_c$ we have considered and shown in the fourth column varies. In the model M\Romannum{1}d, we have shown the variation in the light curve due to a variation in the initial time $t_c$ as the critical time corresponding to $\alpha=1$ for the parameter set M\Romannum{1} is same. The late time profile of luminosity is given by $L \propto t^{n}$. The time evolution of $n$ is shown in Fig. \ref{nev}. The $n$ shown in the fifth column is steady value attained at the late time. See section \ref{resmfr1} for details.   }
\label{parset}
\end{table}

\subsection{MFR2: full disruption of a star with stellar rotation}
\label{resmfr2}

\begin{figure}
\begin{center}
\includegraphics[scale=0.33]{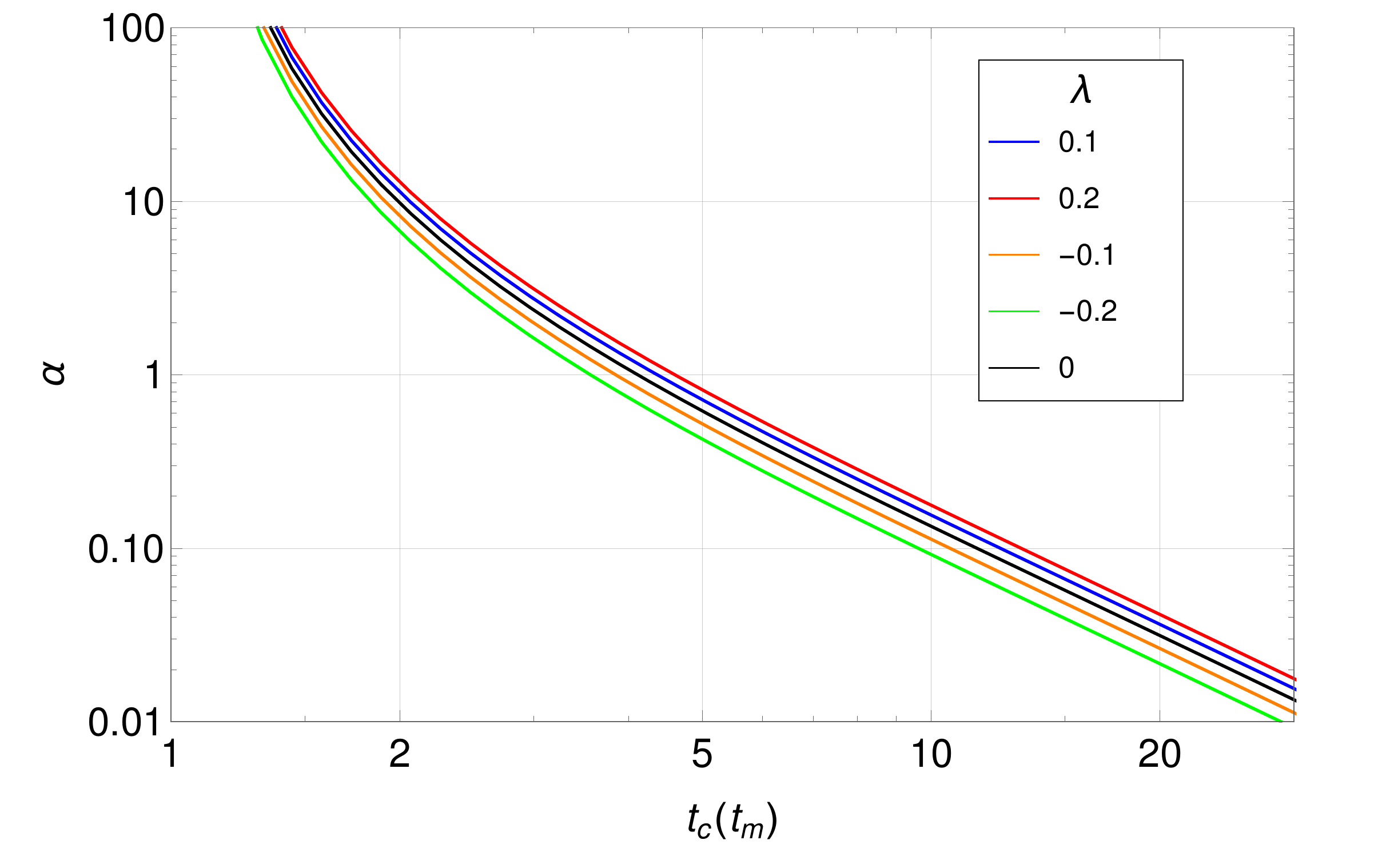}
\end{center}
\caption{The evolution of $\alpha$ as a function of $t_c$ for the mass fallback rate model MFR2 (section \ref{fdsr}) and parameter set M\Romannum{2} given by $\{M_6,~m,~j,\} = \{1,~1,~0.5\}$. The orbital period of innermost bound debris for these $\lambda$ using equation (\ref{tmtr}) is given by $t_m=40.8~{\rm days}~ (\lambda=0),~33.5~{\rm days}~ (\lambda=0.1),~28.1~{\rm days}~ (\lambda=0.2),~51.4~{\rm days}~ (\lambda=-0.1),~67.3~{\rm days}~ (\lambda=0)$. The thin disc model with an $\alpha-$viscosity is valid only if $\alpha \leq 1$. See section \ref{resmfr2} for more details.  }
\label{alfdsr}
\end{figure}

In this case of full disruption of stars and the mass fallback given by MFR2 (section \ref{fdsr}) that includes the initial stellar rotation, we name the prior parameter set to be M\Romannum{2} given by $\{M_6,~m,~j,\} = \{1,~1,~0.5\}$. This results in an outer radius $x_c=90.7$. We consider the stellar spin fractions: $\lambda = (0,~\pm 0.1,~ \pm 0.2)$, where the positive sign corresponds to prograde spin and the negative sign is for a retrograde spin. The orbital period of innermost bound debris for these $\lambda$ and the parameter set M\Romannum{2} is obtained using equation (\ref{tmtr}) and have values $t_m=40.8~{\rm days}~ (\lambda=0),~33.5~{\rm days}~ (\lambda=0.1),~28.1~{\rm days}~ (\lambda=0.2),~51.4~{\rm days}~ (\lambda=-0.1),~67.3~{\rm days}~ (\lambda=-0.2)$. The $\alpha$ decreases with $t_c$ as shown in Fig. \ref{alfdsr} and the critical time $t_{\rm cr}$ corresponding to $\alpha=1$ is given by $t_{\rm cr}= 167.8~{\rm days}~ (\lambda=0,~{\rm black~line}),~146~{\rm days}~ (\lambda=0.1,~{\rm blue~line}),129.3~~{\rm days}~ (\lambda=0.2,~{\rm red~line}),~197.4~{\rm days}~ (\lambda=-0.1,~{\rm orange~line}),~239.6~{\rm days}~ (\lambda=-0.2,~{\rm green~line})$. Thus, the critical time decreases for the prograde spins and increases for the retrograde spins.

\begin{figure}
\begin{center}
\includegraphics[scale=0.38]{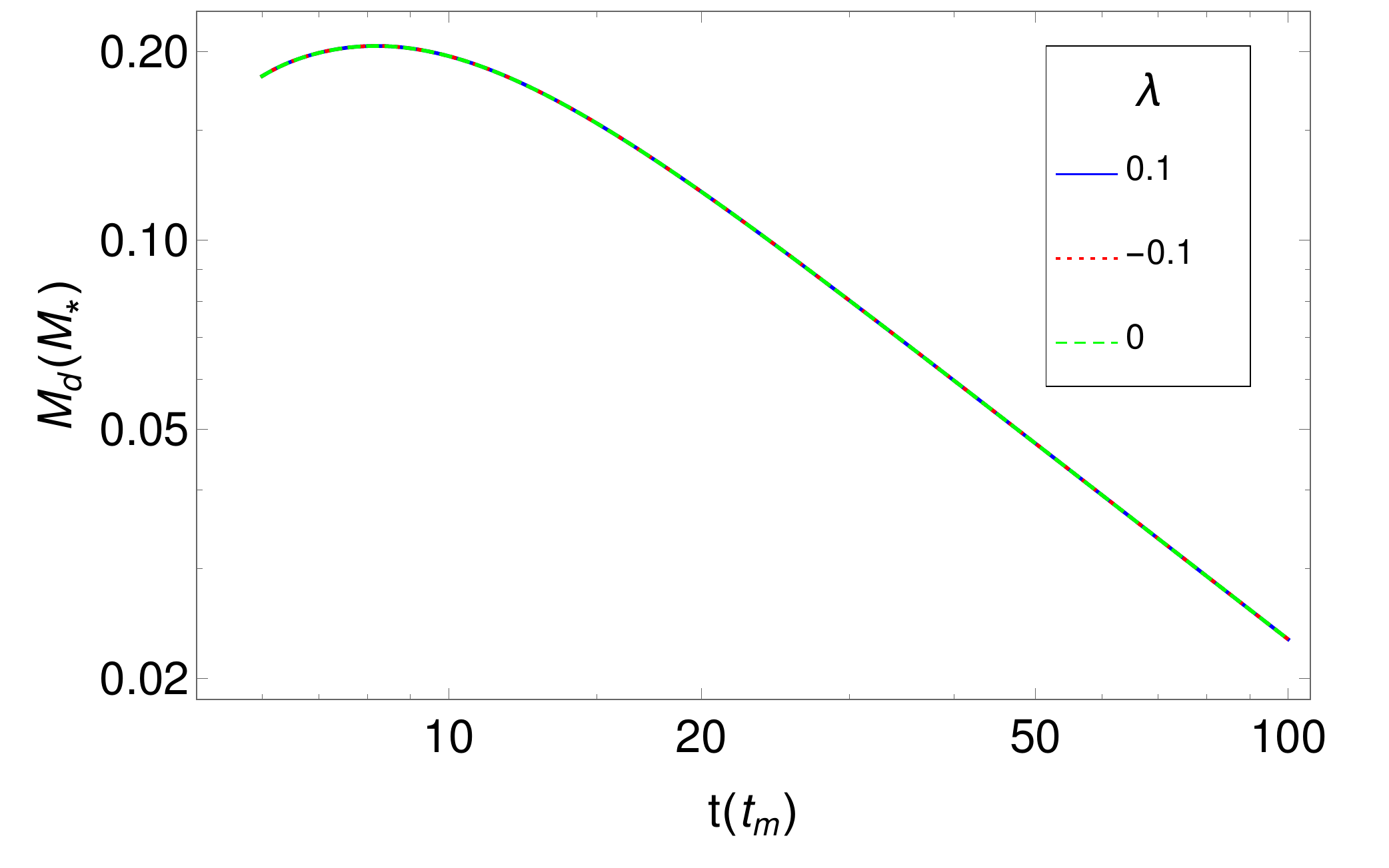}
\end{center}
\caption{The time evolution of the disc mass is shown for the mass fallback MFR2 (section \ref{fdsr}) and the parameter set M\Romannum{2} (section \ref{resmfr2}). The circularization time taken for all the lines are $t_c(t_m)= 6$ where $t_m=~33.5~{\rm days}~{\rm (\lambda=0.1)},~51.4~{\rm days}~{\rm (\lambda=-0.1)}~{\rm and}~40.8~{\rm days}~{\rm (\lambda=0)}$. All the curves are overlapping. The late time of disc mass is $M_{\rm d} \propto t^{p}$ where $p=-1.05$. See section \ref{resmfr2} for details. }
\label{mdisr} 
\end{figure}

\begin{figure}
\begin{center}
\subfigure[]{\includegraphics[scale=0.34]{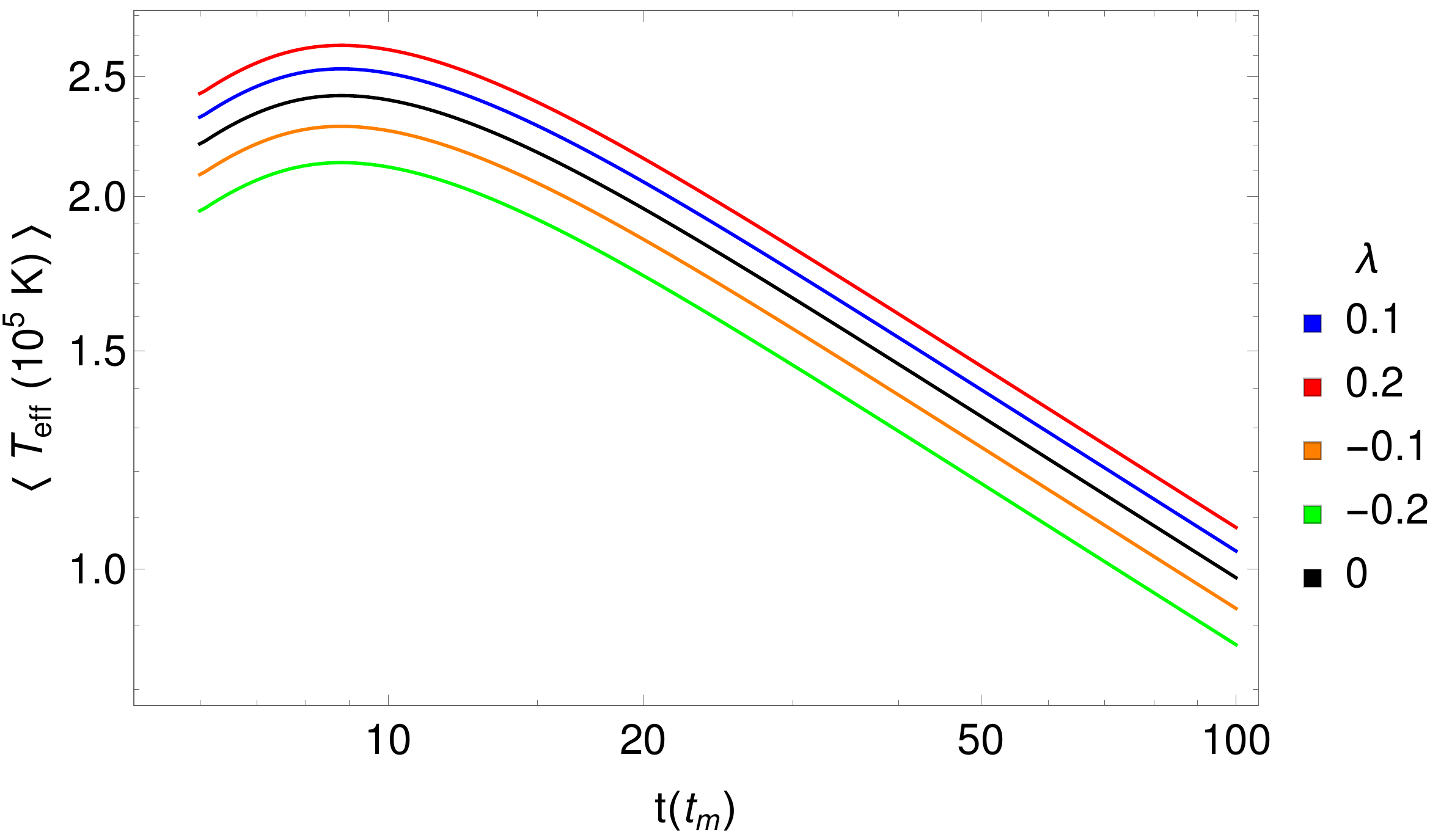}}
\subfigure[]{\includegraphics[scale=0.34]{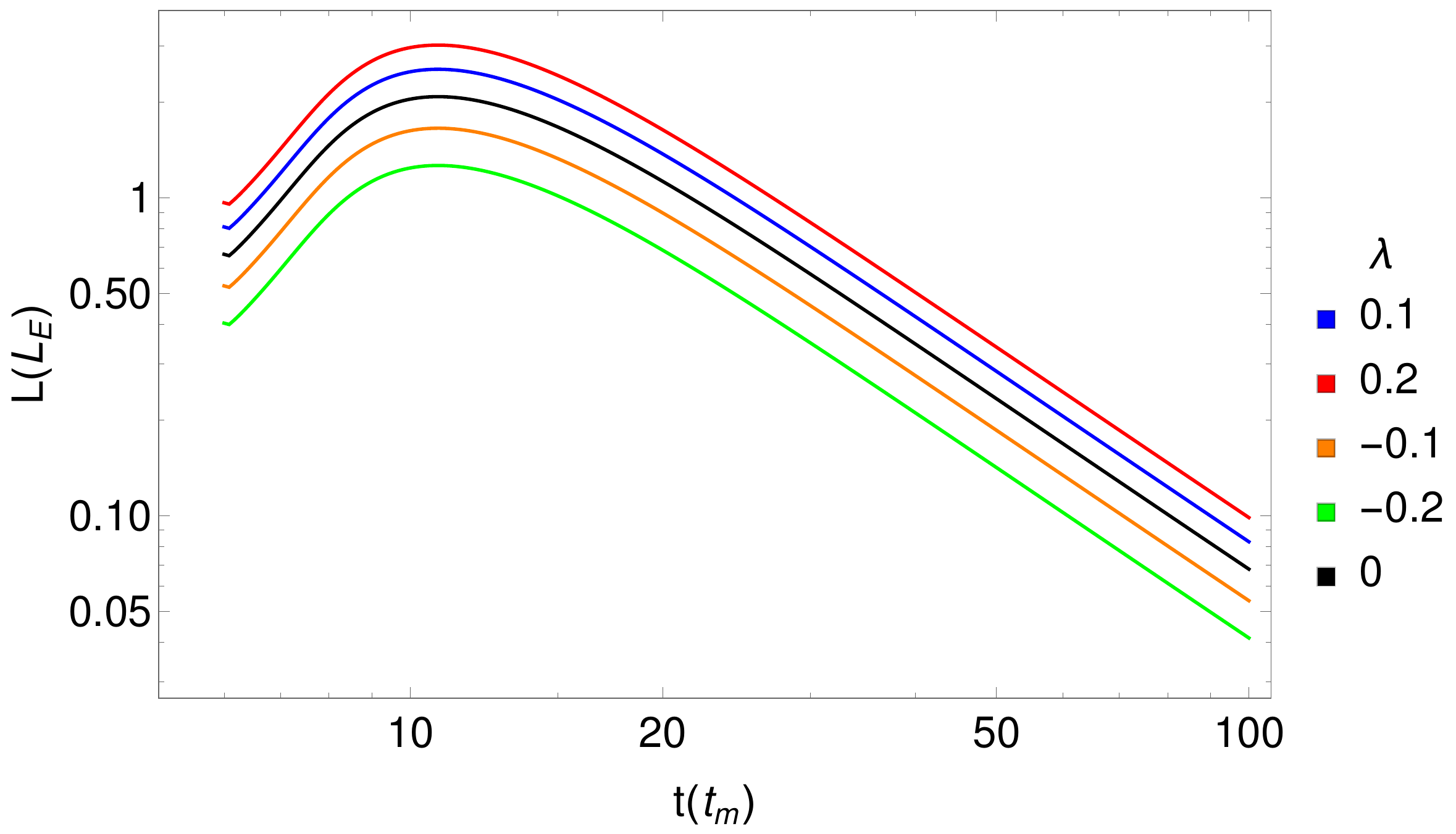}}
\end{center}
\caption{The time evolution of the mean effective temperature in (a) and the bolometric luminosity in (b) are shown for the mass fallback model MFR2 (section \ref{fdsr}) and the parameter set M\Romannum{2} (section \ref{resmfr2}). The circularization time taken for both (a) and (b) is $t_c(t_m)= 6$. The late time evolution of mean temperature is $\left<T_{\rm eff}\right> \propto t^{-0.43}$ and the bolometric luminosity $L \propto t^{-1.78}$. The peak of luminosity increases for a prograde stellar spin and decreases for a retrograde stellar spin.  See section \ref{resmfr2} for details. }
\label{tempr}
\end{figure}

\begin{figure}
\begin{center}
\subfigure[M\Romannum{2}a]{\includegraphics[scale=0.3]{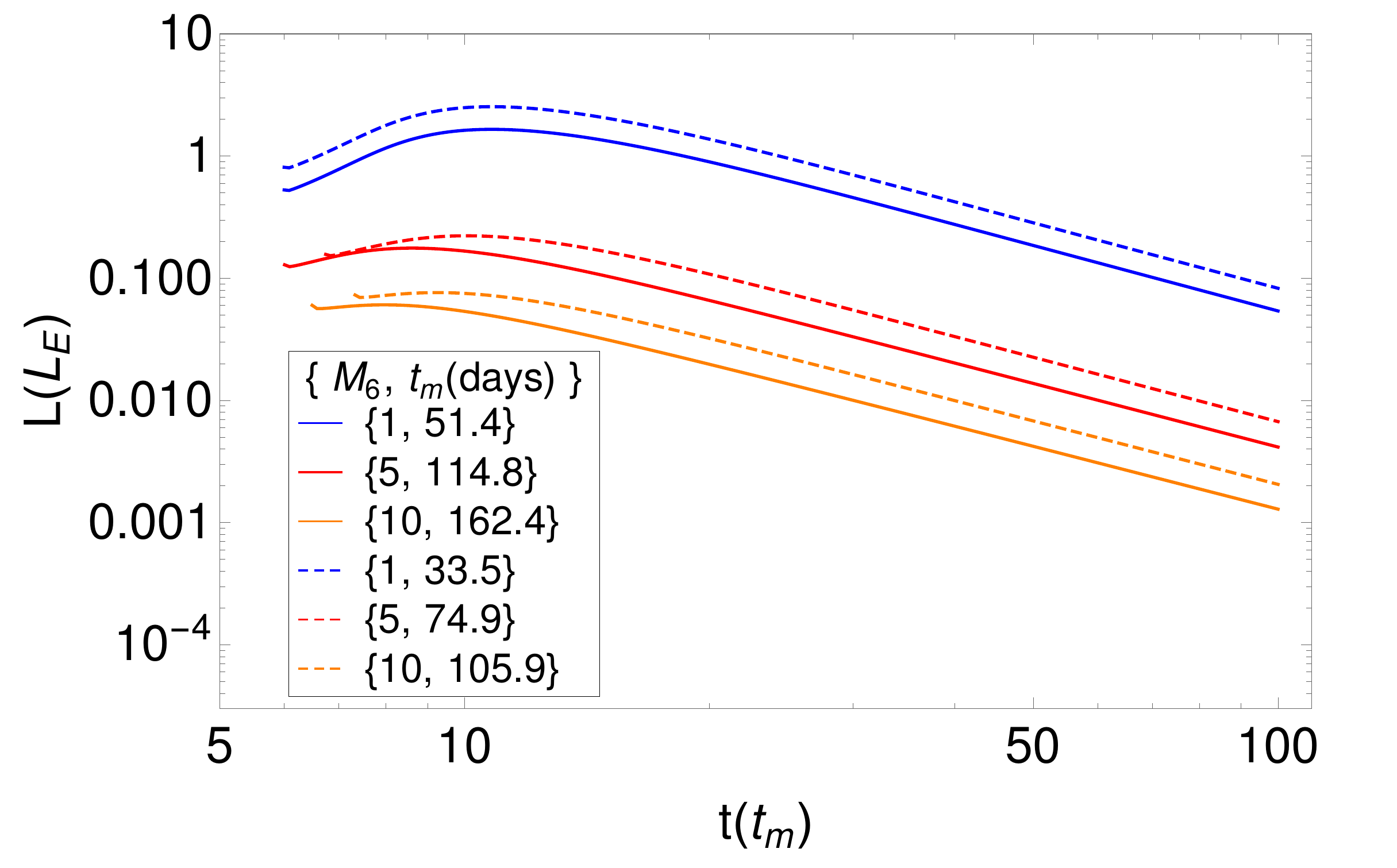}}
\subfigure[M\Romannum{2}b]{\includegraphics[scale=0.3]{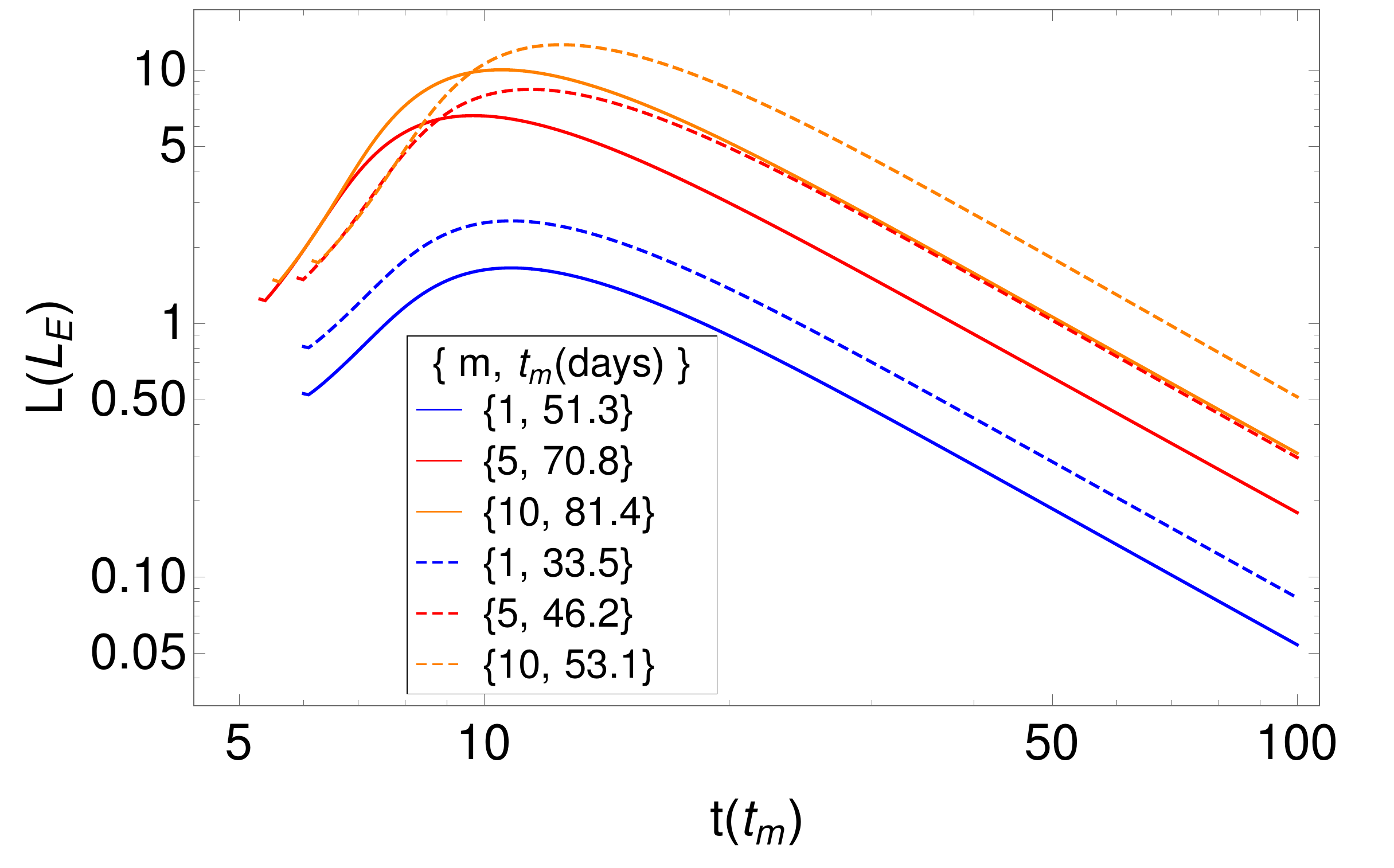}}
\subfigure[M\Romannum{2}c]{\includegraphics[scale=0.3]{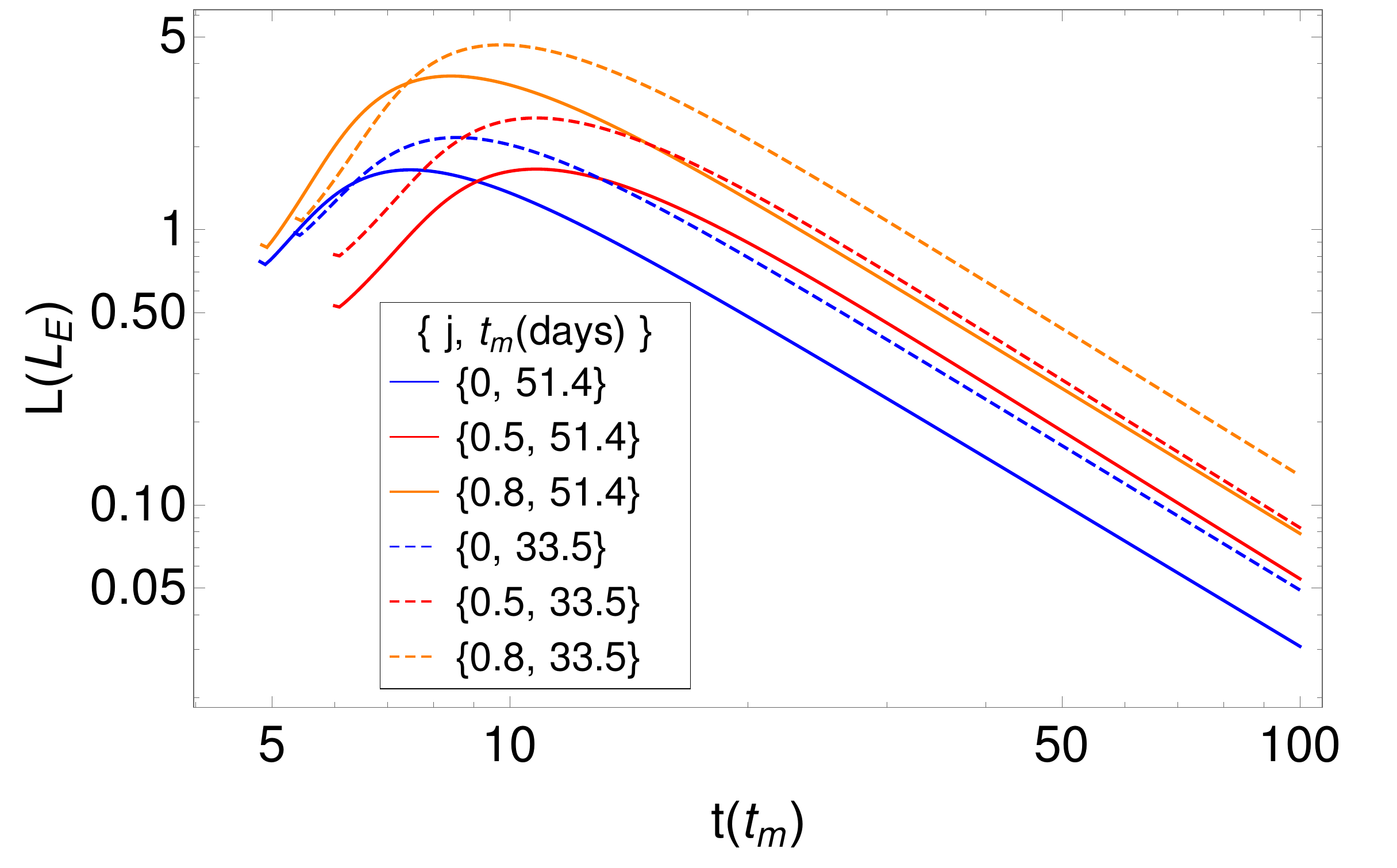}}
\end{center}
\caption{The time evolution of luminosity obtained for the mass fallback model MFR2 (section \ref{fdsr}) with variation in the black hole mass $M_6=M_{\bullet}/[10^6 M_{\odot}]$ in (a), stellar mass $m= M_{\star}/M_{\odot}$ in (b), and black hole spin $j$ in (c). While changing one parameter, we have kept all other parameters same that take the values of parameter set M\Romannum{2} given by $\{M_6,~m,~j \} = \{1,~1,~0.5\}$ and stellar spin fraction $\lambda = \pm 0.1$. The solid lines correspond to a retrograde stellar spin ($\lambda = -0.1$) and the dashed lines correspond to a prograde stellar spin ($\lambda = -0.1$). The late time profile of luminosity is given by $L \propto t^{n}$, where $n$ at late time for various cases have been shown in Table \ref{parsetr}. The time evolution of $n$ is shown in Fig. \ref{nevr}. The Eddington luminosity is given by $L_E = 4 \pi G M_{\bullet} c/\kappa$, where $\kappa$ is the opacity due to Thomson scattering. See section \ref{resmfr2} for details. }
\label{lvaryr}
\end{figure}

\begin{figure}
\begin{center}
\subfigure[M\Romannum{1}a]{\includegraphics[scale=0.3]{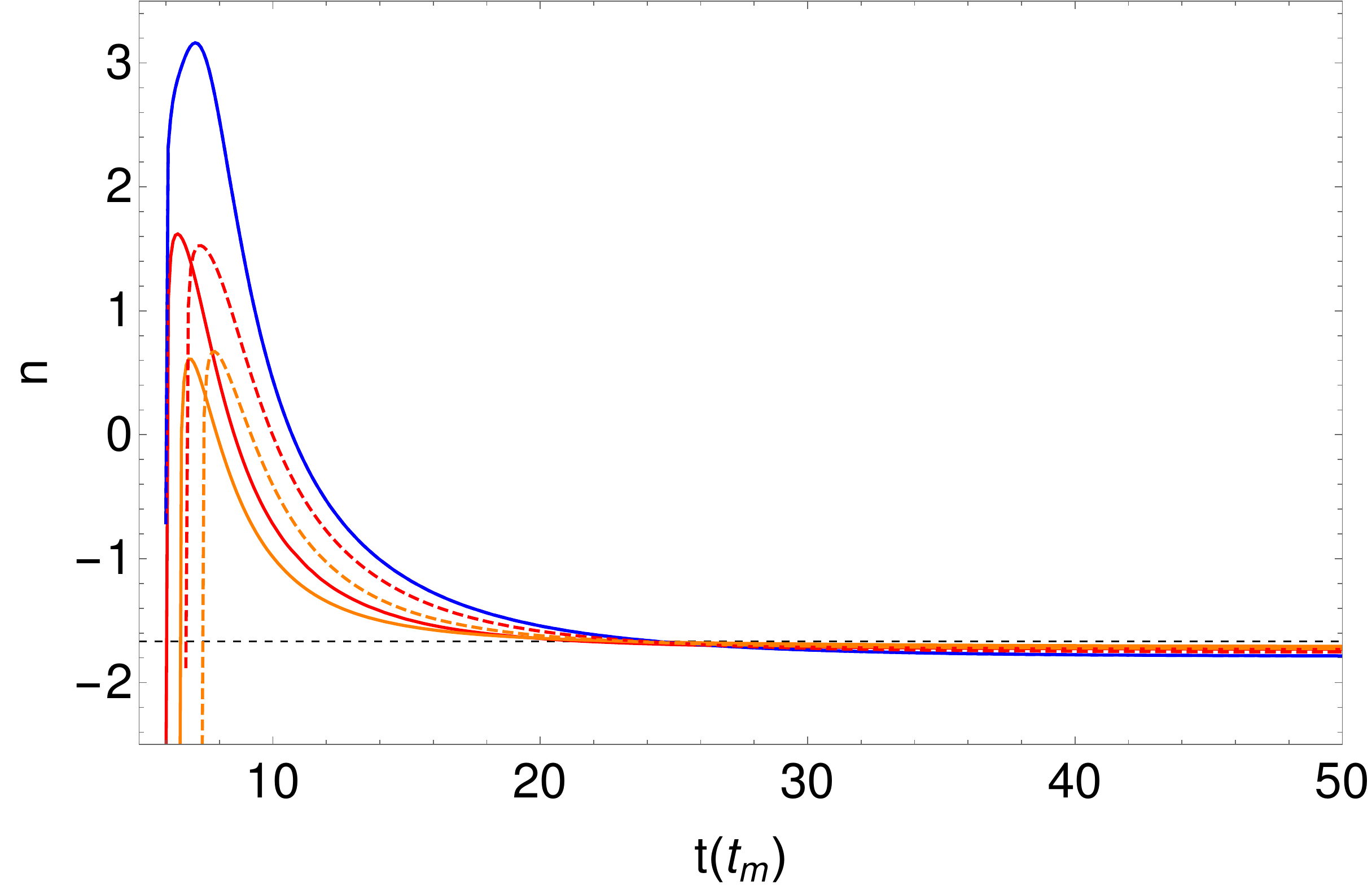}}
\subfigure[M\Romannum{1}b]{\includegraphics[scale=0.3]{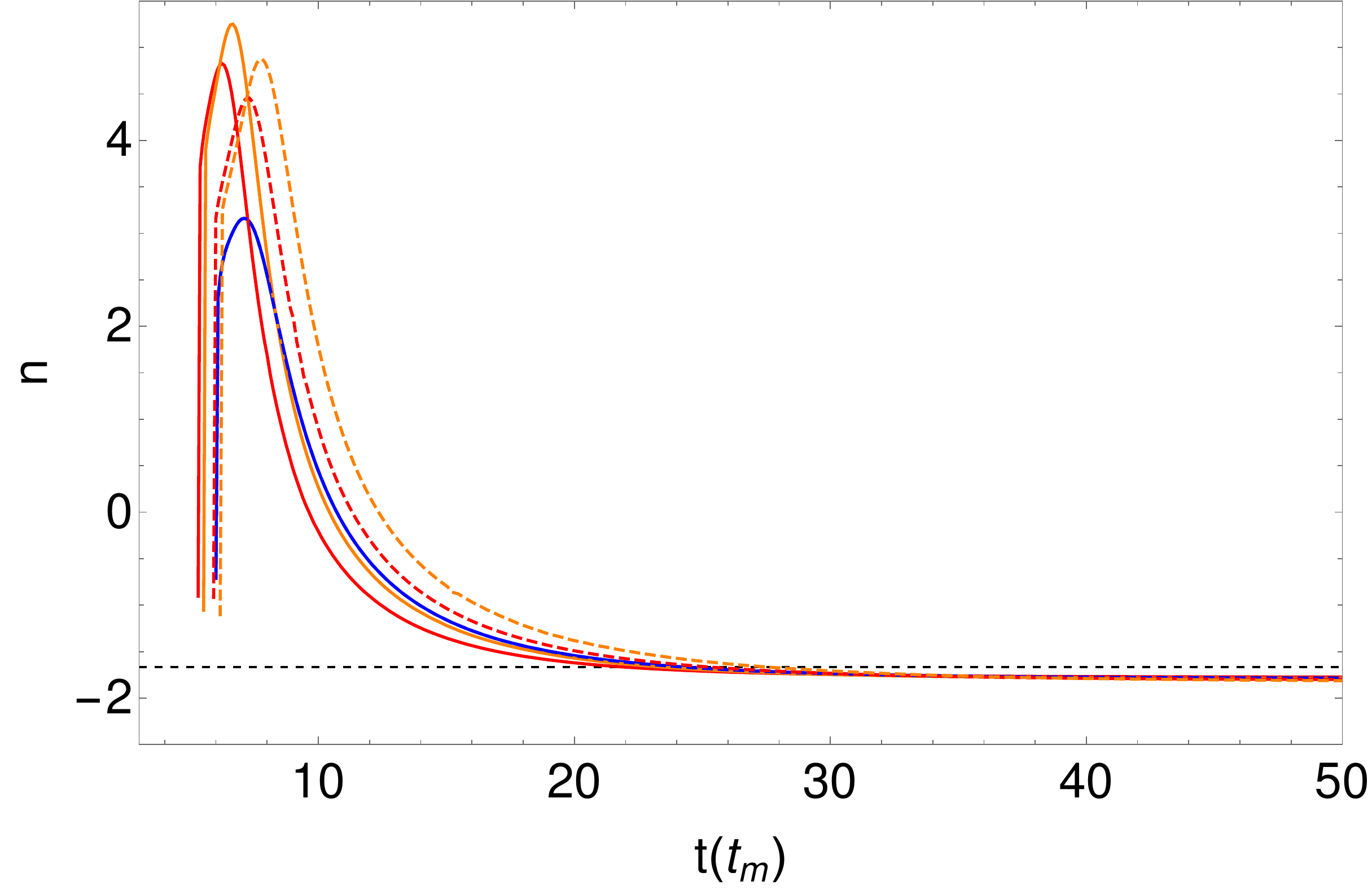}}
\subfigure[M\Romannum{1}c]{\includegraphics[scale=0.3]{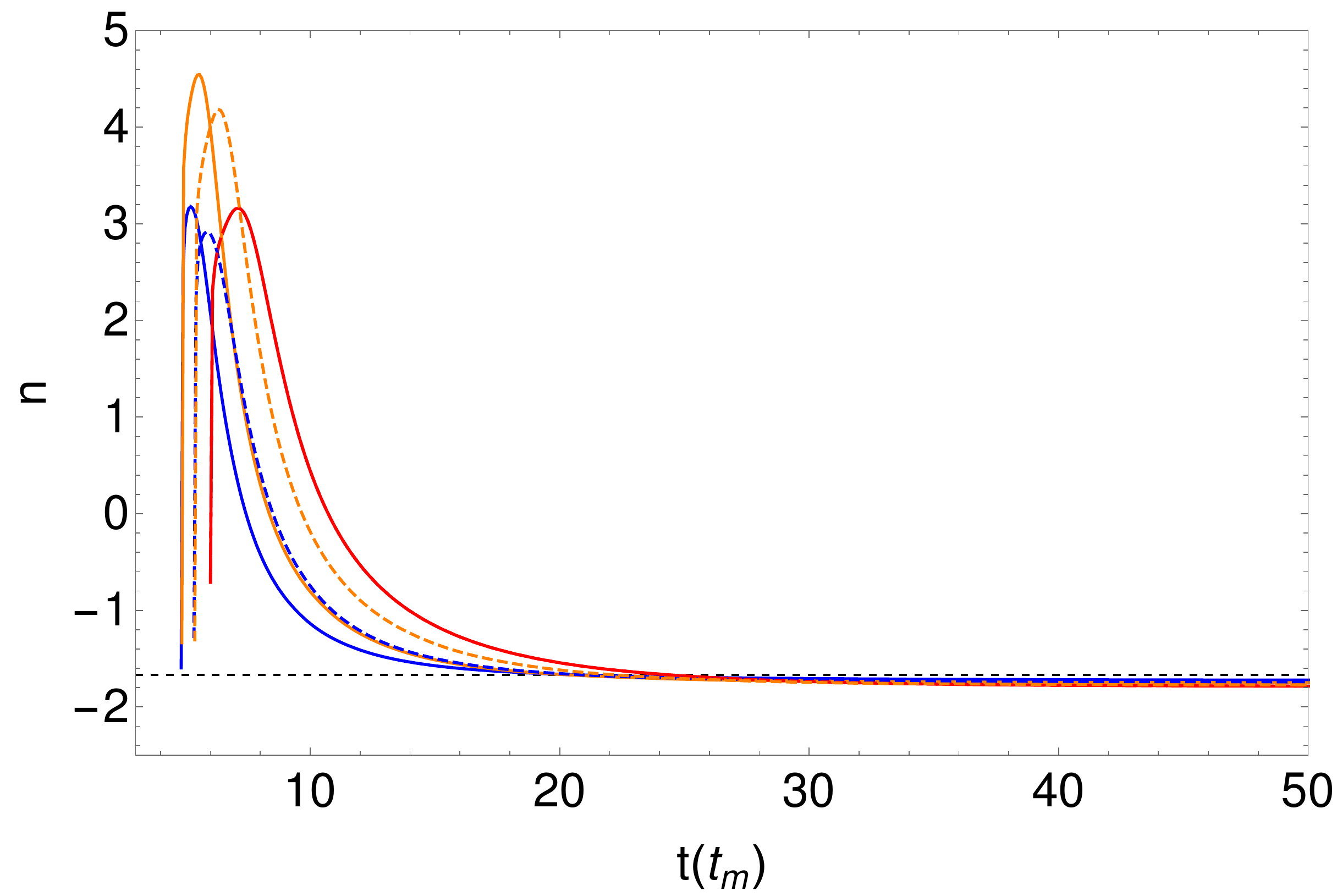}}
\end{center}
\caption{The time evolution of $n = \diff \ln L/\diff \ln t$ obtained for the mass fallback model MFR2 (section \ref{fdsr}) with variation in the black hole mass $M_6=M_{\bullet}/[10^6 M_{\odot}]$ in (a), stellar mass $m= M_{\star}/M_{\odot}$ in (b), and black hole spin $j$ in (c). The solid lines are for a retrograde stellar spin and the dashed lines correspond to a prograde stellar spin. The blue and blue dashed lines in (a) and (b) are overlapping as the corresponding luminosity shows identical time evolution. The luminosity and the physical parameters corresponding to various lines are given in the Fig. \ref{lvaryr} and Table \ref{parsetr}. The dashed black line corresponds to standard declining rate of $n = -5/3$. See section \ref{resmfr2} for details. }
\label{nevr}
\end{figure}

For a given value of stellar spin fraction $\lambda$, the energy of the disrupted debris is $E_d \propto M_{\bullet}\Delta R/r_t^2$ (see equation \ref{edmfr2}) which is similar to that for full disruption of a star with zero initial stellar rotation (section \ref{fds}). The dependence of the orbital time and mass fallback rate on black hole mass and star mass are same for both the models MFR1 (section \ref{fds}: equation \ref{tmt}) and MFR2 (section \ref{fdsr}: equation \ref{tmtr}). The surface density and disc height evolve in a manner similar to that shown for model MFR1 in Figs. \ref{surf} and \ref{hrp} respectively. The time evolution of disc mass for the various stellar spins is shown in Fig. \ref{mdisr} and shows an identical evolution with late-time given by $M_d \propto t^{-1.05}$. We have also verified this for stellar spin $\lambda = \pm 0.2$ but not included in the Fig. \ref{mdisr} as it overlaps with other curves. With an increase in $\lambda$, the energy of the disrupted debris increases which results in an increase in the mass fallback rate and thus the luminosity. The bolometric luminosity for the various stellar spin is shown in Fig. \ref{tempr}b and the bolometric luminosity increases for prograde stellar spins ($\lambda > 0$) and decreases for retrograde stellar spins ($\lambda < 0$). The bolometric luminosity increases with an increase in the star mass and black hole spin but decreases with an increase in the black hole mass as can be seen from Fig. \ref{lvaryr}. This is similar to the results we obtained for MFR1 (see section \ref{resmfr1}) and an explanation for the variation in luminosity with physical parameters is discussed there. The luminosity increases reaching a peak value and then decreases with a late time profile given by $L \propto t^{n}$ where $n \sim -1.75$ as can be seen from Table \ref{parsetr}. For the luminosity shown in Fig \ref{lvaryr}, the time evolution of $n$ is shown in Fig \ref{nevr}. The late time luminosity decline is higher than the standard $t^{-5/3}$ evolution.

\begin{table*}
\begin{tabular}{|c|ccccccc|}
\hline
&&&&&&&\\
Model & Varying parameter & Values &  \multicolumn{2}{c|}{$t_c(t_m)$} &  \multicolumn{2}{c|}{$n$} & Figure   \\
&&& Prograde & Retrograde & Prograde & Retrograde & \\
\hline
&&&&&&&\\
M\Romannum{2}a & $M_6$ & 1 & 6 & 6 & -1.78 & -1.78 &\\
& (Black hole mass & 5 & 6.8 & 6 & -1.75 & -1.73 & Fig. \ref{lvaryr}a\\
& $M_{\bullet} = M_6 10^6 M_{\odot}$ ) & 10 & 7.2 & 6.5 & -1.72 & -1.71 &\\
&&&&&&\\
\hline
&&&&&&\\
M\Romannum{2}b & $m$ & 1 & 6 & 6 & -1.78 & -1.78 &\\
& (Stellar mass & 5 & 6 & 5.3 & -1.76 & -1.74  & Fig. \ref{lvaryr}b\\
& $M_{\star}= m M_{\odot}$)& 10 &  6.2 & 5.5 & -1.74 & -1.74 &\\ 
&&&&&& \\
\hline
&&&&&&\\
M\Romannum{2}c & $j$ & 0 & 5.3 & 5 & -1.74 & -1.72 &\\
& (Black hole prograde & 0.5 & 6 & 6 & -1.78 & -1.78 & Fig. \ref{lvaryr}c\\
& spin)& 0.8 & 5.4 & 5 & -1.78 & -1.78 &\\
&&&&&& \\
\hline
\end{tabular}
\caption{The parameter set M\Romannum{2} given by $\{M_6,~m,~j\} = \{1,~1,~0.5\}$ for the mass fallback MFR2 (section \ref{fdsr}) is used with stellar spin fraction $\lambda= \pm 0.1$, and one parameter is varied to observe the variation in light curve (see Fig. \ref{lvaryr}). The critical value of $t_c$ above which $\alpha < 1$ is a function of black hole mass and spin, and star mass. Thus, the value of $t_c$ we have considered and shown in the fourth column varies. The late time profile of luminosity is given by $L \propto t^{n}$. The time evolution of $n$ is shown in Fig. \ref{nevr}. The $n$ shown in the fifth column is steady value attained at the late time. See section \ref{resmfr2} for details. }
\label{parsetr}
\end{table*}

\subsection{MFR3: partial disruption of a star}
\label{resmfr3}
 
\begin{figure*}
\begin{center}
\includegraphics[scale=0.45]{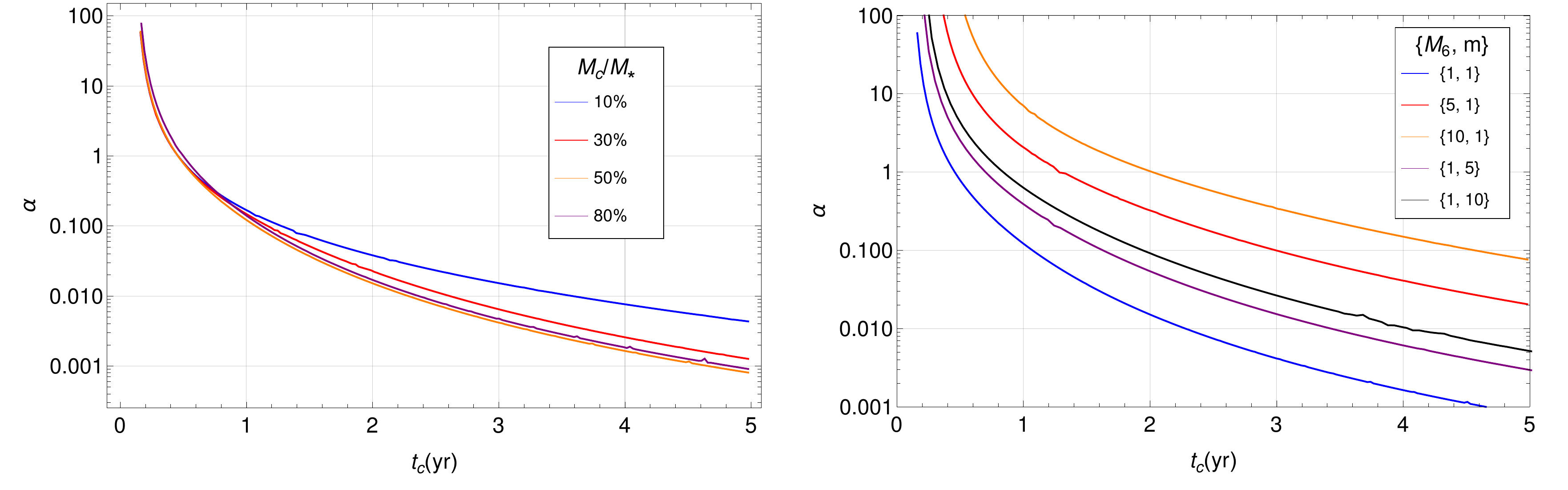}
\end{center}
\caption{The evolution of $\alpha$ as a function of $t_c$ for the mass fallback rate model MFR3 (section \ref{pds}) obtained for partial TDEs. In left plot, we have taken the parameter set M\Romannum{3} given by $\{M_6,~m,~j,\} = \{1,~1,~0.5\}$ and various values of core mass $M_c$ in terms of star mass. In right plot, the value of core mass is $M_c=0.5 M_{\star}$ and the black hole spin is $j=0.5$. The thin disc model with an $\alpha-$viscosity is valid only if $\alpha \leq 1$. See section \ref{resmfr3} for more details. }
\label{alpds}
\end{figure*}

\begin{figure*}
\begin{center}
\includegraphics[scale=0.45]{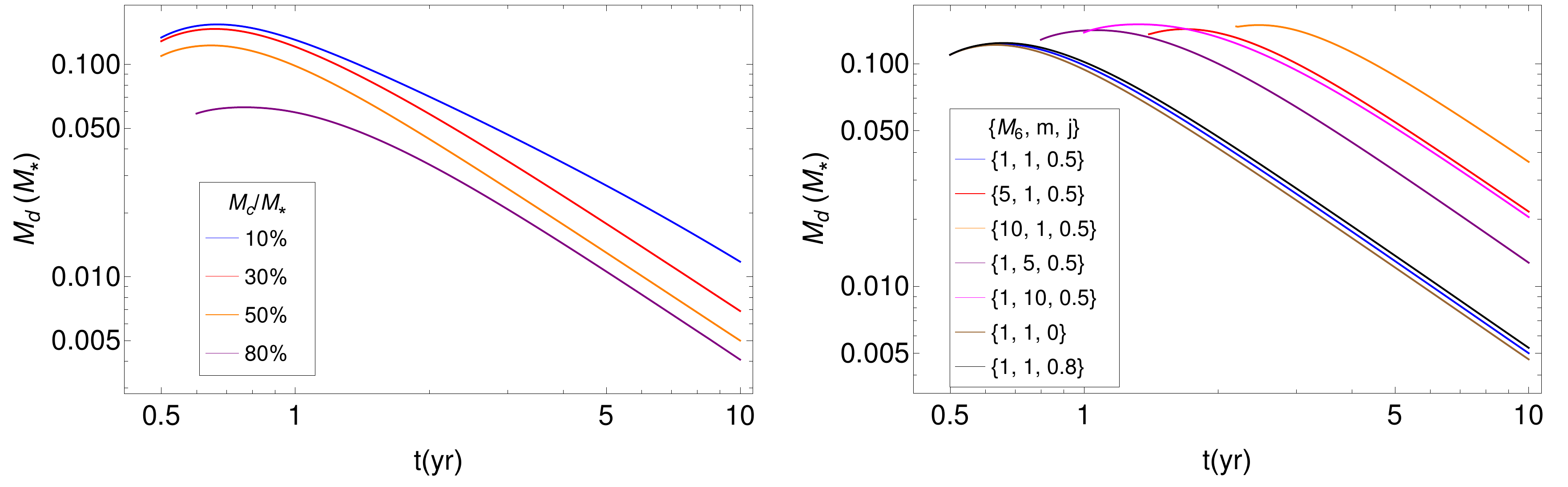}
\end{center}
\caption{The time evolution of the disc mass is shown for the mass fallback model MFR3 (section \ref{pds}) that is obtained for partial TDEs. (Left) The parameter set is taken to be M\Romannum{3} (section \ref{resmfr3}) and the circularization time taken for various lines are $t_c=~0.5~{\rm yr}~{\rm (blue)},~0.5~{\rm yr}~{\rm (red)},~0.5~{\rm yr}~{\rm (orange)}~{\rm and}~0.6~{\rm yr}~{\rm (purple)}$. The late time of disc mass is $M_{\rm d} \propto t^{p}$ where $p=~-1.37~{\rm (blue)},~-1.37~{\rm (red)},~-1.38~{\rm (orange)}~{\rm and}~-1.39~{\rm (purple)}$. (Right) The core mass is take to be $M_c= 0.5 M_{\star}$ and the circularization time taken for various lines are $t_c=~0.5~{\rm yr}~{\rm (blue)},~1.4~{\rm yr}~{\rm (red)},~2.2~{\rm yr}~{\rm (orange)},~0.8~{\rm yr}~{\rm (purple)},~1~{\rm yr}~{\rm (magenta)},~0.5~{\rm yr}~{\rm (brown)}~{\rm and}~0.5~{\rm yr}~{\rm (black)}$. The critical time corresponding to $\alpha =1$ varies with black hole mass and star mass as can be seen from Fig \ref{alpds} and thus we have taken the circularization time to have $\alpha <1$. The late time decline denoted by $p$ for various curves are $p=~-1.38~{\rm (blue)},~-1.37~{\rm (red)},~-1.35~{\rm (orange)},~-1.39~{\rm (purple)},~-1.37~{\rm (magenta)},~-1.37~{\rm (brown)}~{\rm and}~-1.37~{\rm (black)}$. See section \ref{resmfr3} for details. }
\label{mdisp} 
\end{figure*}

The mass fallback rate and its late time slope in case of partial TDEs depends on the surviving core mass $M_c$ (see equation \ref{ninf}). The mass fallback rates for various values of $M_c$ are shown in Fig. \ref{mfpTDE} and an approximated late time mass fallback rate is given by $\dot{M}_f \propto t^{-9/4}$ (see section \ref{pds}). We name the prior parameter set to be M\Romannum{3} given by $\{M_6,~m,~j,~M_c/M_{\star}\} = \{1,~1,~0.5,~0.5\}$. The evolution of $\alpha$ with $t_c$ is shown in Fig. \ref{alpds} and the critical time $t_{\rm cr}$ corresponding to $\alpha=1$ increases with an increase in the core mass. However, the increment in $t_{\rm cr}$ with an increase in the core mass is small. The increase in black hole mass and star mass increases the critical time and thus delays the onset of disc accretion. The surface density and disc height evolve in a manner similar to that shown for model MFR1 in Figs. \ref{surf} and \ref{hrp} respectively. The time evolution of disc mass is shown in Fig. \ref{mdisp} and the disc mass at late time decreases as $M_d \propto t^{p}$, where $p \sim -1.38$ which is higher than the declining rate obtained for full disruption TDEs (see sections \ref{resmfr1} and \ref{resmfr2}). This is because the mass fallback rate declines faster in case of partial TDEs as compared to full disruption TDEs.

\begin{figure}
\begin{center}
\includegraphics[scale=0.34]{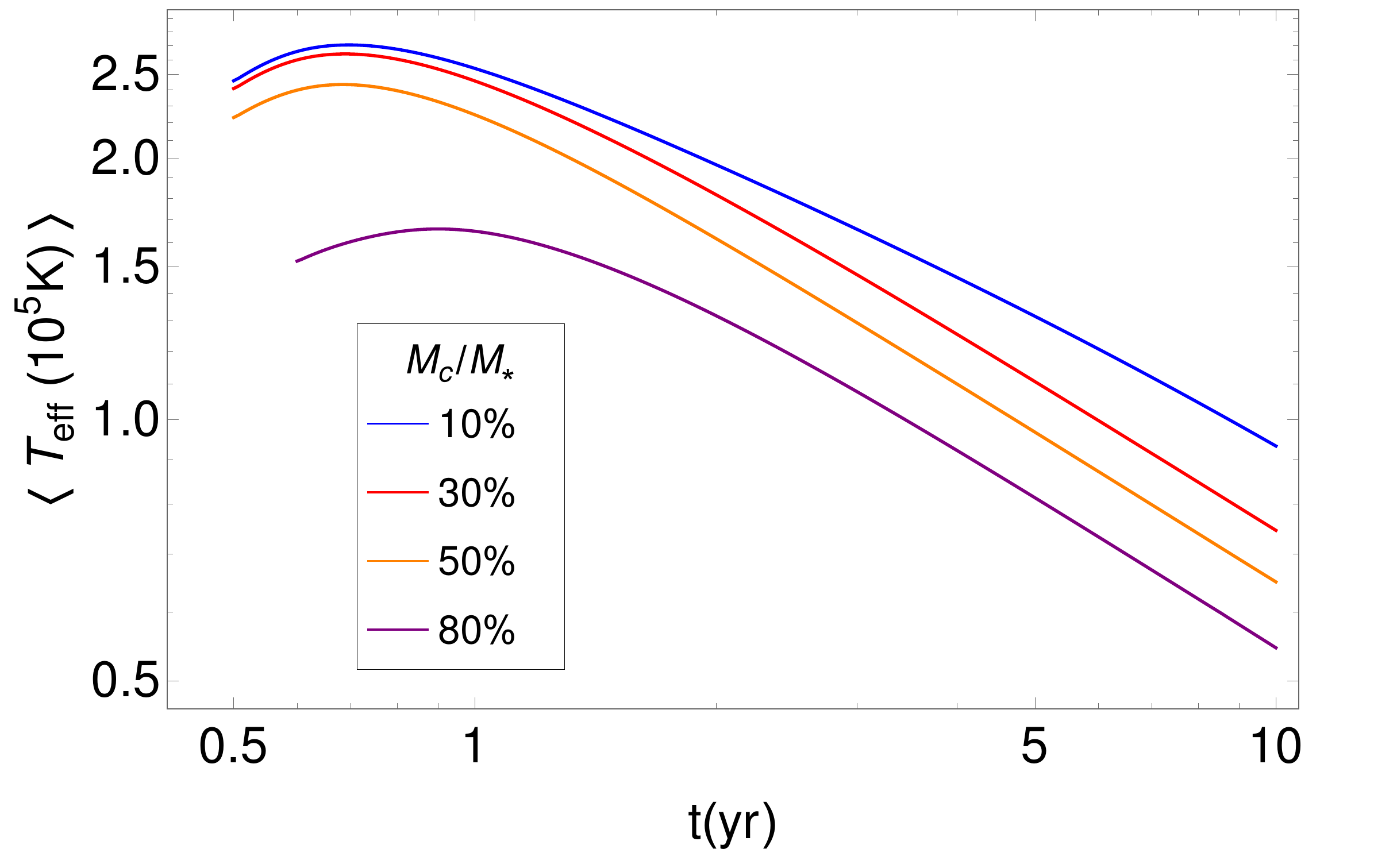}
\end{center}
\caption{The time evolution of mean effective temperature for the mass fallback rate given in MFR3 (section \ref{pds}) obtained for partial TDEs and the parameter set M\Romannum{3} (section \ref{resmfr3}) with various values for core mass. The circularization time is same as that shown in Fig \ref{mdisp}. The late time evolution of mean temperature is $\left<T_{\rm eff}\right> \propto t^{p_1}$ where $p_1= -0.56~{\rm (blue)},~-0.56~{\rm (red)},~-0.57~{\rm (orange)}~{\rm and}~-0.58~{\rm (purple)}$. See section \ref{resmfr3} for details. }
\label{temppds}
\end{figure}

\begin{figure}
\begin{center}
\subfigure[]{\includegraphics[scale=0.32]{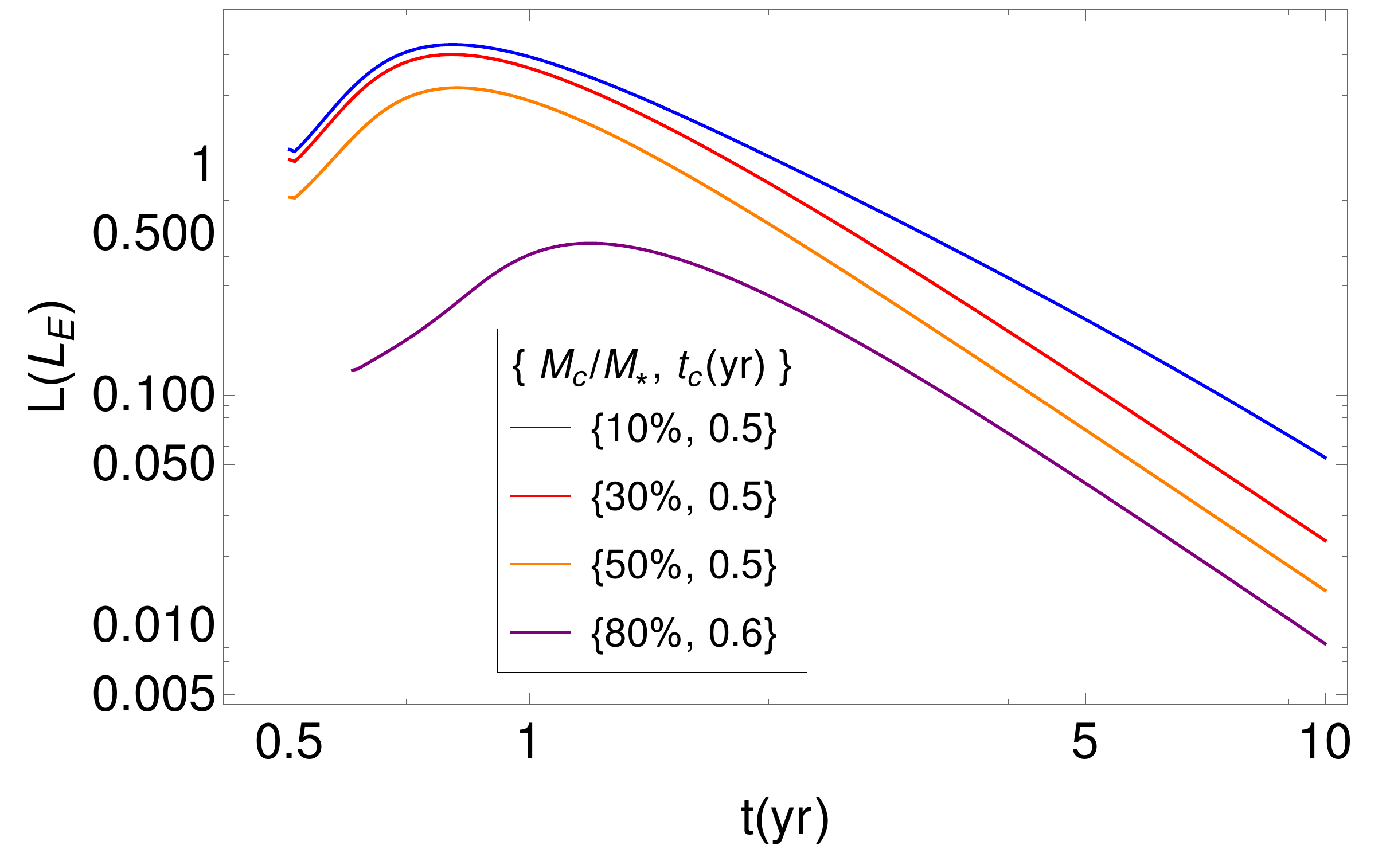}}
\subfigure[]{\includegraphics[scale=0.25]{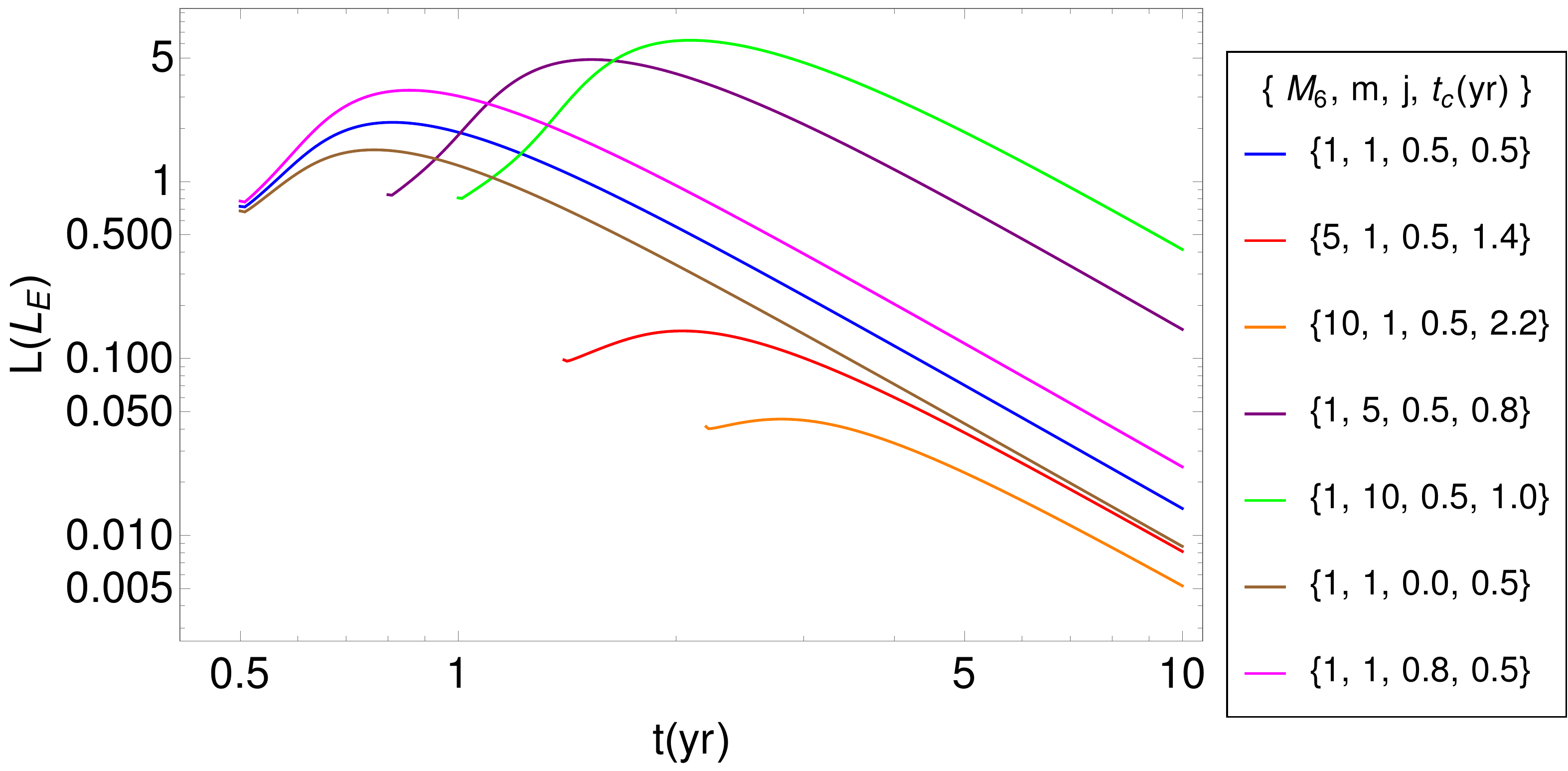}}
\end{center}
\caption{The time evolution of luminosity obtained for the mass fallback MFR3 (section \ref{pds}) corresponding to partial TDEs. (a) The parameter set is taken to be M\Romannum{3} (section \ref{resmfr3}) and the circularization time taken is shown in the legend. (b) The core mass is taken to be $M_c= 0.5 M_{\star}$. The late time profile of luminosity is given by $L \propto t^{n}$, where $n$ for various parameters have been shown in Table \ref{parsetp}. The Eddington luminosity is given by $L_E = 4 \pi G M_{\bullet} c/\kappa$, where $\kappa$ is the opacity due to Thomson scattering. See section \ref{resmfr3} for details.  }
\label{lvaryp}
\end{figure}

\begin{figure}
\begin{center}
\subfigure[]{\includegraphics[scale=0.26]{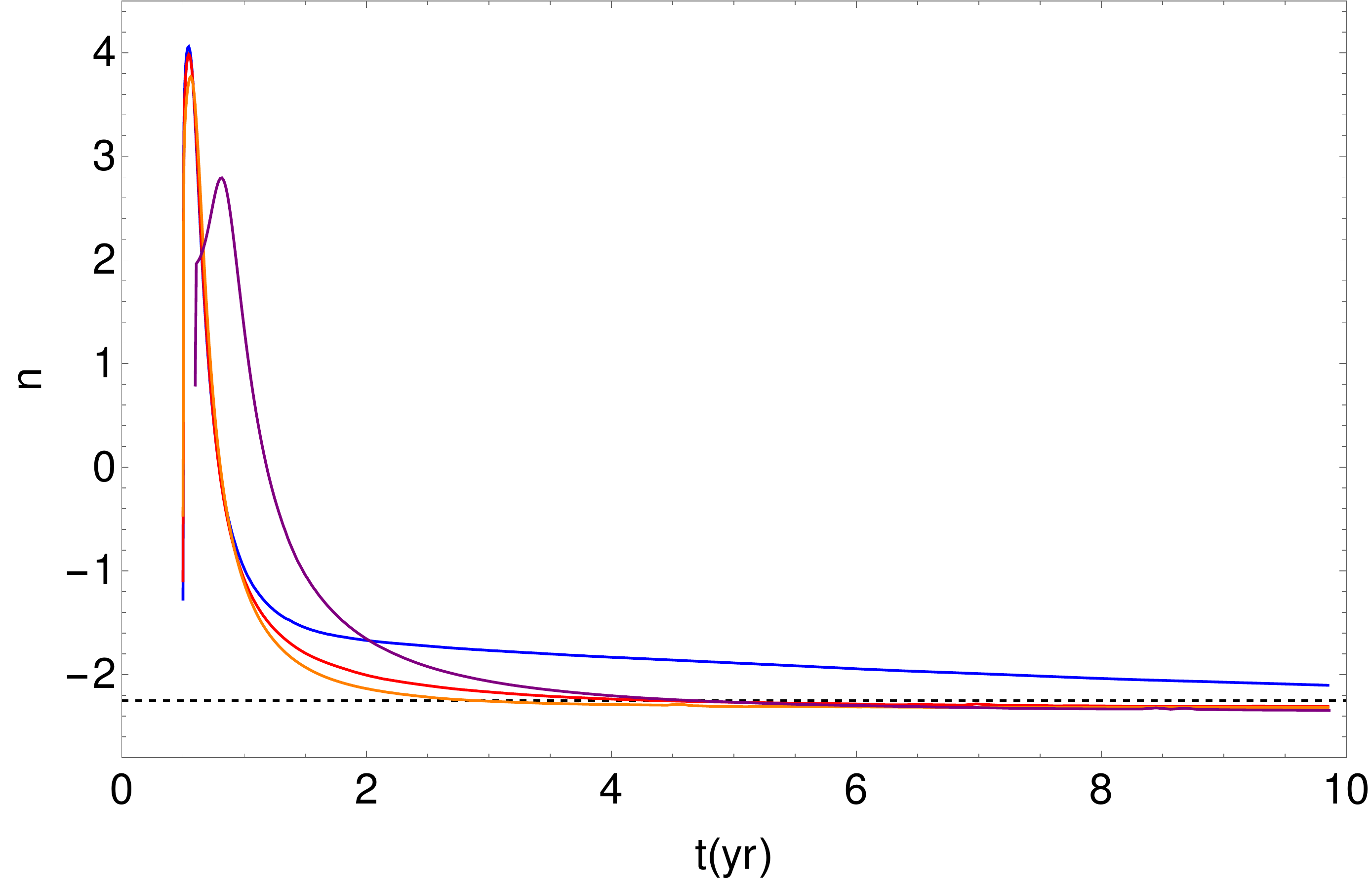}}
\subfigure[]{\includegraphics[scale=0.26]{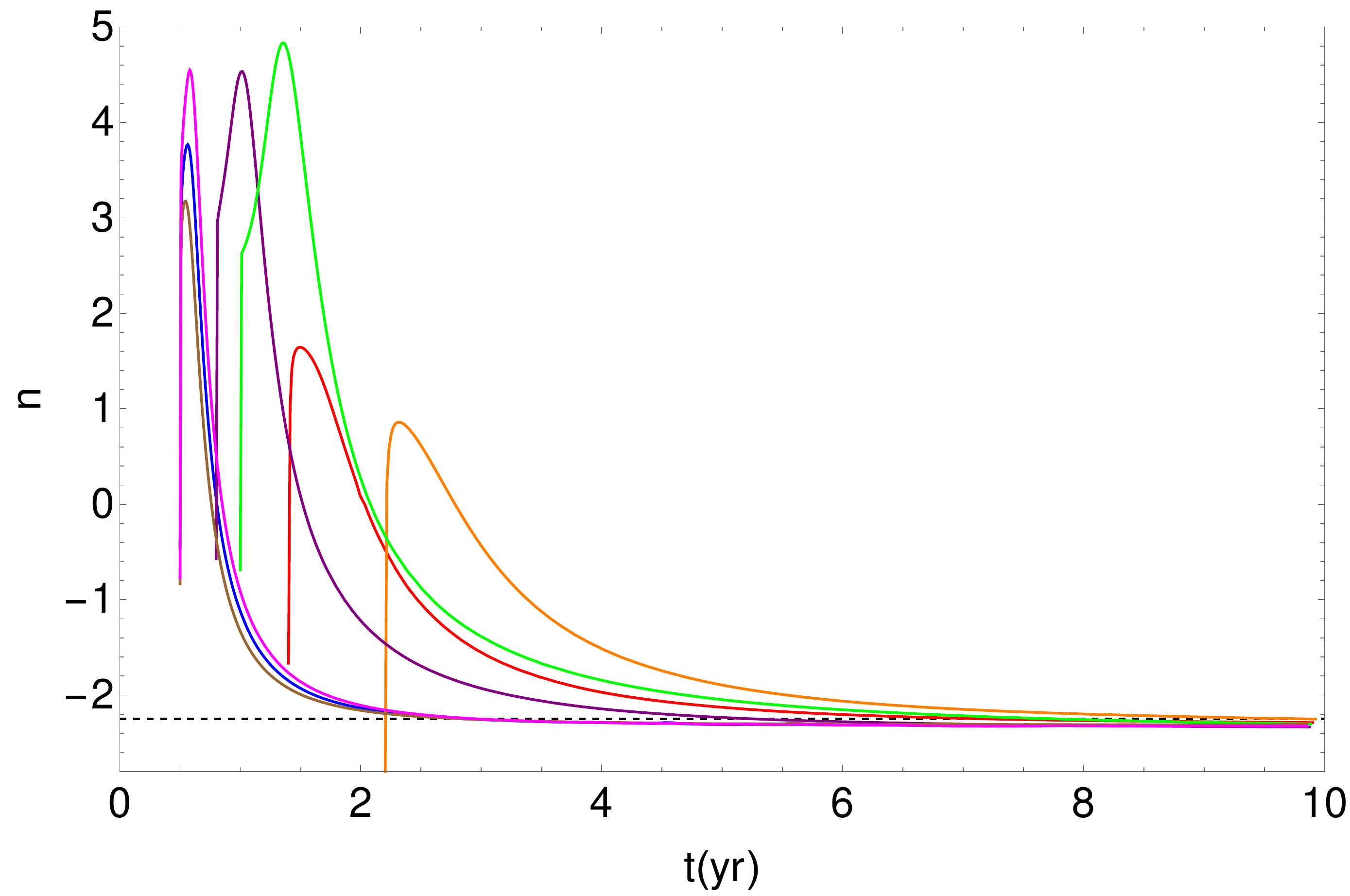}}
\end{center}
\caption{The time evolution of $n = \diff \ln L/\diff \ln t$ obtained for the mass fallback MFR3 (section \ref{pds}) corresponding to partial TDEs. (a) The parameter set is taken to be M\Romannum{3} (section \ref{resmfr3}) and the circularization time taken is shown in the legend of Fig. \ref{lvaryp}. The orange and purple lines are very close.  (b) The core mass is taken to be $M_c= 0.5 M_{\star}$. The black dashed line corresponds to $n=-9/4$ (see section \ref{pds}). The luminosity and the physical parameters corresponding to various lines are given in the Fig. \ref{lvaryp} and Table \ref{parsetp}. }
\label{nevp}
\end{figure}

The increase in core mass results in a decrease in total debris mass for accretion which results in a decrease in the mass fallback rate (see Fig. \ref{mfpTDE}). This results in a decrease in the surface density and thus the effective temperature as can be seen from Fig. \ref{temppds}. The late time decline in mean effective temperature is $\left<T_{\rm eff}\right> \propto t^{-0.57}$ and the power index varies slightly with an increase in the core mass. The bolometric luminosity also decreases with an increase in the core mass as can be seen from Fig. \ref{lvaryp}a due to a decrease in the debris mass for accretion resulting in low mass fallback rate and the declining rate of luminosity at late times increases with an increase in the core mass (see Table \ref{parsetp}). The bolometric luminosity increases with an increase in the star mass and black hole spin but decreases with an increase in the black hole mass. The late time declining rate $n$ shows insignificant variations with the black hole mass and spin, and the star mass (see Table \ref{parsetp}); the late time luminosity can be approximate to $L \propto t^{-2.3}$. For the luminosity shown in Fig. \ref{lvaryp}, the time evolution of $n$ is shown in Fig. \ref{nevp}. At late times, $n$ goes below the declining rate of mass fallback rate and this implies that the luminosity at late times decline faster than $L \propto t^{-9/4}$.

\begin{table}
\scalebox{0.837}{
\begin{tabular}{|c|ccccc|}
\hline
&&&&&\\
Model & Varying parameter & Values & $t_c(yr)$ & $n$ & Figure   \\
&&&&& \\
\hline
&&&&&\\
M\Romannum{3}a & $M_c/M_{\star}$ & 0.1 & 0.5 & -2.28 &\\
& (Mass of the & 0.3 & 0.5 & -2.3 & Fig. \ref{lvaryp}a\\
&  surviving core) & 0.5 & 0.5 & -2.32 &\\
&& 0.8 & 0.6 & -2.34 & \\
&&&&\\
\hline
&&&&&\\
M\Romannum{3}b & $M_6$ & 0.1 & 0.5 & -2.3 & Fig. \ref{lvaryp}b\\
& (Black hole mass & 5 & 1.4 & -2.24 & (blue, red and \\
& $M_{\bullet} = M_6 10^6 M_{\odot}$ ) & 10 & 2.2 & -2.25 & orange lines)\\
&&&&\\
\hline
&&&&\\
M\Romannum{3}c & $m$ & 0.1 & 0.5 & -2.3 & Fig. \ref{lvaryp}b\\
& (Stellar mass & 5 & 0.8 & -2.3  & (blue, purple and\\
& $M_{\star}= m M_{\odot}$)& 10 & 1 & -2.33 & green lines)\\ 
&&&& \\
\hline
&&&&\\
M\Romannum{3}d & $j$ & 0 & 0.5 & -2.3 &  Fig. \ref{lvaryp}b\\
& (Black hole prograde & 0.5 & 0.5 & -2.3 & (brown, blue and\\
& spin)& 0.8 & 0.5 & -2.32 & magenta lines)\\
&&&& \\
\hline
\end{tabular}
}
\caption{The parameter set M\Romannum{3} given by $\{M_6,~m,~j\} = \{1,~1,~0.5\}$ for the mass fallback MFR3 (section \ref{pds}) is used, and one parameter is varied to observe the variation in light curve (see Fig. \ref{lvaryp}). The critical value of $t_c$ above which $\alpha < 1$ is a function of black hole mass and spin, and star mass (see Fig. \ref{alpds}). Thus, the value of $t_c$ we have considered and shown in the fourth column varies. The late time profile of luminosity is given by $L \propto t^{n}$ and the time evolution of $n$ is shown in Fig. \ref{nevp}. The $n$ shown in the fifth column is steady value attained at the late time. See section \ref{resmfr3} for details.}
\label{parsetp}
\end{table}

\section{Discussion}
\label{discus}

A TDE is a complicated phenomenon that includes the consumption of stars through the stellar dynamical process, and the formation and accretion of the disc. The formation of an accretion disc involves stream-stream interactions that result in an exchange of angular momentum for circularization. \citet{2016MNRAS.461.3760H} have shown through numerical simulations that a thin disc is formed if the debris during circularization is radiatively efficient else a thick disc is formed. We assume the circularization to be radiatively efficient such that the geometry of the formed seed disc is thin. To avoid complexity in circularization timescale, we have considered it to be a free parameter whose variation indeed affects the disc evolution and luminosity. The increase in the circularization time decreases the bolometric luminosity.

We obtain the vertically integrated mass and momentum conservation equations in an equatorial plane and assuming a thin disc geometry $H/R \ll 1$, we obtain the time evolution of surface density in the disc. \citet{2018MNRAS.481.3348B} constructed a thin disc model where all the debris has formed an accretion disc with a Gaussian density profile that evolves via accretion and the outer radius evolves with time. We include the mass fallback at the constant outer radius that provides mass to the disc. We consider both the full and partial disruption TDEs that have different time evolution. The mass fallback rate is obtained assuming the star is unperturbed until it reaches the pericenter where an impulse of short duration breaks the star. In case of full disruption TDEs, the debris returns to the pericenter following a Keplerian orbit with a mass fallback rate that at the late time is given by $\dot{M}_f \propto t^{-5/3}$. A star is tidally deformed before reaching the pericenter which affects the stellar density structure and thus the stellar density structure within the star at the pericenter is different from the original polytrope. \citet{2009MNRAS.392..332L} showed through numerical simulations that the ratio $\diff M/ \diff E_d$ evolves at early times after disruption and settles to a distribution later which is wider than the $\diff M/ \diff E_d$ obtained through an impulse approximation (see equation \ref{med}). Similar results are also seen in the numerical simulations by \citet{2019ApJ...872..163G} where they have included the stellar spin. This results in a difference in the mass fallback rate obtained via analytic and numerical calculations at early times but the late time decline is nearly same \citep{2009MNRAS.392..332L,2019ApJ...872..163G}.  Even though this difference in the mass fallback rate can affect the initial disc luminosity, the late time luminosity decline will be similar to that we have obtained using the analytic model of mass fallback rate.

The mass fallback rate in case of partial TDEs depends on the surviving core mass $M_c$ and with an increase in $M_c$, the mass fallback rate decreases (see Fig \ref{mfpTDE}). The mass fallback rate for TDEs with various penetration factor $\beta = r_t/r_p$ is simulated by \citep{2013ApJ...767...25G,2017A&A...600A.124M} and a full disruption occur if $\beta \geq \beta_d$, where $\beta_d \sim 0.9~(\Gamma = 5/3)~{\rm and}~2.0~(\Gamma = 4/3)$, and $\Gamma$ is the polytrope index of star. For $\beta < \beta_d$, they have showed that $n_{\infty}$ ($\dot{M} \propto t^{n_{\infty}}$ at late times; see section \ref{pds}) is smaller than $-5/3$ implying a higher decline in the mass fallback rate. Thus, the pericenter for partial TDEs can be higher than $r_t$ and \citep{2013ApJ...767...25G} showed that for $\Gamma=5/3$, $n_{\infty}= -2.2$ for $\beta \sim 0.7$. We use the simple approximated model of \citet{2019ApJ...883L..17C} in obtaining the mass fallback rate for our calculation as it generates both rising and declining phases of the mass fallback rate.

The viscosity constant $\alpha$ decreases with an increase in the circularization time $t_c$ as can be seen from Figs. (\ref{alfds}, \ref{alfdsr}, \ref{alpds}). The critical circularization time $t_{\rm cr}$ corresponding to $\alpha=1$  (see Figs. \ref{alfds}, \ref{alfdsr} and \ref{alpds}) increases with black hole mass and star mass. The increment is higher for the black hole mass than the stellar mass. This implies a delay in the onset of disc accretion with an increase in the black hole mass and the star mass. The $\alpha \leq 1$ provides a constrain on the circularization time of the disc and this constrain is crucial as it provides a minimum time required to form a valid and thermally stable thin accretion disc. The time evolution of surface density in the disc for fallback model MFR1 is shown in Fig. \ref{surf} which increases at the initial time and then decreases as the mass fallback rate decreases with time. Other mass fallback models (MFR2 and MFR3) also results in a similar evolution. The addition of mass at the outer radius increases the surface density at the lower radius but close to the disc inner radius, the density shows a significant drop.

In equation (\ref{dllk}), at inner radius which is taken to be ISCO, we have $x^2 - 6 x \pm 8 j \sqrt{x} -3 j^2=0$, that results in the partial derivative of angular momentum at inner radius $\partial \ell_{\rm K} /\partial x =0$. The surface density evolution given by equation (\ref{beq}) have a solution at the inner radius if $\bar{S}_{\phi}^R=0$ or $(\partial/\partial x)[x \bar{S}_{\phi}^R]=0$. The first case ($\bar{S}_{\phi}^R=0$) corresponds to a vanishing ISCO stress that implies surface density at inner radius $\Sigma=0$, and the second case ($\partial[x \bar{S}_{\phi}^R]/\partial x=0$) is for a finite ISCO stress that implies $\partial \Sigma/\partial x =0$. \citet{2018MNRAS.481.3348B} has constructed the thin disc model without fallback for both the cases. However, the solution we have obtained for a disc with fallback resulted in a vanishing ISCO stress and the declining surface density implies that any mass entering at the ISCO radius is accreted instantaneously. The surface density at the beginning is assumed to be Gaussian and as the time progresses, the mass added to the disc by infalling debris is transported inside which results in an increase in the surface density at lower radii. The decrease in mass fallback rate results in a decrease in the mass added to the disc and since the accretion is ongoing, the disc mass decreases.

The time evolution of disc mass for the mass fallback models MFR1, MFR2 and MFR3 is shown in Figs. (\ref{mdis}, \ref{mdisr}, \ref{mdisp}) respectively. The late time evolution of disc mass follows a power law given by $M_d \propto t^{p}$, where $p \sim -1.05$ for a full disruption TDE (see Figs \ref{mdis} and \ref{mdisr}) and $p \sim -1.38$ for a partial disruption TDE with a core mass of $M_c = 0.5 M_{\star}$ (see Fig \ref{mdisp}). The mass fallback rate at late time declines faster in partial TDEs ($\dot{M}_f \propto t^{-2.25} $) compared to full disruption TDEs ($\dot{M}_f \propto t^{-5/3} $). Thus, an increase in the mass fallback declining rate increases the declining rate of disc mass at late times. The power index $p$ decreases with an increase in $M_c$ implying a faster evolution but the variation in $p$ is small. 
If we assume the disc mass $M_{\rm d} = \bar{\Sigma}(t) \diff \mathcal{A}$, where $\bar{\Sigma}(t)$ is the averaged surface density, the constant inner and outer radii implies a constant disc area that results in $\bar{\Sigma}(t) \propto M_{\rm d}$. For a full disruption TDE, $\bar{\Sigma}(t) \propto t^{-1.05}$ and for a partial TDE with $M_c = 0.5 M_{\star}$, $\bar{\Sigma}(t) \propto t^{-1.38}$. The surface density due to mass fallback at the outer radius is $\Sigma_c(t) \propto \dot{M}_f^{3/5}$ (see equation \ref{sxc}). At late times, $\Sigma_c(t) \propto t^{-1}$ for a full disruption TDE and $\Sigma_c(t) \propto t^{-1.35}$ for a partial TDE. Thus, the late-time evolution of the average disc density $\bar{\Sigma}(t)$ is close to the evolution of surface density added by the infalling debris at the outer radius.

The viscous stress converts the mechanical energy of the fluid in heat that is emitted as radiation. The effective temperature of the disc evolves with the surface density (see above equation \ref{lumeq}) and is shown in Fig. \ref{temp}. The temperature shows an initial rise due to an increase in the surface density and later decreases as the disc surface density decreases. The mean effective temperature at late times decreases as $\left<T_{\rm eff}\right> \propto t^{-0.43}$ for a full disruption TDE (see Figs. \ref{temp} and \ref{tempr}) and $\left<T_{\rm eff}\right> \propto t^{-0.57}$ for a partial disruption TDE with a core mass of $M_c = 0.5 M_{\star}$ (see Fig \ref{temppds}). The effective temperature obtained using a steady thin disc accretion model is given by $T \propto \dot{M}_f^{1/4}$ \citep{2011MNRAS.410..359L} and decreases as $T \propto t^{-5/12}$ for a full disruption TDE ($\dot{M}_f~ \propto t^{-5/3}$) and $T \propto t^{-9/16}$ for a partial disruption TDE ($\dot{M}_f~\propto t^{-9/4}$). Thus, the late time evolution of the relativistic thin disc mean temperature is close to the temperature evolution in a steady thin disc. The peak of effective temperature is $\left<T_{\rm eff}\right>_{\rm peak} \approx ~ {\rm few}~\times ~10^5~{\rm K}$. The X-ray temperature obtained from a blackbody fit to the X-ray observations of the source such as ASAS-SN 14li is $\sim 10^{5}~{\rm K}$ \citep{2016MNRAS.455.2918H}, XMMSL1 J061927.1-655311 is $\sim 1.4 \times 10^6~{\rm K}$ \citep{2014A&A...572A...1S}, Abell-1795 is $\sim 1.2 \times 10^{6}~{\rm K}$ \citep{2013MNRAS.435.1904M} and NGC-3599 is $\sim 1.1 \times 10^6~{\rm K}$ \citep{2008A&A...489..543E}. Thus, the peak of mean temperature we have obtained is of the order of blackbody temperature expected from the X-ray observations. 

The blackbody fit to optical and UV observations shows a temperature that is ten times smaller than the mean temperature. We are calculating the emission from the disc only, whereas the optical and UV emissions can also arise from outflowing winds or the reprocessing of X-ray emission. The presence of outflowing winds leads to a thick disc whose effective disc temperature may be smaller than the thin disc due to low radiative efficiency. The observed emission in such a disc is a sum of both disc and wind emissions leading to a mean blackbody temperature different from the thin disc \citep{2009MNRAS.400.2070S,2014ApJ...784...87S}. The X-ray emissions from the disc can be reprocessed by either an outflowing wind or the debris unbound at the moment of disruption. The reprocessing depends on the emission from the centre, the velocity and the element compositions in the reprocessing layer which decides the optical depth \citep{2014ApJ...783...23G,2016ApJ...827....3R}. This reprocessing of emissions modifies the spectral luminosities and their evolution profiles such as for ASAS-SN 15oi \citep{2016MNRAS.463.3813H}.   

The relativistic thin disc accretion model without fallback by \citet{2019MNRAS.489..132M} has resulted in a bolometric luminosity given by $L \propto t^{-1.14}$ which is close to $L \propto t^{-1.2}$  obtained for non-relativistic discs \citep{1990ApJ...351...38C}. However, we include the mass fallback rate at the constant outer radius and the luminosity is affected by the physical parameters $\{M_6,~m,~j,~\lambda,~M_c \}$. The effect on the late time evolution of luminosity ($L \propto t^n$) for variations in the physical parameters is given in Tables (\ref{parset}, \ref{parsetr}) for mass fallback MFR1 and MFR2 (full disruption) and in Table \ref{parsetp} for mass fallback MFR3 (partial disruption). The initial evolution of luminosity shows a variation with the physical parameters, but the change in late time declining rate (given by $n$) is small. The temporal evolution of $n$ for mass fallback models MFR1, MFR2 and MFR3 are shown in Figs. (\ref{nev}, \ref{nevr}, \ref{nevp}) respectively. The $n$ can be approximate to $-1.8$ for full disruption TDEs and $-2.3$ for partial disruption TDEs. \citet{2011ApJ...736..126M} constructed a time-dependent non-relativistic accretion model with a fallback at a constant outer radius and without any outflowing wind; they showed that the bolometric luminosity at late times decays as $L \propto t^{-5/3}$. The bolometric luminosity in our model declines faster than the luminosity inferred from $L \propto \dot{M}_f$.

The surface density at the initial time is assumed to be Gaussian and the corresponding disc effective temperature increases inward as can be seen from Fig \ref{temp} (blue line). As the disc evolves, the surface density close to the inner radius drops in a short duration (see paragraph 5 in section \ref{discus}) which results in a decrease in the disc temperature close to the inner radius. In the luminosity figures for all the models, we can notice a short duration decline in the luminosity at initial times. This is because of the short duration decline in the temperature close to the inner radius during initial times that results in a decrease in the peak temperature. This results in a decline in the luminosity. With time, the mass from the outer radius is transported inward resulting in an increase in the surface density and thus the temperature reaching a maximum value and then decline with time. This results in a rise and decline in the luminosity evolution.

The luminosity ratio $L/L_E$ increases with an increase in the star mass and black hole spin, but decreases with an increase in the black hole mass. The luminosity exceeds the Eddington luminosity for low mass black holes and thus the disc is super-Eddington. The super-Eddington discs have strong radiation pressure that thickens the disc but we have $H/R \ll 1$ at all times (see Fig. \ref{hrp}). This is because we are not taking the advection which is crucial for the super-Eddington discs. The luminosity ratio reduces with an increase in the circularization time. Thus, a delay in the onset of accretion reduces the luminosity but increases the rise time of luminosity (see Fig. \ref{lvary}d). We have assumed the mass accretion rate at the outer radius to be equal to the mass fallback rate and derived the surface density at the outer radius given by $\Sigma_c(t) \propto \dot{M}_f^{3/5}$. The super-Eddington disc results in an outflowing wind that carries mass and angular momentum which in result affects the radial velocity. The radial velocity in the super-Eddington disc is comparable to the azimuthal velocity and an increase in the radial velocity results in a decrease in the surface density at the outer radius if the mass accretion rate is equal to the mass fallback rate. Then, the time evolution of surface density at the outer radius depends on the mass accretion and outflow rates. We have taken the angular momentum of a circular orbit in our calculations which results in a vanishing ISCO stress (see paragraph 5 in section \ref{discus}). The angular momentum in a super-Eddington disc can deviate from the circular orbit momentum and the deviation increases with an increase in the mass accretion rate \citep{2009ApJS..183..171S}. This can result in a disc evolution with non-vanishing ISCO stress. To study these effects, it is crucial to include the advection and outflow in the energy, mass and momentum conservation equations. Such a disc will resemble the advection dominated inflow-outflow (ADIO) disc whose stability depends on the Bernoulli function \citep{1999MNRAS.303L...1B}. Our accretion model neither includes the advection nor the outflowing wind and does not resemble an ADIO accretion flow. We will simulate an ADIO disc for TDEs in future. However, our model fit to the observations will yield the expected physical parameters such as black hole mass and spin, star mass and the circularization time. This fit will also be useful in predicting the late time evolution of the luminosity. Our model is directly applicable to higher mass black holes and low mass stars.

A hyperaccretion model for TDEs where the disc is highly super-Eddington is constructed by \citet{2014ApJ...781...82C}, where the accretion flow is taken to be a ZEro-BeRnoulli Accretion (ZEBRA) flow. This results in a highly inflated envelope which is marginally bound and any excess energy added will start unbinding material. The highly critical accretion close to the inner radius results in bipolar jets that carry away excess energy. Using a realistic mass fallback rate obtained through numerical simulations in ZEBRA model, \citet{2018MNRAS.478.3016W} calculated the jet luminosity which declines as $t^{-5/3}$ at late times. The relativistic simulation for TDEs without fallback by \citet{2019MNRAS.483..565C} showed that the super-Eddington disc launch a radiation-driven outflow with a radiative efficiency $\eta_r \lesssim 0.01$. In case of a high spin rotating black hole with a weak magnetic field, a non-relativistic jet is launched and for a high magnetic field, the jets are relativistic.

We consider the disc to be a thin disc dominated by gas pressure at all times and we can see from Figs. \ref{lvary}b (blue line) and \ref{hrp} that even though the disc luminosity is super-Eddington, the low height to radius ratio implies a thin disc structure. The time-dependent ZEBRA model is governed by the global conservation of mass and angular momentum \citep{2014ApJ...781...82C}, whereas we consider the mass and momentum conservations locally and derived the time evolution of surface density. The mass enclosed in the ZEBRA envelope increases with time and the jet luminosity given by $L_j \propto \dot{M}_{\rm acc}$ decreases with a decrease in the mass accretion rate. In our accretion model, the mass of the disc increases initially and decreases later following a power law given by $M_d \propto t^{p}$, where $p$ is $-1.05$ for full disruption TDEs and $-1.38$ for partial disruption TDEs. Both ZEBRA and our model shows the decrease in luminosity ratio $L/L_E$ with an increase in the black hole mass. The ratio $L/L_E$ is smaller in our model compared to the ZEBRA model for the same black hole and star masses. In ZEBRA model, the disc is super-Eddington for $10^7 M_{\odot}$ black hole, whereas, in our model, the disc is sub-Eddington (see Fig. \ref{lvary}a). This is because the critical circularization time corresponding to viscosity constant $\alpha=1$ increases with an increase in the black hole mass (see Fig. \ref{alfds}) resulting in a delay in the onset of accretion. This delay in the initial time results in a decrease in the disc luminosity but an increase in the rising duration (see Fig. \ref{lvary}d). The late time luminosity in our model shows a faster decline than the luminosity obtained using $L \propto \dot{M}_f$. Based on the luminosity, the two models can be distinguished by the peak bolometric luminosity, the duration of the rising phase and the late time decline in the luminosity which is slightly higher in our model (see Figs. \ref{nev}, \ref{nevr} and \ref{nevp}). 

We have also computed the luminosity in various spectral bands assuming a blackbody disc emission and the optical and UV luminosity shows a slow evolution compared to the X-rays. However, some TDEs show spectral evolution that varies from the blackbody emission and could be attributed to various other mechanisms such as non-thermal emissions, reprocessing and winds. Our accretion model is good for TDEs where the non-thermal emission or other non-disc processes are absent and the light curves are from disc emissions. We can fit our model to the observed light curves to extract the physical parameters. Here, we have not considered in deep a relativistic radiative transfer model for the spectral emissions and will be taken up later. However, this simple emission model considered here is good enough as most of the equatorial disc assumed are in the Newtonian region and the relativistic effects dominate close to the ISCO radius.

\section{Summary and conclusions}
\label{sac}

We have developed a time-dependent relativistic accretion model in an equatorial plane with the pressure dominated by gas pressure and a mass fallback at the constant outer radius. We studied the accretion dynamics for both full and partial disruption TDEs. The earlier relativistic accretion models for TDEs does not include the mass fallback. We have derived an explicit form for the vertical height of the disc in terms of the angular momentum and the radial velocity (see equation \ref{height}), that reduce to the non-relativistic result of thin disc \citep{2002apa..book.....F} for $j=0$ and $R \gg 3R_g$ (see below equation \ref{heigd}). We have considered the disc structure to be thin and is satisfied by our accretion model (see Fig. \ref{hrp}). 

Our time-dependent accretion model results in a vanishing ISCO stress and any mass entering the ISCO is immediately accreted by the black hole resulting in a decline in the surface density close to the inner radius. The disc mass initially grows and then decreases with a late time evolution given by $M_{\rm d} \propto t^{-1.05}$ for full disruption TDEs and $M_{\rm d} \propto t^{-1.38}$ for partial TDEs. The late time evolution of average disc surface density closely resemble the time evolution of surface density at the outer radius.

The effective temperature of the disc shows an evolution similar to the surface density which increases initially and follows a gradual decline later. The peak temperature at any time lies close to the inner radius implying high energy emissions dominate near the inner radius. The mean effective temperature is of the order of $10^5~{\rm K}$ which is expected and observed from the X-ray observations of the TDEs. The mean effective temperature shows a power decline at late times given by $\left<T_{\rm eff}\right> \propto t^{-0.43}$ for full disruption TDEs and $\left<T_{\rm eff}\right> \propto t^{-0.57}$ for partial TDEs, which is close to $ T \propto \dot{M}_f^{1/4}$ obtained for a steady structured thin disc model with accretion rate equal to the mass fallback rate \citep{2011MNRAS.410..359L}.

The bolometric luminosity shows an initial rise and then decreases with time following a power-law decline at late times given by $L \propto t^{n}$. The obtained values of $n$ for various physical parameters is $n \sim -1.8$ for full disruption TDEs and $n \sim -2.3$ for partial TDEs. The late time luminosity decline is higher than the luminosity obtained assuming $L \propto \dot{M}_f$. The time evolution of $n$ for full and partial disruption TDEs are shown in Figs. \ref{nev} (model MFR1), \ref{nevr} (model MFR2) and \ref{nevp} (model MFR3). The bolometric luminosity increases with an increase in stellar mass $M_{\star}$ but decreases with an increase in the black hole mass. The black hole spin $j$ reduces the inner radius of the disc, and the luminosity increases with $j$. The prograde stellar spin increases the luminosity, whereas the retrograde stellar spin decreases the bolometric luminosity. In case of partial TDEs, the bolometric luminosity decreases with an increase in the core mass. The increase in core mass also results in a faster decline in the bolometric luminosity.  Our spectral emissions are from the disc only as we have assumed the disc to be blackbody and thus can be applied to the observations where the spectral emissions show a gradual decline with time. Our model is useful in extracting the physical parameters such as black hole mass and spin, stellar mass and the circularization time from observations. Our model is also useful in obtaining the emission from the central relativistic disc to study the effect in reprocessing dynamics due to the mass fallback rate.

We have used an impulse approximation in calculating the mass fallback rate for our accretion model. In reality, the star is deformed before reaching the pericenter that changes the density structure within the star. This affects the mass fallback rate at initial times (see paragraph 2 in section \ref{discus}). \citet{2018MNRAS.478.3016W} applied the mass fallback rate from the numerical simulation in their super-Eddington disc model ZEBRA. We will apply the realistic mass fallback rate obtain through numerical simulations for both partial and full disruption TDEs in future. The difference in the mass fallback rate obtained through simulation and analytic model is higher at the initial time but the late time evolution follows a similar decline. Thus, the late time decline in disc luminosity will be same for both numerical and analytic mass fallback (see Fig. 7 in \citealp{2018MNRAS.478.3016W}). 

The relativistic accretion disc we have constructed here is equatorial where the relativistic effects are close to the inner radius only. We have included the mass fallback rate at the outer radius derived from the Newtonian dynamics and neglected the deep TDE encounters. We will study the effect of the relativistic mass fallback rate on the disc evolution in the future. Since our disc is equatorial, it lacks the relativistic precession effects that may impact the accretion dynamics and the emission. \citet{2019MNRAS.487.4965Z} have studied the effect of fallback on an inclined disc that is assumed to have a disc with a steady-state accretion rate. The relativistic precession on a time-dependent inclined disc with fallback may affect the emission light curves and result in a time lag in spectral band emissions due to difference in precession at various radii. 

\section*{Acknowledgements}

We thank the referee for insightful suggestions that have improved our paper.




\bibliographystyle{mnras}
\bibliography{ref} 




\appendix

\section{Derivation of basic equations}
\label{dbe}

We transform the Kerr metric in Boyer-Lindquist coordinate $\{t,~r,~\theta,~\phi\}$  to cylindrical coordinate $\{t,~R,~\phi,~z\}$, using $R=r \sin\theta$ and $z=r \cos\theta$. The space-time metric in the geometrical units ($c=G =1$) with the signature ($- + + +$), is given by 

\begin{multline}
\diff S^2 = -\left[1-\frac{2 M (R^2+z^2)^{3/2}}{(R^2+z^2)^2+a^2 z^2}\right] \diff t^2 - \frac{4 M a R^2 \sqrt{R^2+z^2}}{(R^2+z^2)^2+a^2 z^2} \diff t \diff \phi \\ + \frac{(R^2+z^2)^2+a^2 z^2}{(R^2+z^2)^2}\left[\frac{R^2}{R^2+z^2- 2 M \sqrt{R^2+z^2} +a^2}+ \frac{z^2}{R^2+z^2}\right] \diff R^2 + \\ \frac{(R^2+z^2)^2+a^2 z^2}{(R^2+z^2)^2} \left[\frac{z^2}{R^2+z^2- 2 M \sqrt{R^2+z^2} +a^2}+ \frac{R^2}{R^2+z^2}\right] \diff z^2 + \\ \frac{2 R z [(R^2+z^2)^2+a^2 z^2]}{(R^2+z^2)^2} \left[ \frac{1}{R^2+z^2- 2 M \sqrt{R^2+z^2} +a^2}- \frac{1}{R^2+z^2}\right] \diff R \diff z + \\ \frac{R^2}{R^2+z^2} \left[R^2+z^2+a^2 + \frac{2 M a R^2 \sqrt{R^2+z^2}}{(R^2+z^2)^2+a^2 z^2}\right] \diff \phi^2.
\end{multline}

In the limit of thin disc $z \ll R$, the metric tensors are given by 

\begin{eqnarray}
g_{\rm t t}&=& -1+ \frac{2 M }{R}- \frac{M (2 a^2+ R^2) z^2}{R^5}, \\ 
g_{\rm t R}&=& g_{\rm R t} = 0, \\
g_{\rm t \phi} &=& g_{\rm \phi t} =  -\frac{2 M a}{R} + \frac{M (2 a^3 + 3 a R^2)z^2}{R^5}, \\
g_{\rm t z} &=& g_{\rm z t} = 0, \\
g_{\rm R R} &=& \frac{R^2}{R^2 -2 M R + a^2} + \nonumber \\
 &&  \frac{[2 a^4 + 3 a^2 (R^2-2M R)+ M (4 M R^2-3 R^3) ] z^2}{R^2(R^2- 2 M R +a^2)^2}, \label{grr} \\
g_{\rm R \phi} &=& g_{\rm \phi R} = 0, \\
g_{\rm R z} &=& g_{\rm z R} = 2 \left(-1+ \frac{R^2}{R^2 - 2 M R + a^2}\right) \frac{z}{r}, \label{gpp}\\
g_{\rm \phi \phi} &=& \frac{R^4+ a^2 R^2 + 2 M a^2 R}{R^2} - \nonumber \\ 
&&  \frac{a^2(R^4 + M R (2 a^2 + 5 R^2)) z^2}{R^6}, \\
g_{\rm \phi z} &=& g_{\rm z \phi} = 0, \\
g_{\rm z z} &=& 1 + \left[\frac{a^2- R^2}{R^2} + \frac{R^2}{R^2- 2 M R + a^2} \right] \frac{z^2}{R^2},
\end{eqnarray}

\noindent which is the same as the metric tensor given in \citet{2015PhyU...58..527Z}. At the equatorial plane ($z=0$), the space-time metric reduces to 

\begin{equation}
\diff S^2 = - \left(\frac{R- 2 M}{R}\right) \diff t^2 - \frac{4 M a }{R} \diff t \diff \phi + \frac{R^2}{\Delta} \diff R^2 + \frac{A}{R^2} \diff \phi^2 + \diff z^2,
\end{equation}

\noindent where $\Delta = R^2- 2 M R +a^2$ and $A= R^4+ a^2 R^2 + 2 M a^2 R$. The stress-energy tensor is given by \citep{1973grav.book.....M}

\begin{equation}
T^{\alpha \beta} = \rho u^{\alpha} u^{\beta} + P h^{\alpha \beta} + S^{\alpha \beta} + \tau^{\alpha \beta},
\end{equation}

\noindent $\rho$ is the density, $u^{\alpha}$ is the contravariant components of four velocity, $P$ is the total pressure (sum of radiation and gas pressure), $S^{\alpha \beta}$ is the viscous stress tensor, $h^{\alpha \beta}= g^{\alpha \beta} + u^{\alpha} u^{\beta}$ is the projection tensor and $\tau^{\alpha \beta}= q^{\alpha} u^{\beta} + u^{\alpha} q^{\beta}$ with radiative energy flux $q^{\alpha}$. The total energy density of the fluid is given by $\varepsilon= \rho c^2 + \Pi$, where $\Pi$ is the internal energy. Following \citet{1996ApJ...471..762A}, in non-relativistic approximations, $\Pi \ll \rho c^2$ and the pressure $P \ll \rho c^2$, such that the stress-energy tensor reduces to 

\begin{equation}
T^{\alpha \beta} = \rho u^{\alpha} u^{\beta} + P g^{\alpha \beta} + S^{\alpha \beta} + \tau^{\alpha \beta}.
\end{equation}

In Boyer-Lindquist coordinates using orthonormal tetrad in the local non-rotating frame (LNRF) \citep{1972ApJ...178..347B}, the contravariant components of four velocities are given by \citep{2011ApJS..195....7X} 

\begin{eqnarray}
u^t &=& \frac{\gamma_L A^{1/2}}{r \Delta^{1/2}}, \\
u^r &=& \frac{V}{\sqrt{1-V^2}} \frac{\Delta^{1/2}}{r}, \\
u^{\phi} &=& \frac{r^2 \mathcal{L}}{A} + \gamma_L \omega \frac{A^{1/2}}{r \Delta^{1/2}}, \\
u^{\theta} &=& \frac{\gamma_L}{r} U \cos\theta,
\end{eqnarray}

\noindent where $A$ and $\Delta$ are same as defined earlier with $r$ replacing $R$, $\gamma_L$ is the Lorentz factor in LNRF near the equatorial plane, $\omega= 2 M a r/A$, $V$ is the radial velocity in the co-rotating frame, $U$ is the vertical velocity and $\mathcal{L}$ is the angular momentum per unit mass. The Lorentz factor near the equatorial plane is given by 

\begin{equation}
\gamma_L^2 = \frac{1}{1-V^2} + \frac{r^2 \mathcal{L}^2}{A}.
\end{equation}

\noindent By using a coordinate transformation, the four velocities in cylindrical coordinate near the equatorial plane is given by 

\begin{eqnarray}
u^t &=& \frac{\gamma_L A^{1/2}}{R \Delta^{1/2}}, \label{utu}\\
u^R &=& \frac{V}{\sqrt{1-V^2}} \frac{\Delta^{1/2}}{R}, \label{uru} \\
u^{\phi} &=& \frac{R^2 \mathcal{L}}{A} + \gamma_L \omega \frac{A^{1/2}}{R \Delta^{1/2}}, \label{upu} \\
u^z &=& \frac{z}{R} \frac{V}{\sqrt{1-V^2}} \frac{\Delta^{1/2}}{R} - \frac{\gamma_L U z}{R}, \label{uzu}
\end{eqnarray}

\noindent where $r^2 = R^2 + z^2 \simeq R^2$ near the equatorial plane. For a thin disc $z \ll R$ and assuming there is no significant vertical flow such that $U \simeq 0$, we have $u^z \simeq 0$. The covariant four velocities near the equatorial plane are then given by

\begin{eqnarray}
u_t &=& - \frac{\gamma_L R \Delta^{1/2}}{A^{1/2}} -\omega \mathcal{L}, \label{utl}\\
u_R &=&  \frac{R}{\Delta^{1/2}} \frac{V}{\sqrt{1-V^2}}, \label{url}\\
u_{\phi} &=& \mathcal{L}, \label{upl}\\
u_z &=& 0. \label{uzl}
\end{eqnarray} 

In the upcoming sections, we will derive the mass and momentum conservation equations in an equatorial plane using the four velocity given by equations( \ref{utu}, \ref{uru}, \ref{upu}, \ref{uzu}, \ref{utl}, \ref{url}, \ref{upl}, \ref{uzl}).

\subsection{Mass conservation}

The mass conservation is given by 

\begin{equation}
(\rho u^{i})_{; i} =0,
\end{equation}

\noindent and after the vertical integration, it reduces to 

\begin{equation}
\frac{\partial}{ \partial t}(\Sigma u^t) + \frac{1}{R} \frac{\partial}{\partial R} (R \Sigma u^R) = 0.
\label{massr}
\end{equation}

\subsection{Radial conservation equation}

The radial momentum conservation is given by 

\begin{equation}
T^{i R}_{; R} = 0.
\label{radeq}
\end{equation}

For a fluid where the internal energy and the pressure are less than the rest mass energy density of the fluid, $P \ll \rho c^2$, equation (\ref{radeq}) using four velocity results in 

\begin{multline}
\frac{\gamma_L A^{1/2}}{R^2 (1-V^2)^{3/2}} \frac{\partial V}{\partial t} + \frac{V}{(1-V^2)^2} \frac{\Delta}{R^2} \frac{\partial V}{\partial R} + \frac{V^2}{1-V^2} \frac{\Delta}{R^2} \frac{1}{\rho} \frac{\partial P}{\partial R} + \\ K(\mathcal{L},~R)=0,
\label{radlr}
\end{multline}

\noindent where 

\begin{eqnarray}
K(\mathcal{L},~R)&=& \frac{\gamma_L^2 M A}{R^6} \frac{(\Omega- \Omega_K^{+})(\Omega- \Omega_K^{-})}{\Omega_K^{+} \Omega_K^{-}}, \\
\Omega &=& \frac{u^{\phi}}{u^t} = \omega + \frac{R^3 \Delta^{1/2} \mathcal{L}}{\gamma_L A^{3/2}}, \\
\Omega_K^{\pm}&=& \pm \frac{M^{1/2}}{R^{3/2} \pm a M^{1/2}}. \label{omek}
\end{eqnarray}

\noindent $\Omega_K^{\pm}$ is the angular velocity of a circular orbit where the positive sign is for prograde orbits and negative sign for retrograde orbits.

\subsection{Angular momentum conservation}

The angular momentum conservation is given by 

\begin{equation}
\left(T^i_k \xi^k \right)_{; i}=0,
\label{azieq}
\end{equation}

\noindent where $\xi^k \equiv \delta_{(\phi)}^k$ is the azimuthal Killing vector. By solving equation (\ref{azieq}), using four velocity and after vertical integration, we obtain

\begin{equation}
\Sigma \left[ \frac{\gamma_L A^{1/2}}{R \Delta^{1/2}} \frac{\partial \mathcal{L}}{\partial t} + \frac{V}{\sqrt{1-V^2}} \frac{\Delta^{1/2}}{R} \frac{\partial \mathcal{L}}{\partial R} \right] + \frac{1}{R} \frac{\partial}{\partial R} (R \bar{S}_{\phi}^R) = 0,
\label{azir}
\end{equation}

\noindent where $\bar{S}_{\phi}^R$ is the vertically integrated viscous stress. The viscous stress tensor is given by $S_{\beta}^{\alpha}= -2 \eta  \sigma_{\beta}^{\alpha}$, where $\sigma_{\beta}^{\alpha}= g^{\alpha i} \sigma_{i \beta}$, and $\sigma_{i \beta}$ is given by

\begin{equation}
\sigma_{i \beta} = \frac{1}{2} \left[\frac{\partial u_i}{\partial x^{\beta}} + \frac{\partial u_\beta}{\partial x^{i}} - 2 \Gamma^{\lambda}_{i \beta} u_{\lambda} + u^{\nu} \left(u_i u_{\beta}\right)_{; \nu} \right]- \frac{1}{3} u^{\nu}_{; \nu} \left[g_{i \beta} + u_i u_{\beta}\right],
\label{sigeq}
\end{equation}

\noindent where $x$ represent coordinates and $\Gamma$ is the Christoffel symbol. We assume that the viscous stress is dominated along $\{R,~\phi\}$ direction and by solving equation (\ref{sigeq}), we obtain

\begin{multline}
\sigma_{\phi}^R = \frac{1}{2} \frac{\Delta^{1/2} A^{3/2} \gamma_L^3}{R^5} \frac{\partial \Omega}{\partial R} + \frac{V \sqrt{1-V^2} \gamma_L^4 A^2}{2 R^5 \Delta^{1/2}} \frac{\partial \Omega}{ \partial t} + \\ \frac{1}{2} \frac{\Delta}{R^2} \left[\frac{\gamma_L A^{1/2}}{\Delta} \frac{1+V^2}{(1-V^2)^{3/2}} \mathcal{L} \frac{\partial V}{\partial t} + \frac{2 V \mathcal{L} }{(1-V^2)^2} \frac{\partial V}{\partial R} + \frac{V^2}{1-V^2} \cdot \right. \\ \left. \left\{ \frac{2 \gamma_L a M (a^2+ 3 R^2)}{R \Delta^{1/2} A^{1/2}} + 2 (R^3 - a^3 M)\frac{\mathcal{L}}{A} \right\} \right] - \frac{1}{2} \frac{V}{\sqrt{1-V^2}} \mathcal{L} \left[ \frac{\partial V}{\partial t} \cdot \right. \\ \left. \left\{ \frac{A^{1/2}}{R^2 \gamma_L} \frac{V}{(1-V^2)^2} + \frac{V \mathcal{L}^2}{\gamma_L A^{1/2} (1-V^2)} \right\} +\frac{\gamma_L^2 \mathcal{L} A}{R^3 \Delta^{1/2}}(1-V^2) \frac{\partial \Omega}{\partial t} + \right. \\ \left. \frac{\Delta}{R^2} \frac{1}{(1-V^2)^{3/2}} \frac{\partial V}{\partial R}+ \frac{V}{\sqrt{1-V^2}} \frac{R-M}{R^2} \right].
\label{sigrp}
\end{multline}

\noindent For a subsonic flow such that angular velocity is smaller than the sound speed $\sim \sqrt{P/\rho}$, and assuming the radial flow and its evolution to be small and weak, the first term on the RHS of equation (\ref{sigrp}) dominates. Thus, we neglect the other terms and consider $\sigma_{\phi}^R$ given by

\begin{equation}
\sigma_{\phi}^R = \frac{1}{2} \frac{\Delta^{1/2} A^{3/2} \gamma_L^3}{R^5} \frac{\partial \Omega}{\partial R}.
\end{equation}  

\noindent The dynamic viscosity $\eta= \nu \rho$, where $\nu$ is the kinetic viscosity. The viscous stress is then given by $S_{\phi}^R = -2 \nu \rho \sigma_{\phi}^R$, and after vertical integration, it is given by 

\begin{equation}
\bar{S}_{\phi}^R = - \nu \Sigma \frac{\Delta^{1/2} A^{3/2} \gamma_L^3}{R^5 } \frac{\partial \Omega}{\partial R},
\label{sphir}
\end{equation}

\noindent which is the same as the viscous stress obtained by \citet{1994ASIC..417..341L} in Boyer-Lindquist coordinates for a thin disc. 

\subsection{Height of the disc}
\label{htd}

To calculate the half-thickness of the disc $H$, we solved the vertical momentum equation given by

\begin{equation}
\left(T_z^{i}\right)_{; i} = 0.
\end{equation}

\noindent Using the four velocity and restricting to first order in $z/R$, we get

\begin{equation}
\frac{\gamma_L^2}{\Delta A} \frac{z}{R^7} \Lambda_1 + \frac{2 \gamma_L \mathcal{L}}{\Delta^{1/2} A^{3/2}} \frac{z}{R^4} \Lambda_2 + \frac{\mathcal{L}^2}{A^2} \frac{z}{R} \Lambda_3 + \frac{1}{\rho} \frac{\partial P}{\partial z} =0,
\label{zeq}
\end{equation}

\noindent where 

\begin{eqnarray}
\Lambda_1 &=& M A^2 [4 a^2 + R (R- 4 M)] - 4M^2 a^2 R A [4 a^2 + R (3 R - 4M)] + \nonumber \\
&& 4 M^2 a^2 R^2 [4 a^4 M + 4 M R^4 - a^2 R (4 M^2 -5 M R + R^2)], \label{Lambda1} \\
\Lambda_2 &=& - a M A [4 a^2 + R (3 R - 4M)]+ 2 M a R [4 a^4 M +4 M R^4 - \nonumber \\
&& a^2 R (4 M^2-5 M R + R^2)], \label{Lambda2} \\
\Lambda_3 &=& 4 a^4 M +4 M R^4 -a^2 R (4 M^2-5 M R + R^2). \label{Lambda3}
\end{eqnarray}

Following the vertical integration of equation (\ref{zeq}), the height of the disc is given by

\begin{equation}
\left(\frac{H}{R}\right)^2 = \frac{P}{\rho} \frac{1}{R^2} \left[\frac{\gamma_L^2}{\Delta A} \frac{\Lambda_1}{R^7} + \frac{2 \gamma_L \mathcal{L}}{\Delta^{1/2} A^{3/2}} \frac{\Lambda_2}{R^4} + \frac{\mathcal{L}^2}{A^2} \frac{\Lambda_3}{R}\right]^{-1}. 
\label{height}
\end{equation}

\subsection{Conservation equations in dimensionless form}
\label{cmdf}

Now, we write the conservation equations in dimensionless form. The dimensionless quantities are

\begin{equation}
x = \frac{R}{R_g},~~j = \frac{a}{M},~~{\rm and}~~\ell = \frac{\mathcal{L}}{G M_{\bullet}/c}.
\end{equation} 

\noindent In terms of above variables, we obtain $A = R_g^4 A_k$ where $A_k = x^4 + x^2 j^2 +2 x j^2$, $\Delta= R_g^2 \Delta_k$ where $\Delta_k= x^2 -2 x +j^2$, $\Omega = [G M_{\bullet}/(c R_g^2)] \omega_1(x,~j,~V,~\ell)$ where

\begin{eqnarray}
\omega_1(x,~j,~V,~\ell) &=& \frac{2 j x}{A_k} + \frac{x^3 \Delta_k^{1/2}}{A_k^{3/2}} \frac{\ell}{\gamma_L}, \label{omega1}\\
\gamma_L^2 &=& \frac{1}{1-V^2} + \frac{x^2 \ell^2}{A_k}. \label{gammald}
\end{eqnarray}

The dimensionless angular velocity of a circular orbit using equation (\ref{omek}) is given by 

\begin{equation}
\omega_K^{\pm} = \pm \frac{1}{x^{3/2} \pm j}.
\end{equation}

The mass, radial and azimuthal conservations given by equations (\ref{massr}, \ref{radlr}, \ref{azir}) in terms of dimensionless variables are given by

\begin{eqnarray}
\frac{\partial \Sigma}{\partial t} &=& - \frac{\Sigma}{\gamma_L^2}\left[\frac{V}{(1-V^2)^2} \frac{\partial V}{\partial t} + \frac{x^2}{A_k} \ell \frac{\partial \ell}{\partial t} \right]- \frac{c}{R_g} \frac{\Delta_k^{1/2}}{A_k^{1/2}} \frac{1}{\gamma_L} \cdot \nonumber \\
&& \frac{\partial}{\partial x}\left(\Delta_k^{1/2} \Sigma \frac{V}{\sqrt{1-V^2}} \right), \label{masscd} \\
\frac{\partial V}{\partial t} &=& -\frac{c}{R_g} \frac{V}{\sqrt{1-V^2}} \frac{\Delta_k}{\gamma_L A_k^{1/2}} \frac{\partial V}{\partial x} -\frac{1}{R_g c}\frac{\Delta_k (1-V^2)^{3/2}}{\gamma_L A_k^{1/2}} \frac{1}{\rho}\frac{\partial P}{\partial x} + \nonumber \\
&& \frac{R_g}{c}\left(\frac{G M_{\bullet}}{c R_g^2}\right)^2 \gamma_L (1-V^2)^{3/2} \frac{A_k^{1/2}}{x^4} \cdot \nonumber \\
&& (x^3-j^2)(\omega_1-\omega_K^{+})(\omega_1-\omega_K^{-})\\
\frac{\partial \ell}{\partial t} &=& -\frac{c}{R_g} \frac{V}{\sqrt{1-V^2}}\frac{\Delta_k}{A_k^{1/2} \gamma_L} \frac{\partial \ell}{\partial x}- \frac{c}{G M_{\bullet} R_g} \frac{\Delta_k^{1/2}}{A_k^{1/2} \gamma_L} \cdot \nonumber \\
&& \frac{1}{\Sigma} \frac{\partial }{\partial x}\left(x \bar{S}_{\phi}^R\right), \label{angscd}
\end{eqnarray}

\noindent where $\bar{S}_{\phi}^R$ given by equation (\ref{sphir}) in dimensionless form is 

\begin{equation}
\bar{S}_{\phi}^R= -R_g \left(\frac{G M_{\bullet}}{c R_g^2}\right) \nu \Sigma \frac{\Delta_k^{1/2} A_k^{3/2} \gamma_L^3}{x^5} \frac{\partial \omega_1}{\partial x}.
\end{equation}

The height of the disc given by equation (\ref{height}) reduces to 

\begin{equation}
H = \left(\frac{R_g}{c}\right)^2 \frac{2 P }{\Sigma} \chi_1^{-1},
\label{heightd}
\end{equation}

\noindent where $\chi_1$ is given by

\begin{equation}
\chi_1 = \frac{\gamma_L^2}{\Delta_k A_k} \frac{\lambda_1}{x^7} + \frac{2 \gamma_L \ell}{\Delta_k^{1/2} A_k^{3/2}} \frac{\lambda_2}{x^4} + \frac{\ell^2 \lambda_3}{x A_k^2},
\end{equation}

\noindent and $\lambda_1,~\lambda_2$ and $\lambda_3$ are the dimensionless form of equations (\ref{Lambda1}, \ref{Lambda2}, \ref{Lambda3}) given by

\begin{eqnarray}
\lambda_1 &=& A_k^2 [4 j^2 + x(x-4)]- 4 j^2 x A_k [4 j^2 + x(3 x -4)] + \nonumber \\
 && 4 x^2 j^2 [4 j^4 + 4 x^4 - j^2 x (4 -5 x+ x^2)], \\
\lambda_2 &=& -j A_k [4 j^2 + x(3 x-4)]+ 2 j x [4 j^4 + 4 x^4 - \nonumber \\
&&  j^2 x (4 -5 x+ x^2)], \\
\lambda_3 &=& 4 j^4 + 4 x^4 - j^2 x (4 -5 x+ x^2).
\end{eqnarray}

In solving the vertical momentum equation, we have restricted to first order in $z/R$ which is generally applied to a thin disc theory. In such a disc, the thermal pressure is smaller than the rotational energy term and is neglected. Assuming the angular momentum in the disc is equal to the angular momentum of circular orbit $\ell_{\rm K}$ given by \citep{1972ApJ...178..347B}

\begin{equation}
\ell_{\rm K}= \pm \frac{x^2 \mp 2 j \sqrt{x} + j^2}{x^{3/4} (x^{3/2}-3 \sqrt{x} \pm 2 j)^{1/2}},
\end{equation}

\noindent where the upper sign is for the prograde orbits and the lower sign represent the retrograde orbits, the equation (\ref{angscd}) results in 

\begin{equation}
\Sigma V \sqrt{\Delta_k} = -\frac{1}{G M_{\bullet}} \left(\frac{\partial \ell_{\rm K}}{\partial x}\right)^{-1} \frac{\partial}{\partial x} \left(x \bar{S}_{\phi}^R\right).
\label{svdk}
\end{equation}

\noindent Substituting this in the mass conservation equation (\ref{masscd}), we get

\begin{equation}
\frac{\partial \Sigma}{\partial t} = \frac{c}{G M_{\bullet} R_g} \frac{\Delta_k^{1/2}}{A_k^{1/2}} \frac{1}{\gamma_L} \frac{\partial}{\partial x} \left[\left(\frac{\partial \ell_{\rm K}}{\partial x}\right)^{-1} \frac{\partial}{\partial x} \left(x \bar{S}_{\phi}^R\right)\right],
\end{equation}

\noindent which has a form of the diffusion equation.


\bsp	
\label{lastpage}
\end{document}